\documentclass[twocolumn,aps,prb,floatfix,superscriptaddress,showpacs]{revtex4}
\usepackage{graphicx}
\usepackage{dcolumn}
\usepackage{bm}
\usepackage{amsfonts}
\usepackage{amssymb}
\usepackage{amsmath}

\usepackage{epsfig}

\DeclareMathAlphabet{\mathitb}{OT1}{cmr}{bx}{sl}

\begin{document}
\pagenumbering{arabic}
  \title{Screening properties and plasmons of Hg(Cd)Te quantum wells}
  \author{Stefan~Juergens}
  \affiliation{Institute of Theoretical Physics and Astrophysics, University of W\"urzburg, D-97074 W\"urzburg, Germany}
  \author{Paolo~Michetti}
   \affiliation{Institute of Theoretical Physics, TU Dresden, D-01062 Dresden, Germany}
  \author{Bj\"orn~Trauzettel}
  \affiliation{Institute of Theoretical Physics and Astrophysics, University of W\"urzburg, D-97074 W\"urzburg, Germany}

  \date{\today}

  \begin{abstract}
Under certain conditions, Hg(Cd)Te quantum wells (QWs) are known to realize a time-reversal symmetric, two-dimensional topological insulator phase.
Its low-energy excitations are well-described by the phenomenological Bernevig-Hughes-Zhang (BHZ) model that interpolates between Schr{\"o}dinger and Dirac fermion physics.
We study the polarization function of this model in random phase approximation (RPA) in the intrinsic limit and at finite doping.
While the polarization properties in RPA of Dirac and Schr{\"o}dinger particles are two comprehensively studied problems, our analysis of the BHZ model bridges the gap between these two limits, shedding light on systems with intermediate properties.
We gain insight into the screening properties of the system and on its characteristic plasma oscillations. 
Interestingly, we discover two different kinds of plasmons that are related to the presence of intra- and interband excitations. 
Observable signatures of these plasmons are carefully analyzed in a variety of distinct parameter regimes, including the experimentally relevant ones for Hg(Cd)Te QWs.
We conclude that the discovered plasmons are observable by Raman or electron loss spectroscopy.
\end{abstract}

 \maketitle

  \section{Introduction}

Topological insulators (TIs) are amongst the most actively investigated systems in condensed matter physics \cite{hasan2010,qi2011,budich2013}. 
In reality, there is evidence for their existence in two \cite{konig2007} and three \cite{hsieh2008} spatial dimensions. Due to bulk-boundary 
correspondence, non-trivial topological states of matter have edge states at their boundaries with peculiar transport and optical properties. 
For instance, the two-dimensional (2D), time-reversal symmetric quantum spin Hall state -- that is realized in Hg(Cd)Te quantum wells (QWs) -- is known 
to come along with helical edge states that are protected against elastic backscattering of non-magnetic impurities \cite{xu2006,wu2006}. However, 
not only the edge state physics of these systems is interesting but also the 2D bulk physics bears exciting novelties. 
The low-energy excitations of Hg(Cd)Te QWs are described by a model -- the Bernevig-Hughes-Zhang (BHZ) model \cite{bernevig2006} -- that interpolates between the limiting cases of Schr{\"o}dinger 
and Dirac fermions.
This interplay between Schr{\"o}dinger and Dirac physics constitutes an opportunity for new phenomena to emerge. 
We have, for instance, recently discovered collective charge excitations at zero doping, i.e. 
intrinsic plasmons, in this system which are absent in both separate limits \cite{juergens2014}.

In this article, we complement our study of the screening properties and the collective charge excitations of Hg(Cd)Te QWs on the basis of random phase approximation (RPA),
and hence present a comprehensive analysis of its polarization function in the static and full dynamic limit, at zero and finite doping. 
Continuously tuning the parameters of the BHZ model, we reproduce the limits of pure Dirac and pure Schr{\"o}dinger fermions and explore intermediate regimes, in order to understand how 
analogies and differences emerge.
We support our numerical calculations of the polarization functions with analytical expressions derived by f-sum rules.
In the static limit, we calculate the screening properties due to the intrinsic system and at finite doping, analyzing the induced charge density (with Friedel oscillations) in response to a charged impurity. 
%
Different to the Dirac fermion system graphene, where static screening in the intrinsic limit is momentum independent and can therefore be absorbed into an 
effective dielectric constant \cite{wunsch2006,hwang2007}, the BHZ model shows a significant momentum dependence that translates into a finite 
extent of the induced charge density. 
In the dynamic limit, we are particularly interested in a better understanding of the plasmon excitations of this system away from zero doping where we previously 
found a new plasmon due to the interplay between Schr{\"o}dinger and Dirac fermion physics~\cite{juergens2014}. 
At finite doping, under certain conditions specified below that are e.g. applicable to Hg(Cd)Te QWs, 
we find a coexistence between this novel (interband) plasmon and an ordinary (intraband) plasmon. 
Both plasmons can be rather weakly 
damped by single-particle excitations and should therefore be observable. 
Interestingly, the two plasmons respond to the topology of the bandstructure with a distinctive behavior.
They seem to merge one into the other in a normal insulating phase, 
while they remain clearly resolved when the system realizes a topological insulator.

Generally, RPA is known to provide reliable predictions at large densities and in systems with a large number of fermionic degrees of freedom.
While its validity was indeed questioned for the intrinsic Dirac limit, where the system is unable to screen the Coulomb interaction and strong renormalization effects are expected~\cite{kotov2012},
RPA has been shown to yield a quantitative description of many-body effects in graphene~\cite{barnes2014,hofmann2014}. 
It has been widely used for the study of plasmons in the Dirac model, including various forms of (multilayer) graphene and TI surface states, see Ref.~\onlinecite{stauber2014} for a comprehensive review. Closely related to our work, the intraband plasmons of black phosphorous have been studied on the basis of RPA and an extended version of the BHZ model including anisotropy~\cite{low2014}. A similar study has been done for MoS$_2$~\cite{scholz2013}. 

Our article is organized as follows. 
In Sec.~\ref{sec_model}, we introduce the BHZ model and present the general formalism we employ to calculate the static and dynamical dielectric function and the induced charge density. 
The nature of the nontrivial pseudospin, the origin of possible interband plasmons, experimentally relevant parameters and the different contributions to the f-sum rule are also discussed here. 
Subsequently, in Sec.~\ref{sec_undoped}, we present the static screening properties, the dynamical excitation spectrum (new interband plasmon) and the f-sum rule in the undoped 
regime. 
Here we revisit and go beyond the results from Ref.~\onlinecite{juergens2014}.
In Sec.~\ref{sec_doped}, this analysis is extended to the case of finite doping where inter- and intraband excitations equally matter. 
We begin by discussing the ability of the BHZ model to interpolate between Dirac and Schr\"odinger physics. 
Afterwards, we have a closer look at parameters which are experimentally relevant for Hg(Cd)Te QWs, see Sec.~\ref{sec: Hg(Cd)Te quantum wells}. 
In this limit, we find a coexistence of inter- and intraband plasmons occuring for energies and momenta which are suitable for Raman or electron loss spectroscopy. 
We close this chapter by investigating the influence of a non-trivial topology on the plasmonic excitation spectrum. 
Finally, in Sec.~\ref{sec_con}, a conclusion and a brief outlook are given.

\section{Model and Formalism} \label{sec_model}
  The BHZ Hamiltonian~\cite{bernevig2006} for a two-dimensional electron gas (2DEG) near the $\Gamma$-point
has the form
  \begin{align}
    H = & \left(
      \begin{array}{cc}
	h\left(\boldsymbol{k}\right) & 0\\
	0 & h^{*}\left(\boldsymbol{-k}\right)
      \end{array}\right),
   \label{eq:Hamiltonian}\\
    h\left(\boldsymbol{k}\right) =& \ V(k) + \boldsymbol{d}_{\boldsymbol{k}}\cdot\vec{\sigma},\nonumber \\
    \boldsymbol{d}_{\boldsymbol{k}} =& \left(\begin{array}{ccc}
    Ak_{x}, & Ak_{y}, & M\left(k\right)\end{array}\right). \nonumber
  \end{align}
 Here $\vec{\sigma}$ are the Pauli matrices associated with the
band-pseudospin degree of freedom (band $E_{1}$ and $H_{1}$ in
Hg(Cd)Te quantum wells (QWs)), $V(k)=C-Dk^{2}$, $M(k)=M-Bk^{2}$ with $B,D<0$.  The
system possesses time-reversal symmetry and $H$ is block diagonal
in the Kramer's partner or spin degree of freedom. Restricting
ourselves to the block $h\left(\boldsymbol{k}\right)$,
the results can be extended to the other one by applying the time
reversal operator. $h\left(\boldsymbol{k}\right)$ describes fermions
with intermediate properties between a Dirac and a conventional 2DEG
system. The off-diagonal term ($A$ parameter) is typical for a Dirac
system ($A \mathrel{\widehat{=}\hbar v_f}$ in graphene), with $M$ the Dirac mass (corresponding to a gap of $2\left|M\right|$).
We consider positive and negative masses, where the latter one corresponds
to an inversion of the bandstructure and the system is topologically
non-trivial~\cite{bernevig2006}. For simplicity, we restrict ourselves to a bandstructure
 with a minimum at the $\Gamma$-point, which limits the mass to $M>-\frac{1}{2}\frac{A^{2}}{|B|}$.
In analogy to a 2DEG, the diagonal elements bear kinetic
energy elements which preserve ($B$ parameter)  and break ($D$ parameter)
particle-hole (p-h) symmetry ($-B\mp D\mathrel{\widehat{=}} \frac{1}{2m}$ for Schr\"odinger fermions with $m$ the quasi-particle mass).

The eigenstates of Eq.~(\ref{eq:Hamiltonian})
are described by the following dispersion and pseudospin
	\begin{eqnarray}
    E_{k,\lambda} = & V(k) + \lambda~|\boldsymbol{d}_{\boldsymbol{k}}| \; ,
    \label{eq:Dispersion}\\
    \lambda \hat{\boldsymbol{d}}_{\boldsymbol{k}} = & \langle\boldsymbol{k}, \lambda|~\vec{\sigma}~|\boldsymbol{k}, \lambda\rangle
  \end{eqnarray}
 with $\lambda=\pm$ for valence and conduction band. Note that
we consider electrons to be perfectly localized on the 2D X-Y plane
and therefore we neglect the real shapes of the envelope functions
due to the quantum confinement along $Z$ direction~\cite{michetti2012}.

\subsection{Energy and momentum scales}

The BHZ model is characterized by intrinsic scales for momentum, $q_{0}=\frac{A}{\left|B\right|}$, and energy, $E_{0}=Aq_{0}$, which reflect the interpolating character
of the model between Dirac ($A$ parameter) and Schr\"odinger ($B$ parameter) system. 
Fermi momentum $k_{f}$ and chemical potential $\mu$ provide externally tunable momentum and energy scales, 
which we call Fermi scales in the following. 
We expect the ratio between Fermi and intrinsic scales to govern the physics of this system. 
We therefore define the dimensionless quantities
\begin{eqnarray}
\boldsymbol{X}& =& \frac{\boldsymbol{q}}{q_{0}},  \ \Omega = \frac{\omega}{E_{0}},  X_{f} = \frac{k_{f}}{q_{0}},\\
\Omega_{f} & =& \frac{\mu}{E_{0}},\ \xi_{M} = \frac{M}{E_{0}},\ \xi_{D} = \frac{D}{\left|B\right|}\label{eq:dim-parameters} \nonumber
\end{eqnarray}
 where we set $\hbar=1$ in the following. $\Omega_{f}$ is defined to be the energy to the wave vector $X_{f}$, such that $\Omega_{f}=\pm\left|\xi_{M}\right|$ if $X_{f}=0$.
 For $X\sim1$, we therefore expect intermediate physics, while in the limit $X,\Omega\rightarrow0$ ($X,\Omega\rightarrow\infty$)
the Dirac (2DEG) physics should be recovered.

\subsection{Polarization function}

The linear response of an homogeneous system to an external applied
potential is described by the density-density generalized susceptibility
or retarded polarization function $\Pi^{R}\left(\boldsymbol{q},\omega\right)$.
This response comprises two main phenomena: screening, described by
the real part $\Re\left[\Pi^{R}\left(\boldsymbol{q},\omega\right)\right]$, and
dissipation by single-particle excitations (SPEs), given by the imaginary
part $\Im\left[\Pi^{R}\left(\boldsymbol{q},\omega\right)\right]$.

The polarization function in RPA yields the expression
\begin{equation}
\Pi^{R}\left(X,\Omega\right)=\frac{g_{s}}{\left|B\right|}\underset{\lambda,\lambda'}{\sum}\int\frac{d^{2}\tilde{X}}{4\pi^{2}}
\mathcal{F}^{\lambda,\lambda'}_{\boldsymbol{\tilde{X}},\boldsymbol{\tilde{X}'}}\frac{f\left(\epsilon_{\tilde{X},\lambda}\right)-f
\left(\epsilon_{\tilde{X}',\lambda'}\right)}{\Omega+i{\rm 0^{+}+\epsilon_{\tilde{X},\lambda}-\epsilon_{\tilde{X}',\lambda'}}},
\label{eq:Pi_par}
\end{equation}
 with $\boldsymbol{\tilde{X}'}=\boldsymbol{\tilde{X}}+\boldsymbol{X}$,
$0^{+}$ a positive infinitesimal, $g_{s}=2$ for spin degeneracy,
$\epsilon_{\tilde{X},\lambda}=E_{q_{0}\tilde{X},\lambda}/E_{0}= -\xi_{D}X^2 + \lambda \sqrt{\left(\xi_{M}+X^2\right)^2+X^2}$ and
$f\left(\epsilon\right)=\frac{1}{e^{\bar{\beta}\left(\epsilon-\Omega_{f}\right)}+1}$
the Fermi-Dirac function with $\bar{\beta}=\frac{E_{0}}{k_{B}T}$ and $k_{B}$
the Boltzmann constant. In the following we will assume zero temperature, $T=0$. The overlap factor is given by
\begin{equation}
\mathcal{F}^{\lambda,\lambda'}_{\boldsymbol{X},\boldsymbol{X'}}=\left|\langle \boldsymbol k,\lambda|\boldsymbol{k'},\lambda'\rangle\right|^2=
\frac{1}{2}\left[1+\lambda\lambda'\hat{\boldsymbol{d}}_{q_{0}\boldsymbol{X}}\cdot\hat{\boldsymbol{d}}_{q_{0}\boldsymbol{X'}}\right].\label{eq:F Factor}\end{equation}
Eq.~($\ref{eq:Pi_par}$) implies that $\left|B\right|\Pi^{R}\left(X,\Omega\right)$
is only a function of the reduced dimensionless variables $X$ and
$\Omega$ and parametrically depends on $\xi_{M}$, $\xi_{D}$ and
$X_{f}$.


\subsection{Overlap factor}

In the massless Dirac limit ($B=M=0$), eigenspinors are characterized
by their helicity and consequently the overlap factor $\mathcal{F}^{\lambda,\lambda'}_{\boldsymbol{k},\boldsymbol{k'}}=
\frac{1}{2}\left(1+\lambda\lambda'\cos\theta\right)$
only depends on the angle $\theta$ between $\boldsymbol{k}$ and $\boldsymbol{k'}$.
It is strictly one (zero) for states with the same (opposite) helicity.

\begin{figure}
\includegraphics[width=4.2cm]{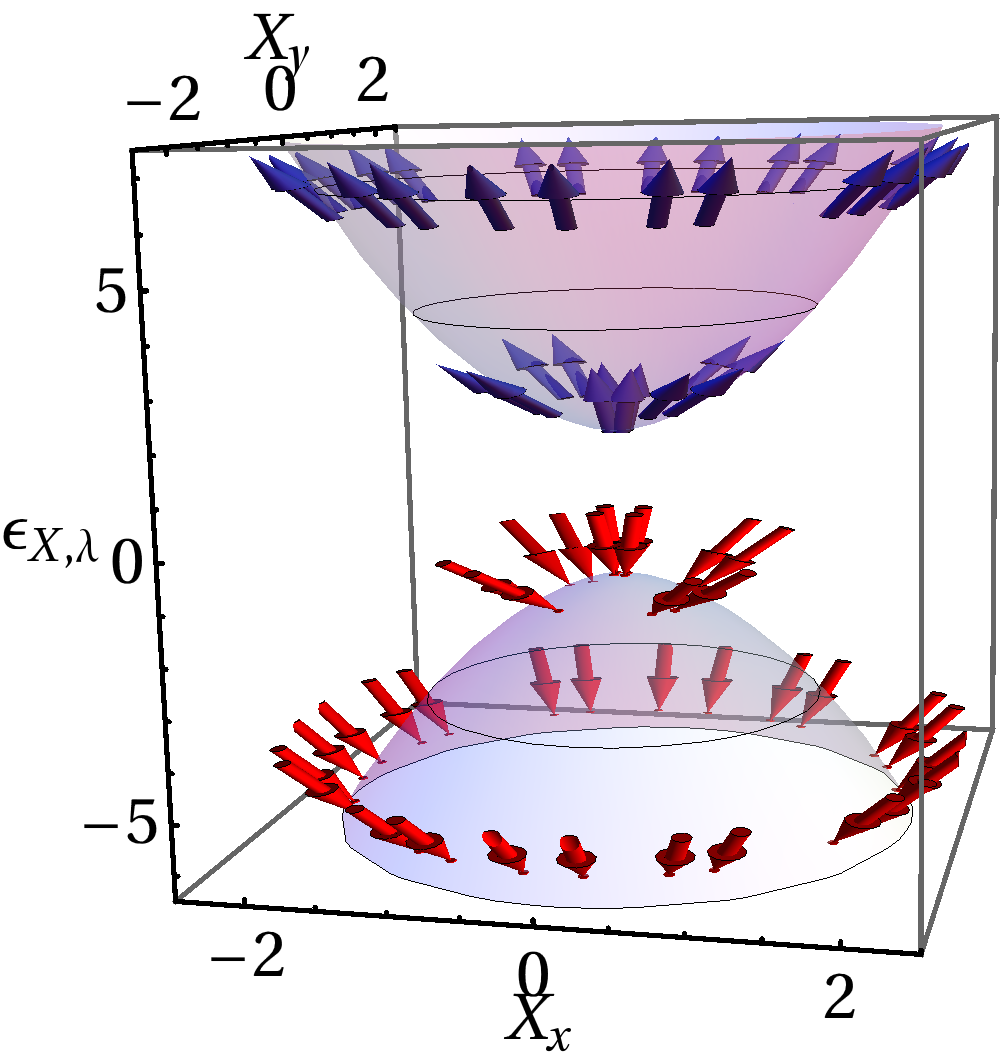}
\includegraphics[width=4.2cm]{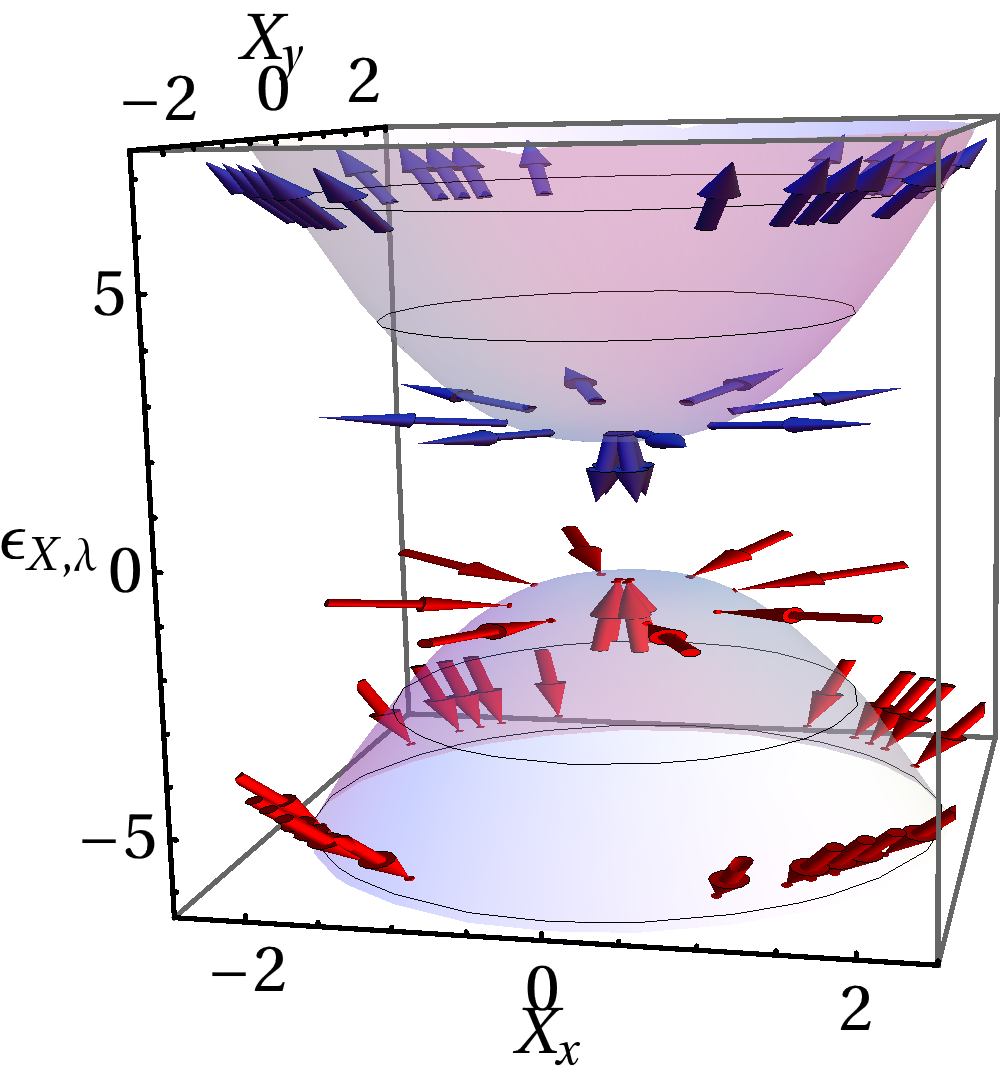}

\caption{(Color online) Dispersion relation and pseudospin of a NI (a), $\xi_{M}=\frac{4}{9}$,
and a TI phase (b), $\xi_{M}=-\frac{4}{9}$. The bands are separated
by an additional $2\epsilon_{X,\lambda}$ for better illustration
of the pseudospin. \label{fig: overlap}}

\end{figure}

In the BHZ model, the quadratic terms have the effect of turning the
pseudospin of the eigenstates out of plane in opposite directions
for conduction and valence bands at large $X$, see Fig.~\ref{fig: overlap}.
This results in a decay of the overlap factor down to $0$ in the limit of a conventional
2DEG system ($A\rightarrow0$ or $X\rightarrow\text{\ensuremath{\infty}}$).

A finite mass $\xi_{M}\neq0$ has a similar effect, but in the limit
of $X\leq\left|\xi_M\right|$. The pseudospin turns in the same (opposite) direction as for the quadratic term
for positive (negative) mass, see Fig.~\ref{fig: overlap}. 
This has the direct consequence that for a normal insulator (NI) phase the interband overlap factor
is reduced, while it is increased for a TI phase.
On the contrary, a positive (negative) mass enhances (diminishes) the intraband overlap factor. 
This picture is also confirmed in section~\ref{F-sum rule introduction} by calculating the f-sum rule.


\subsection{Coulomb interaction}


The bare Coulomb interaction $v\left(q\right)=\frac{e^{2}}{2\varepsilon_{0}q}$ in an electron gas is modified by 
screening into the effective interaction $v_{eff}\left(q,\omega\right)=\frac{v\left(q\right)}{\varepsilon\left(q,\omega\right)}$.
There, screening is described by the dynamical dielectric function. Employing dimensionless units, it acquires the form
\begin{eqnarray}
\frac{\varepsilon\left(X,\Omega\right)}{\varepsilon_{r}} &=& 1-\alpha g\left(X,\Omega\right),\label{eq:epsi}
\end{eqnarray}
where we have introduced the interaction strength parameter $\alpha$ (effective Dirac fine structure constant~\cite{kotov2011}) and the dimensionless function $g\left(X,\Omega\right)$ 
\begin{eqnarray}
\alpha &=& \frac{1}{A}\frac{e^{2}}{4\pi\varepsilon_{0}\varepsilon_{r}}\\
g\left(X,\Omega\right) &=& 2\pi\frac{\left|B\right|}{X}\Pi^{R}\left(X,\Omega\right).
\end{eqnarray}
In graphene one finds~\cite{grigorenko2012} $\alpha=2.2/\varepsilon_r$, while in Hg(Cd)Te QWs it is of
the order $\alpha\approx 4/\varepsilon_r$~\cite{buttner2011,schmidt2009}.
Here, $\varepsilon_{r}$ is the background dielectric constant, accounting
for screening of internal electronic shells, while $-\alpha g\left(X,\Omega\right)$
gives the dynamic screening due to electrons in the low energy bands. 
Zeros of $\varepsilon\left(X,\Omega\right)$ describe a density-density
(longitudinal) perturbation of the system that it is able to sustain
itself, which forms a collective mode called plasmon. It is defined by
\begin{equation}
\varepsilon\left(X,\Omega_{p}-i\Gamma\right)=0\label{eq:Plasmon Equation}\end{equation}
with the plasma frequency $\Omega_{p}$, and the finite imaginary
part $\Gamma=\frac{\gamma}{E_{0}}$
accounts for the possible damping due to single-particle excitations~\cite{Fetter}.

The dissipation of the interacting system, including both single-particle
excitation and the plasmon mode, is then described by the imaginary part of the interacting polarization
function $\Pi^{RPA}\left(X,\Omega\right)=\frac{\Pi^R\left(X,\Omega\right)}{\varepsilon\left(X,\Omega\right)}$. 
In order to compare to the non-interacting one, we will plot the normalized functions
\begin{eqnarray}
\Pi^{Im}_{rpa} \equiv \varepsilon_r \Im\left[\Pi^{RPA}\right], \ \Pi^{Im}\equiv \Im\left[\Pi^{R}\right], \ \Pi^{Re}\equiv \Re\left[\Pi^{R}\right] \nonumber
\end{eqnarray}
in the following, with $\varepsilon_r \Pi^{RPA} \underset{\alpha\rightarrow 0}{=} \Pi^{R}$.


  \subsubsection{(Anti-)Screening and intrinsic plasmons} \label{(Anti-)Screening and intrinsic plasmons}

In RPA, Eq.~(\ref{eq:epsi}) characterizes the screening of the interaction between two electrons exchanging momentum $X$ and energy $\Omega$, by the creation
of electron-hole pairs in the electron gas with the same momentum $X$. 
If these pairs are resonant in energy $\Omega_{eh}=\Omega$, they correspond to a physical
process leading to dissipation and a lowering of the Coulomb interaction - described by
the imaginary part of the polarization function, Eq.~(\ref{eq:Pi_par}). 
When $\Omega_{eh}\neq\Omega$, we have only virtual electron-hole pairs, which either still screen the interaction, if $\Pi^{Re}<0$, or
even enhance it (antiscreening effect), if $\Pi^{Re}>0$.
These effects depend on the energy of the created pair, for $\Omega_{eh}<\Omega$ one finds antiscreening, while $\Omega_{eh}>\Omega$ leads to a screening of the bare Coulomb interaction. 
This can be directly seen from the definition of the polarization function, Eq.~(\ref{eq:Pi_par}).
For every allowed excitation, the real part of the integrand in Eq.~(\ref{eq:Pi_par}) becomes
\begin{equation}
\mathcal{F}^{1,\lambda}_{\boldsymbol{\tilde{X}},\boldsymbol{\tilde{X}'}}\frac{2\Omega_{eh}\left[\boldsymbol{\tilde{X}},\boldsymbol{\tilde{X}'}\right]}{\Omega^2-\Omega_{eh}
\left[\boldsymbol{\tilde{X}},\boldsymbol{\tilde{X}'}\right]^{2}}, \label{eq:integrand}
\end{equation}
with $\lambda=1$ ($\lambda=-1$) for intraband (interband) excitations. Therefore every process with energy less than $\Omega$ increases
$\Pi^{Re}$, lowering $\varepsilon$ and thus increasing the interaction.

In the intrinsic Dirac system within RPA one finds $\Pi^{Re}=0$ for all energies $\Omega$ where
electron-hole excitations are allowed~\cite{stauber2014}. Thus the screening effect of virtual excitations with $\Omega_{eh}>\Omega$ cancels
exactly with the one from excitations with $\Omega_{eh}<\Omega$, such that the only screening comes from the resonant process $\Omega_{eh}=\Omega$. 
In the BHZ model, the high energy excitations become less likely as the electron and the hole band get decoupled for large $\Omega$. Additionally their excitation
energy is higher as in the Dirac case for the same momentum $X$, leading to an additional reduction of their
influence on $\Pi^{Re}$ due to the Lorentzian in Eq.~(\ref{eq:integrand}). Further, low energy
excitations become more important, as processes are allowed that where
forbidden in the Dirac system by helicity (see Sec. \ref{sec: BHZ model no mass no D} for details). Combining these effects, one finds the virtual
excitations which increase the Coulomb interaction, $\Omega_{eh}<\Omega$, dominating for larger frequency $\Omega$, leading to
an increased effective interaction and the possibility
of intrinsic plasmons in the BHZ model~\cite{juergens2014}.

More mathematically speaking, the described effects alter the high energy behaviour of $\Pi^{Im}$ from a
decay like $\Omega^{-1}$ in the Dirac case to a  $\Omega^{-2}$ decay in the BHZ model, as is shown in Sec. \ref{sec: Expansion of Pi}. 
Taking the Kramers-Kronig relation $\Pi^{Re}\left(X,\Omega\right)=\frac{1}{\pi}\underset{0}{\overset{\infty}{\int}}
d\Omega'\frac{2\Omega'}{\Omega'^2-\Omega^2}\Pi^{Im}\left(X,\Omega'\right)$ one finds directly that the real part of the polarization 
changes sign for $\Pi^{Im}\propto\Omega^{-2}$, but not for $\Pi^{Im}\propto\Omega^{-1}$. In more general terms, one can 
expect intrinsic interband plasmons to appear in all models for which $\Pi^{Im}$ decays faster as $\Omega^{-1}$ for high energies.


\subsection{Static limit and screening}

The static limit of the polarization function is obtained by sending $\Omega\rightarrow 0$ at finite momentum $X$ in Eq.~(\ref{eq:Pi_par}).
In this limit we can easily analyze the response of the system to the application of a static (or sufficiently slowly varying) external potential.
An important physical problem of this kind is the screening of a charged impurity by the electronic system.

The static polarization is a strictly real function, that we define as
\begin{equation}
\Pi(X) \equiv \Pi^R(X,0)= \Pi_0(X) + \Pi_\mu(X).  \label{eq:Pi_static}
\end{equation}
In a multiband system, like the BHZ model, it is useful to separate the contributions to the static polarization coming from the
intrinsic neutral system, $\Pi_0(X)$ (obtained for $\mu=0$), and the contribution due to a finite charge density, $\Pi_\mu(X)$ (finite $\mu$).
Consistently with the notation of Eq.~(\ref{eq:Pi_static}), the dielectric function, Eq.~(\ref{eq:epsi}), can therefore be rearranged into
\begin{eqnarray}
 \epsilon(X) \equiv \epsilon(X,0) = \epsilon_r \left[1 - \alpha g_0(X) -\alpha g_\mu(X) \right].
\end{eqnarray}

From the static dielectric constant we can find the induced charge density in response to a test charge $Ze$ placed at the origin.
The variation of the electronic charge density in momentum space corresponds to $Ze n(X)$,
where $n(X)$ is given by~\cite{Fetter}
\begin{eqnarray}
 n(X) &=& \frac{1}{\epsilon(X)}-1 = \frac{1}{\epsilon_r[1-\alpha g(X)]}-1= \\
 &=& n_r(X) + n_0(X) + n_\mu(X). \nonumber
\end{eqnarray}
Here the induced charge density can be seen as a sum of three contributions of different physical nature.
The first is due to the background polarization $n_r(X)$ (high energy polarization of the system),
the second to the intrinsic polarization $n_0(X)$ (polarization of the natural system) and the third to the polarization of the finite charge density in the system $n_\mu(X)$, with
\begin{eqnarray}
 n_r(X) &=& \frac{1}{\epsilon_r} -1  \label{eq:n_background}\\
 n_0(X) &=& \frac{1}{\epsilon_r}\frac{\alpha g_0(X)}{1 - \alpha g_0(X)}\\
 n_\mu(X) &=& \frac{1}{\epsilon_r}\frac{1}{1 - \alpha g_0(X)}\frac{\alpha g_\mu(X)}{1-\alpha g(X)}.
\end{eqnarray}

In real space, the density fluctuation (using physical dimensional units) is given by
$$ n(r) = \frac{1}{2 \pi}\int dq~ q~ J_0(q r)~ n(q),$$
with $J_0$ the zero-th order Bessel function.


\subsection{Experimental parameters} \label{sec: experimental_parameters}

  Including Coulomb interaction, we now have a 4-dimensional parameter space consisting of
  $\xi_{M}$, $\xi_{D}$, $X_{f}$ and $\alpha$.
  This parameter space will be explored systematically in the following.
  While the exploration of the different physical behaviors featured by the BHZ model 
  in different regions of this parameter space has a clear theoretical significance, 
  we want to stress that our discussion is also relevant for experiments.
  In particular, realistic parameters for Hg(Cd)Te QW structures~\cite{buttner2011,schmidt2009} are roughly $\xi_{D}\leq-0.5$,
  $q_{0}\approx0.4\ \frac{1}{\mathrm{nm}}$, $E_{0}\approx 140\ \mathrm{meV}$
  and masses $M$ with absolute values up to several meV. 
  The interaction strength is around $\alpha\approx4/\varepsilon_{r}\approx0.3$ 
  with an average $\varepsilon_{r}=15$ from the CdTe substrate ($\varepsilon_{r}=10$) and HgTe ($\varepsilon_{r}=20$).
  Considering the experimental acceptable damping rate for plasmons, we refer to experiments on
  the surface states of a 3D TI~\cite{pietro2013}. 
  There, plasmons with a ratio of $\frac{\Gamma}{\Omega_p}=0.5$ are perfectly resolvable.


\subsection{F-sum rule} \label{F-sum rule introduction}

The f-sum rule for the polarization function provides the total spectral weight of all excitations in the system.
It is identical for the interacting and noninteracting system, as the interaction conserves the number of particles.
%
Thus the sum rule is a powerful tool to check our numerics.
Additionally, it offers a deeper insight concerning the shift of spectral weight between the
inter and intra SPEs as well as the different plasmons in the system.


\subsubsection{Definition and calculation}

The f-sum rule is defined by~\cite{Nozieres}
\begin{equation}
-\frac{2}{\pi}\underset{0}{\overset{\infty}{\int}}d\omega\omega\Im\left[\Pi\left(\mathbf{q},\omega\right)\right]=g_s\left\langle 0\left|
\left[\left[n_{\mathbf{q}},H^0\right],n_{\mathbf{q}}^{\dagger}\right]\right|0\right\rangle\ \label{eq:f-sum def}
\end{equation}
with the density operator $n^\dag_{\mathbf{\mathbf{q}}}=\sum_\mathbf{k}\Psi^\dag_{\mathbf{k}+\mathbf{q}}\Psi_{\mathbf{k}}$ 
and the Hamiltonian $H^0 = \sum_\mathbf{k} \Psi^\dag_{\mathbf{k}} h\left(\boldsymbol{k}\right) \Psi_{\mathbf{k}}$ with 
$h\left(\boldsymbol{k}\right)$ as defined in Eq.~(\ref{eq:Hamiltonian}). 
$\Psi_{\mathbf{k}}$ is a spinor associated with the band-pseudospin degree of freedom (band $E_{1}$ and $H_{1}$ in
Hg(Cd)Te QWs). The spin degree of freedom enters via the degeneracy factor $g_s=2$.
For the calculation we follow the steps outlined in the appendix of Ref.~\onlinecite{sabio2008}, 
where the f-sum rule for the Dirac model is obtained. For the BHZ model the computational 
steps are the same, therefore we only present important intermediate results and differences to the Dirac limit. 
The commutator in Eq.~(\ref{eq:f-sum def}) is given by 
\begin{eqnarray}
\left[\left[n_{\mathbf{q}},H^0\right],n_{\mathbf{q}}^{\dagger}\right]&=&
\sum_\mathbf{k}\left(\Psi^\dag_{\mathbf{k}}H^0_{\mathbf{k},\mathbf{q}}\Psi_{\mathbf{k}}-\Psi^\dag_{\mathbf{k}+\mathbf{q}}H^0_{\mathbf{k}+\mathbf{q},\mathbf{q}}\Psi_{\mathbf{k}+\mathbf{q}}\right)\nonumber\\
&-&2q^2\sum_\mathbf{k}\Psi^\dag_{\mathbf{k}+\mathbf{q}}\left(D\sigma_0+B\sigma_z\right)\Psi_{\mathbf{k}+\mathbf{q}}\label{eq:f-sum commutator}
\end{eqnarray}
with $H^0_{\mathbf{k},\mathbf{q}}=A\mathbf{q}\cdot\boldsymbol{\sigma}-D\mathbf{q}\left(2\mathbf{k}+\mathbf{q}\right)\sigma_0-B\mathbf{q}\left(2\mathbf{k}+\mathbf{q}\right)\sigma_z$. A simple shift of the momentum sums in Eq.~(\ref{eq:f-sum commutator}) would put the first line to zero, but this is not allowed. In the same way as in the Dirac system, the operators are unbounded and one has to work with a large momentum cutoff $\kappa$. While in the Dirac limit one finds simply $H^0_{\mathbf{k},\mathbf{q}}=A\mathbf{q}\cdot\boldsymbol{\sigma}$ and the second line of Eq.~(\ref{eq:f-sum commutator}) would be zero,  now the latter gives rise to a contribution depending on the chemical potential, as one would expect for a 2DEG. The sums in Eq.~(\ref{eq:f-sum commutator}) are then converted into integrals and solved in the limit of large $\kappa$. Care has to been taken when converting the momentum cutoff $\kappa$ into the frequency cutoff $\lambda$, such that both integrals cover the same phase space. 


\subsubsection{Formulas}
For a pure Dirac system one find the f-sum rule~\cite{sabio2008}
\begin{equation}
\underset{0}{\overset{\lambda}{\int}}d\omega\omega\Im\left[\Pi\left(q,\omega\right)\right]=-\frac{g_s q^{2}\lambda}{16} \label{eq:dirac f-sum rule}
\end{equation}
where the cutoff $\lambda$ is needed as the Dirac spectrum is unbounded. In a 2DEG system one finds
\begin{equation}
\underset{0}{\overset{\infty}{\int}}d\omega\omega\Im\left[\Pi\left(q,\omega\right)\right]=\frac{g_s}{4} \left(B\pm D\right) k_{f}^{2}q^{2}=-\frac{\pi Nq^{2}}{2m} \label{eq:2DEG f-sum rule}
\end{equation}
with $N=\frac{g_s}{4\pi}k_{f}^{2}$ the electron density and $\Im\left[\Pi\left(q,\omega\right)\right]\neq0$ only over a finite range of $\omega$. 
Similar to a Dirac system, the BHZ spectrum is unbounded which complicates
the evaluation of the sum rule and makes it necessary to introduce
a high-energy cutoff $\Lambda=\frac{\lambda}{E_0}$. 
We find approximately for $\Lambda\gg1$
\begin{align}
f\sum &\equiv  -\underset{0}{\overset{\Lambda}{\int}}d\Omega\Omega\left|B\right|\Im\left[\Pi\left(X,\Omega\right)\right] \label{eq:BHZ f-sum rule}\\
=  &\frac{g_{s}}{8}X^{2}\Biggl[\ln\left(\frac{2\Lambda e^{-1-2\xi_{M}+2\left|\Omega_{f}\right|}}{1+2X_{f}^{2}\left(1+\gamma\xi_{D}\right)+2\xi_{M}+2\left|\Omega_{f}\right|}\right) \nonumber \\
  +&\frac{1-X^{2}+4\xi_{M}}{\Lambda}-\frac{2X^{4}+\left(1+4\xi_{M}\right)^{2}-4X^{2}\left(2+7\xi_{M}\right)}{4\Lambda^{2}}\Biggl] \nonumber \\
  +& \mathcal{O}\left(\frac{\xi_{D}}{\Lambda^{2}}\right)+\mathcal{O}\left(\frac{1}{\Lambda^{3}}\right) \nonumber \end{align}
with $\gamma=sgn\left[\Omega_{f}\right]$ and Euler's number $e$ , so the leading order term diverges
logarithmically with $\Lambda$. This is due to the fact that $\Im\left[\Pi\left(X,\Omega\right)\right]$
decays like $\Omega^{-2}$ for $\Omega\gg1$, and not as $\Omega^{-1}$ as for a Dirac
system. The sum rule is exact up to order $\Lambda^{-1}$ ($\Lambda^{-2}$) for finite (zero) $\xi_D$.

The f-sum rules for BHZ, Dirac and 2DEG models are always proportional to $q^2 \propto X^2$ in the leading order,
but otherwise distinct from one another. 
Taking the limit $A\rightarrow0$ in the BHZ result, Eq.~(\ref{eq:BHZ f-sum rule}),
gives the 2DEG case, Eq.~(\ref{eq:2DEG f-sum rule})~\footnote{Due to the details of the derivation of the analytical expansion in Eq.~(\ref{eq:BHZ f-sum rule}), 
the same is not possible for the limit $B\rightarrow0$, as there the defined cutoff $\Lambda=\lambda\frac{\left|B\right|}{A^2}$ would go to zero.}.


\subsubsection{Comparing different orders in cutoff $\Lambda$}

We begin our discussion of Eq.~(\ref{eq:BHZ f-sum rule}) by comparing the 
contributions from the different orders $\mathcal{O}\left(\ln\left(\Lambda\right)\right)$, 
$\mathcal{O}\left(\Lambda^{-1}\right)$ and $\mathcal{O}\left(\Lambda^{-2}\right)$. In the limit of $\xi_{M}=\xi_{D}=X_f=0$ 
we find $\frac{f\sum_{\mathcal{O}\left(\Lambda^{-1}\right)}}{f\sum_{\mathcal{O}\left(\ln\left(\Lambda\right)\right)}}=-\frac{X^2-1}{\Lambda\left(\ln\left(2\Lambda\right)-1\right)}$ 
and $\frac{f\sum_{\mathcal{O}\left(\Lambda^{-2}\right)}}{f\sum_{\mathcal{O}\left(\ln\left(\Lambda\right)\right)}}=-\frac{2X^4-8X^2+1}{4\Lambda^2\left(\ln\left(2\Lambda\right)-1\right)}$, 
thus the ratio $\frac{X^2}{\Lambda}$ determines the importance of higher order corrections for $X\gg1$. 
We take $\Lambda=2\left(\beta X\right)^{2}$ for the cutoff in the following. 
Already for $\beta=2$ and a maximal momentum $X=X_{max}=6$, the corrections of order $\mathcal{O}\left(\Lambda^{-1}\right)$ are 2\% of order $\mathcal{O}\left(\ln\left(\Lambda\right)\right)$, while contributions of order $\mathcal{O}\left(\Lambda^{-2}\right)$ are smaller than 0.1\%. 
A modest cutoff $2\le\beta\le5$ works best for comparing
Eq.~(\ref{eq:BHZ f-sum rule}) to numerical data, as the latter one is only given over a finite range of $\Omega$. A larger
 $\Lambda$ makes it necessary to extrapolate the data, providing a
source for errors.


\subsubsection{Influence of finite $\xi_M$, $\xi_D$ and $X_f$}

\begin{figure}
\includegraphics[width=4.2cm]{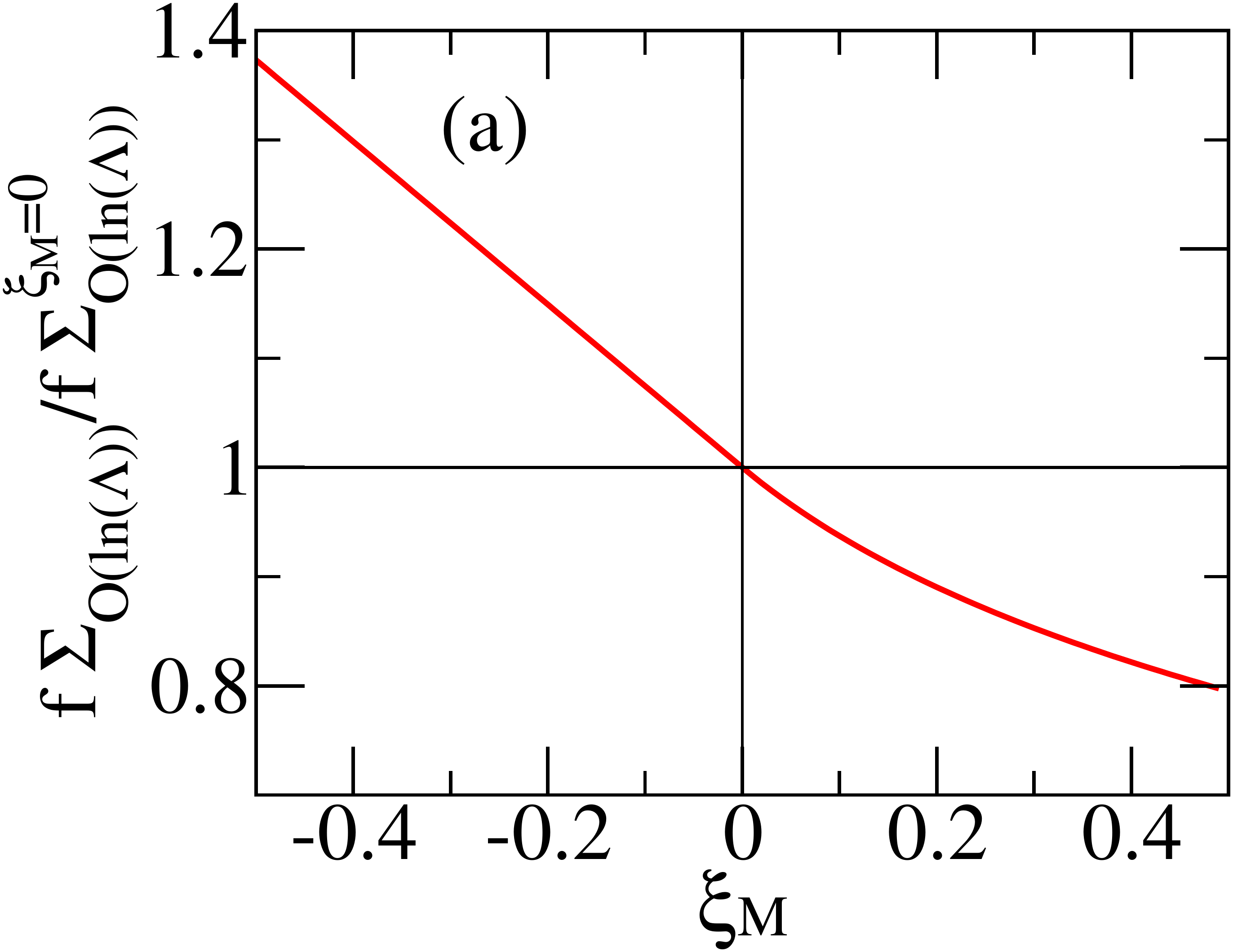}
\includegraphics[width=4.2cm]{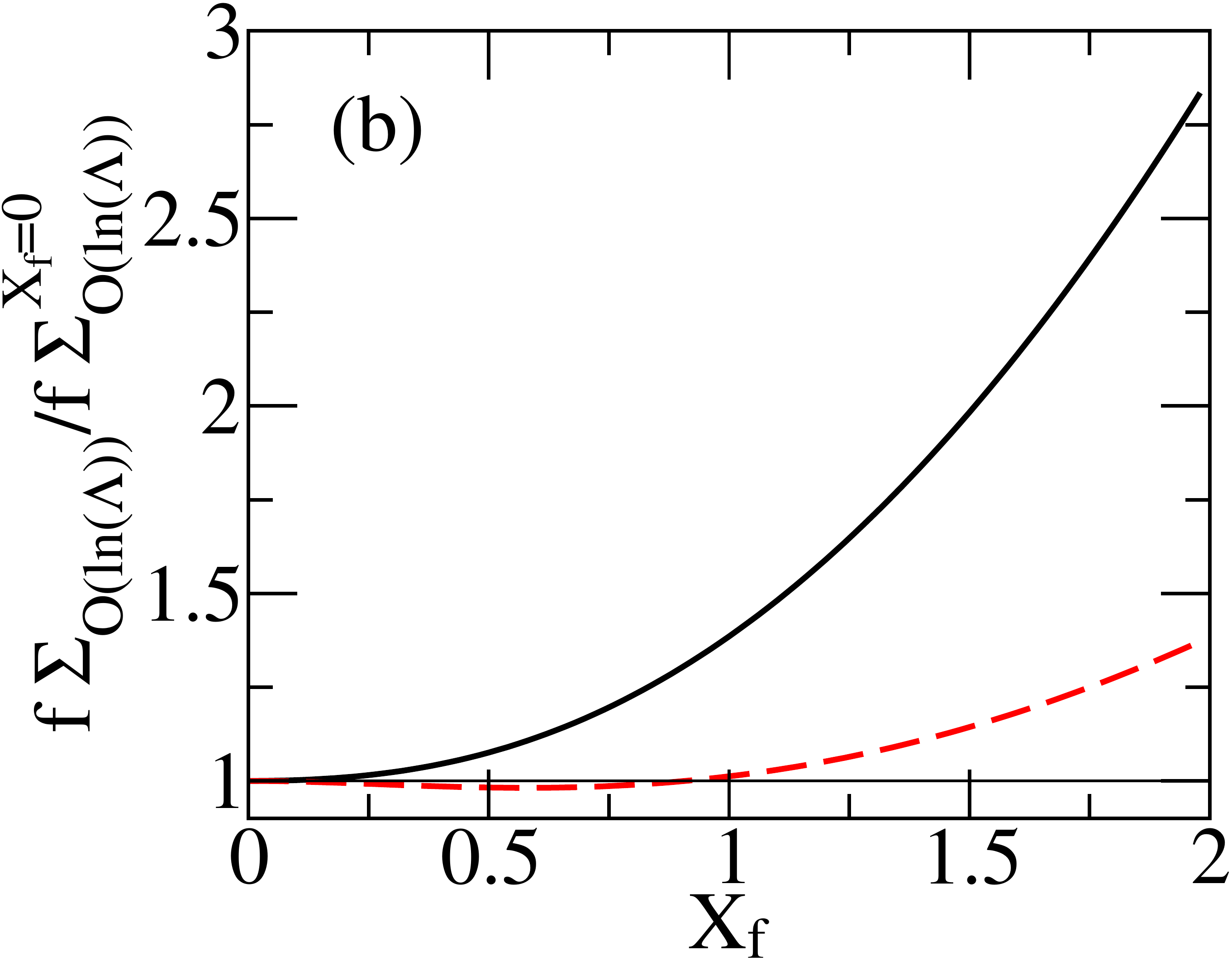}

\caption{(Color online) (a) Ratio $\frac{f\sum_{\mathcal{O}\left(\ln\left(\Lambda\right)\right)}}{f\sum_{\mathcal{O}\left(\ln\left(\Lambda\right)\right)}^{\xi M=0}}$
of the lowest order f-sum rule including mass over the one without
mass. $\xi_{D}=0$ and $X_{f}=0$. (b) Ratio $\frac{f\sum_{\mathcal{O}\left(\ln\left(\Lambda\right)\right)}}{f\sum_{\mathcal{O}\left(\ln\left(\Lambda\right)\right)}^{Xf=0}}$
of the lowest order f-sum rule including finite doping over the one
without doping, $\Omega_{f}>0$ ($\Omega_{f}<0$) as a black, solid (red, dashed)
line. $\xi_{M}=0$ and $\xi_{D}=-0.5$. $\beta=2$ and $X_{max}=6$ in both plots. \label{fig: f-sum rule}}

\end{figure}

Next, we investigate changes to the f-sum rule and therefore to the total spectral weight by varying the mass.
The influence of a finite mass is studied in Fig.~\ref{fig: f-sum rule}
(a) for $\xi_{D}=0$ and $X_{f}=0$. A positive mass lowers the f-sum
rule, while a negative mass increases it linearly. This is a direct consequence from the change of the overlap factor:
a negative mass enhances the coupling between the two bands, while a positive mass diminishes it, as in the latter case the pseudospins
do not match. It is also consistent with the increase in the optical conductivity observed in the undoped limit with negative mass~\cite{juergens2014}.

Last, we consider the effects of finite doping. 
It blocks interband transitions close to the Dirac point, but due to the small density of states, these transitions carry only a small spectral weight. 
On the other hand, doping enables intraband transitions, which carry a large spectral weight due the combined effects of larger overlap factor, 
density of states and smaller excitation energies compared to interband transitions. 
Therefore, a finite doping usually increases the f-sum rule, as seen in Fig.~\ref{fig: f-sum rule} (b), 
where we plot $\frac{f\sum_{\mathcal{O}\left(\ln\left(\Lambda\right)\right)}}{f\sum_{\mathcal{O}\left(\ln\left(\Lambda\right)\right)}^{Xf=0}}$
for positive (black, solid line) and negative (red, dashed line) doping with $\xi_{D}=-0.5$ and $X_{M}=0$.
A finite $\xi_{D}$ adds a term $\pm\frac{g_{s}}{4}\left|\xi_D\right|X_f^2X^{2}$ to the leading order of the f-sum rule, $+$ ($-$)
for positive (negative) doping. It can be seen as an increased (decreased) contribution
from the 2DEG part of the spectrum, Eq.~(\ref{eq:2DEG f-sum rule}), and leads to the slight decrease of the f-sum rule for negative doping in panel (b).

\subsubsection{Comparing the spectral weight of excitations} \label{F-sum rule interpretation exc spectrum}

In order to compare the importance of different excitations in the system, 
one should compare their spectral weight and thus their contribution to the f-sum rule. 
The latter has the benefit of being independent of the Coulomb interaction 
strength and the position of the excitation peaks, in contrast to the polarization function $\Pi^{Im}_{rpa}$. 
As an example, we assume that the excitation spectrum, $\Pi^{Im}_{rpa}$, is governed by a single plasmonic peak following a Lorentzian shape with width $\Gamma$ and peak height $\frac{1}{c\Gamma}$. 
Then the f-sum rule is proportional to $\underset{0}{\overset{\infty}{\int}}d\Omega\Omega\frac{1}{c}\frac{\Gamma}{\Gamma^2+\left(\Omega-\Omega_p\right)^2}=\frac{\Omega_p}{c}\underset{0}{\overset{\infty}{\int}}\frac{d\Omega}{\Omega_p}\frac{\Omega}{\Omega_p}\frac{\frac{\Gamma}{\Omega_p}}{\left(\frac{\Gamma}{\Omega_p}\right)^2+\left(\frac{\Omega}{\Omega_p}-1\right)^2}$. The value of this integral should be independent of $\alpha$ and thus of $\Omega_p$. Therefore we find $c\propto\Omega_p$, such that the peak height of a resonance in $\Pi^{Im}_{rpa}$ naturally has to scale with $1/\Omega_p$ to fulfill the f-sum rule.

We conclude that the importance of a resonance in $\Pi^{Im}_{rpa}$ should be judged by its spectral weight, which can 
be estimated by multiplying the peak height with its position $\Omega_p$. The relevant width of the peak is given 
by $\frac{\Gamma}{\Omega_p}$, with $\Gamma$ being the width of the resonance in $\Pi^{Im}_{rpa}$.


\section{Undoped system} \label{sec_undoped}

In this section, we focus on an intrinsic (undoped limit $\mu=0$) BHZ model system.
First, we analyze the static polarization function and the static screening properties.
Then we consider the long wavelength limit of the dynamical polarization function, providing an analytical expansion.
Finally, we add some complementary arguments elucidating the origin of the new interband plasmon (absent both in the Dirac and 2DEG cases), whose 
appearence for the intrinsic BHZ model has been proposed in Ref.~\onlinecite{juergens2014}.

\subsection{Static limit}
In order to set a reference with a closely related and analytically solvable model, we discuss the static intrinsic polarization
for a massive Dirac limit, given by~\cite{kotov2008,pyatkovskiy2009}
\begin{eqnarray}
  \Pi_0(q) = \frac{-g q}{8\pi A} \hspace{-0.05cm}\left[ \frac{1}{\chi}\hspace{-0.05cm} +\hspace{-0.05cm} 
  \left(\hspace{-0.05cm} 1\hspace{-0.05cm}-\hspace{-0.05cm}\frac{1}{\chi^2}\hspace{-0.1cm}\right)\arctan{\chi} \right]
      \overset{M\rightarrow0}{\longrightarrow} \frac{-g q}{16 A},
  \label{eq:Pi_dirac_static}
\end{eqnarray}
where the index $0$ stands for intrinsic limit $\mu=0$, $g$ account for possible spin and band degeneracy, and $\chi=\frac{Aq}{2M}$.
When the Dirac system is massless ($M=0$), $\Pi_0(q)$ is a linear function of the momentum $q$.
A finite Dirac mass suppresses the polarization for $q\lesssim M/A$, where $\Pi_0(q)$ shows a super-linear behavior.
For $q\gg M/A$, the mass is negligible instead and the result of the massless limit is reproduced.

\begin{figure}
 \centering
 \includegraphics[width=8cm]{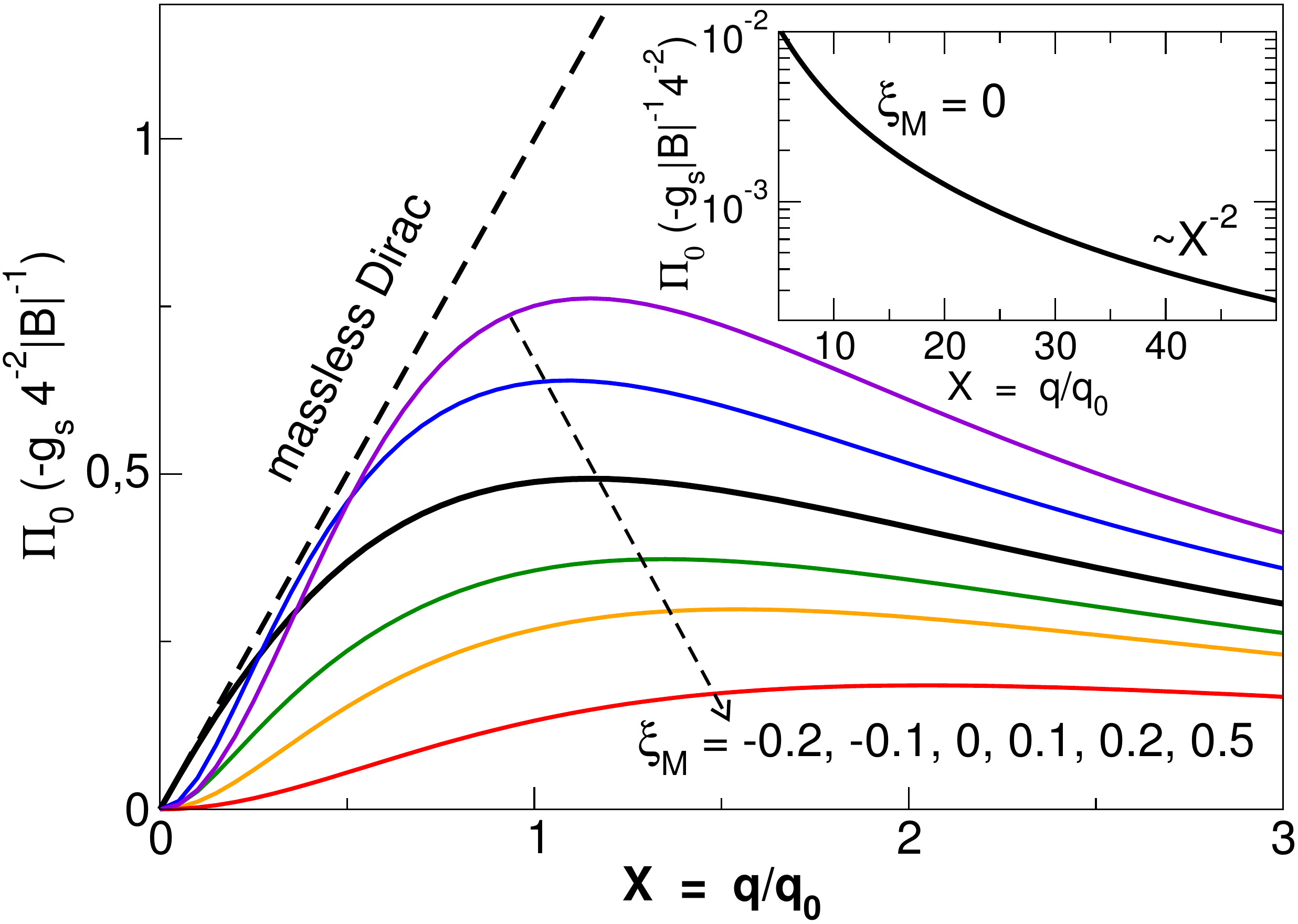}
 \caption{(Color online) Static intrinsic polarization function of the BHZ model for variable $\xi_M$ and $\xi_D=0$.}
 \label{fig:Pi_0_static}
\end{figure}

The static polarization function of the BHZ model is simply obtained by direct numerical
evaluation of Eq.~(\ref{eq:Pi_par}) at zero frequency.
In Fig.~\ref{fig:Pi_0_static}, we show $\Pi_0(X)$ calculated for a particle-hole symmetric BHZ system ($\xi_D=0$).
Note that we obtain the massless Dirac case in the limit $B \rightarrow 0$ (and therefore $X\rightarrow0$), where $\lim_{X\rightarrow0} |B|\Pi_0(X)/X = -\frac{g_s}{16}$.
A finite $B$ parameter determines a fundamental qualitative change with respect to a Dirac system.
Indeed, $\Pi_0(X)$ reaches a maximum at $X\approx 1$ and then decays as $1/X^{2}$ for $x\gg1$
as shown in the inset of Fig.~\ref{fig:Pi_0_static}.
A finite and positive Dirac mass $M$ leads to a general suppression of the polarization function with respect to the massless case.
In the region $X<1$ (where quadratic terms are less important), $\Pi_0(X)$ resembles the massive Dirac case, with a
super-linear increase in the region $X\lesssim\xi_M$, due to the suppression of the interband overlap
factor determining a reduction of the polarization at small momentum.
For intermediate values  $\xi_M\lesssim X\lesssim 1$, analogously to the massive Dirac limit,
$\Pi_0(X)$ is approximatively linear in $X$.
Considering larger momenta $X\gtrsim 1$, the behavior is dominated by the quadratic terms and the polarization eventually vanish for $X\rightarrow\infty$.
In general, the interplay of quadratic terms and a finite Dirac mass shifts the maximum of $\Pi_0(X)$.
When the Dirac mass $M$ is negative (topological insulator phase), we observe a less pronounced suppression of the polarization for $X<\xi_M$,
with respect to a massive Dirac system (normal insulator) with equal modulus of $M$.
Moreover, on the contrary to the $M>0$ case, $\Pi_0(X)$ is enhanced at large $X$ with respect to the massless, particle-hole symmetric limit.
This behavior is due to the enhanced overlap factor between electron and hole bands in the topological insulator phase.

\begin{figure}
 \centering
 \includegraphics[width=8cm]{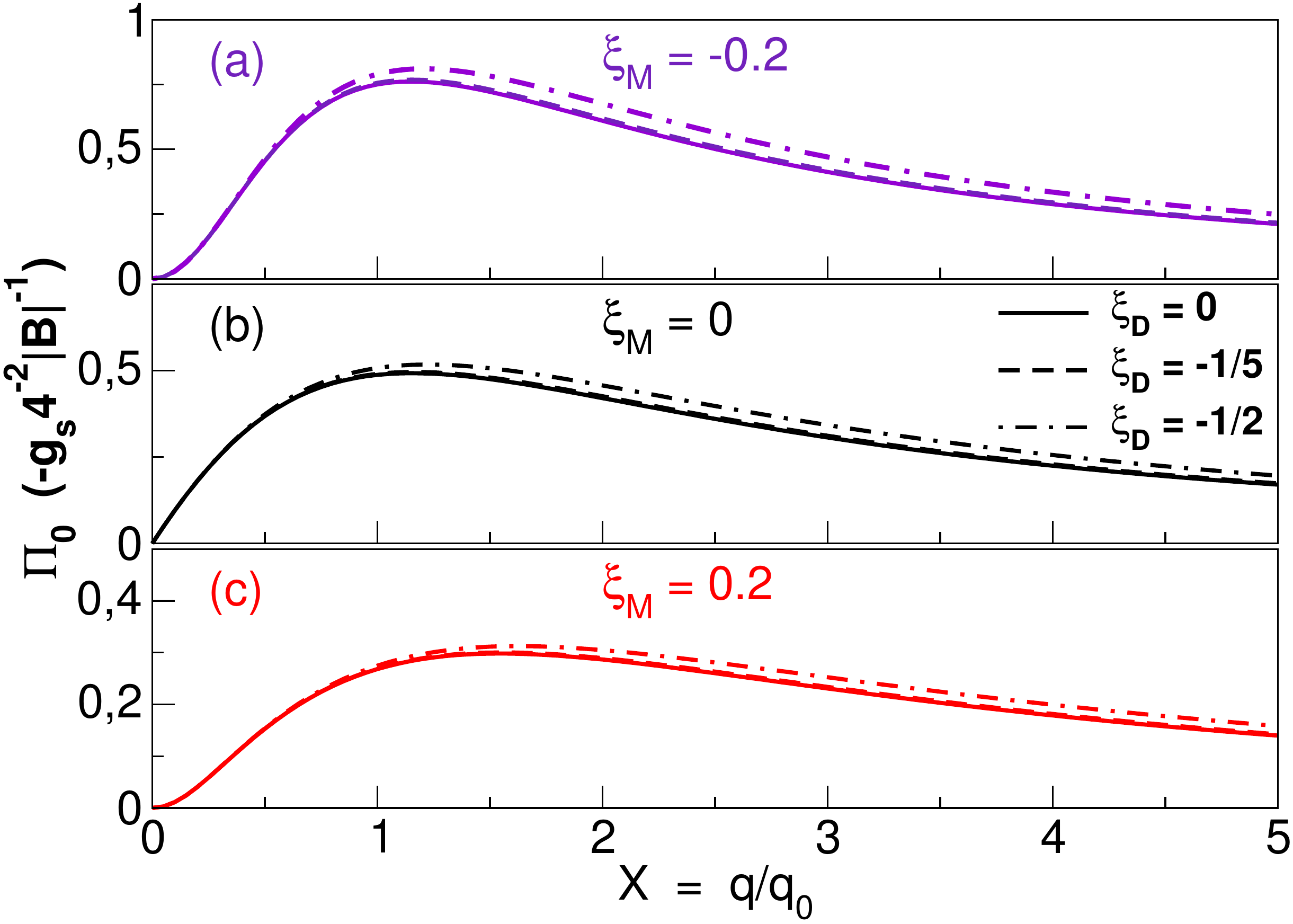}
 \caption{(Color online) Static intrinsic polarization function of the BHZ model for a finite $\xi_D$ value.}
 \label{fig:D}
\end{figure}

In Fig.~\ref{fig:D}(a-c),
we analyze the effects of a finite value of the parameter $\xi_D$ in the BHZ model,
for $\xi_M=-0.2$, $0$ and $0.2$.
A finite $\xi_D$ breaks particle-hole symmetry by changing the effective masses of conduction and valence bands.
We only found quantitative changes to $\Pi_0(X)$, which is progressively reduced for increasing $\xi_D$.
%


\subsection{Screening} \label{sec:screening_int}
In a massless Dirac system, where the static polarization is linear in $q$ [Eq.~(\ref{eq:Pi_dirac_static})], the dielectric function is a constant
\begin{eqnarray}
 \varepsilon(q)= \varepsilon_r \left(1+\frac{g_s g_v\pi}{8}\alpha\right)\equiv\varepsilon, \label{constant screening dirac}
\end{eqnarray}
therefore the intrinsic polarization contribution can be absorbed into an effective background dielectric constant $\epsilon$.
As a consequence, a test charge $Z e$, placed at the origin,
induces a screening electronic density
$$Z e \left[ n_0(\mathitb q)+n_r \right]= Z e \left( \frac{1-\varepsilon}{\varepsilon}\right),$$
which in real space corresponds to a screening image charge [a fraction $(1-\varepsilon)/\varepsilon$ of the external one] placed exactly at the same position
\begin{eqnarray}
n(\mathitb r) = -\left( \frac{1-\varepsilon}{\varepsilon}\right) \delta(\mathitb r). \label{eq:charge_Dirac}
\end{eqnarray}
Note that the screening charge only due to the electronic system (without background contribution) is a fraction $-(\epsilon - \epsilon_r )/ \epsilon \epsilon_r$ of the external one.

In a massive Dirac system, the large $q$ behavior of $\Pi_0(q)$ reproduces the massless limit and therefore a screening charge
given by Eq.~(\ref{eq:charge_Dirac}) is also developed at vanishing distances $r$ in response to an external test charge.
However, in the long wave length limit ($q<M/A$) $\Pi_0(q)$ has a superlinear behavior and thus $n_0(0)\propto \lim_{q\rightarrow0}\Pi_0(q)/q = 0$.
Thus an induced charge density of the same sign as the external charge is developed at finite distances~\cite{kotov2008} [summing up to $Ze(\epsilon - \epsilon_r )/ \epsilon \epsilon_r$],
so that the test charge feels only the background screening over long distances, as expected in an insulator.

For the BHZ model, we find similar to Eq.~(\ref{constant screening dirac})
\begin{eqnarray}
 \lim_{X\rightarrow0}\varepsilon(X)= \varepsilon_r \left(1+\frac{g_s \pi}{8}\alpha\right) \label{constant screening BHZ}
\end{eqnarray}
in the long wavelength limit, but $\lim_{X\rightarrow\infty}\varepsilon(X)=\varepsilon_r$.
In order to understand this, we discuss next the induced charge density in real space for the BHZ model. It is given by
\begin{eqnarray}
  n_0(r) &=& \eta_0 \int dX J_0(X r q_0)~\frac{|B|~\Pi_0(X)}{1 - \alpha g_0(X)}
  \label{eq:n_0_Y}
\end{eqnarray}
with $\eta_0=\frac{\alpha}{\epsilon_r} q_0^2$ a natural charge density constant of the model.
We note that $n_0(r)$ is proportional to $q_0^2$ and $\alpha$, but $n_0(r)$ has an additional dependence on $\alpha$ (and thus on $A$) through its integrand.
It also parametrically depends on $\xi_M$ and $\xi_D$ through $\Pi_0(X)$ and $g_0(X)$.

\begin{figure}
 \centering
 \includegraphics[width=8cm]{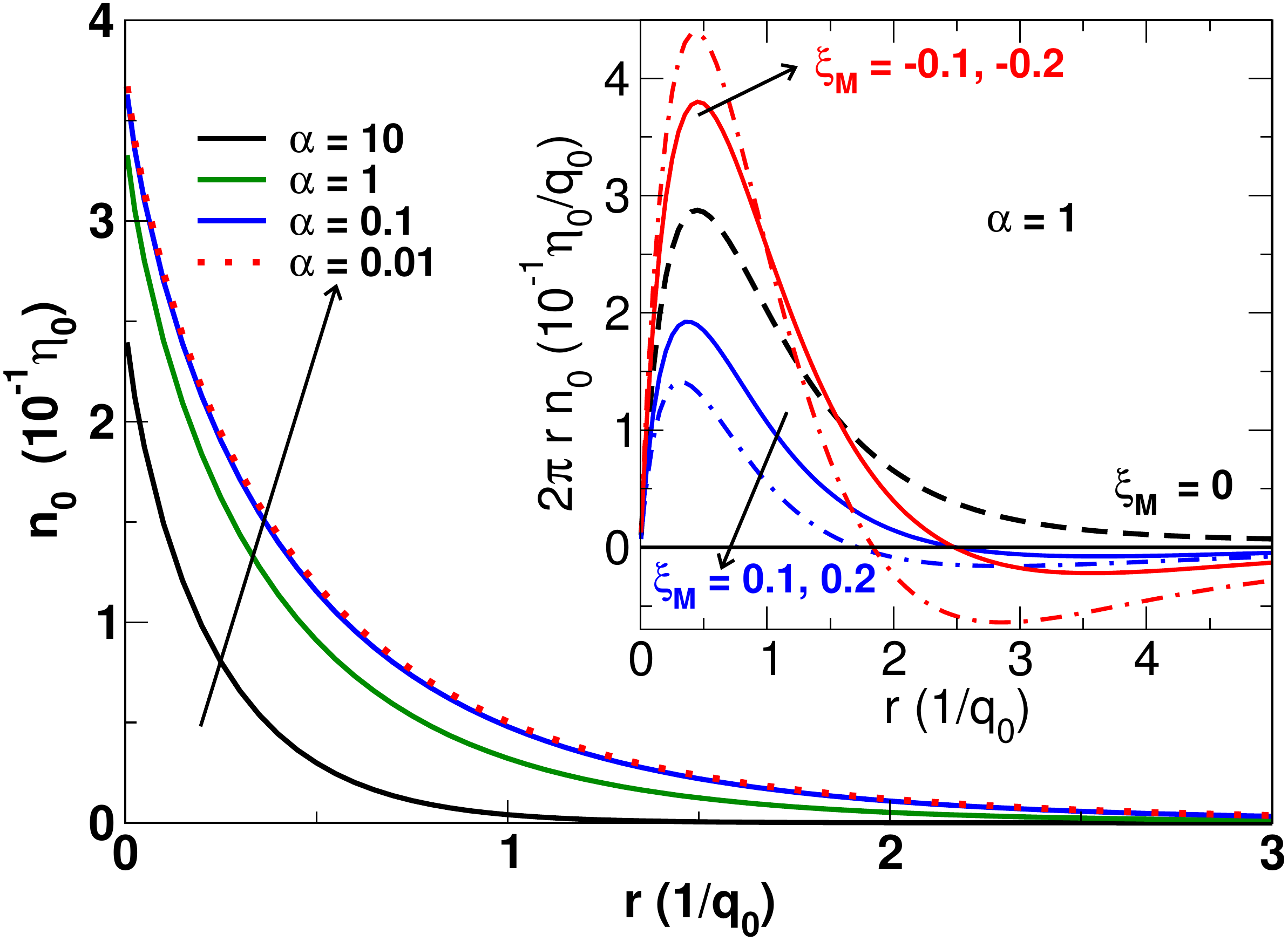}
 \caption{(Color online) Induced charge $n_0(r)$ due to a test charge in the intrinsic limit of the BHZ model for $\xi_D=\xi_M=0$.
 The plot is invariant under a change of $B$ parameter and only depend on the effective fine structure constant $\alpha$.
 In the inset, $r \ n_0(r)$ calculated for $\alpha=1$ and finite Dirac mass.}
 \label{fig:n_int}
\end{figure}

In Fig.~\ref{fig:n_int},
we plot the induced charge density $n_0(r)$ in real space for  $\xi_D=\xi_M=0$ with different values of $\alpha$.
Opposite to a Dirac system, the induced charge density has a finite extent over a distance of the order of $1/q_0$, which is clearly related to the decay of $\Pi_0$ at large
wavevector due to the presence of quadratic $B$ terms.
$n_0(r)$ decays at large distances as $r^{-2}$.
An electron far away from this induced charge, $r\gg1/q_0$, does not see the finite extent of it and is therefore screened in the
same way as in the Dirac system, leading to the similarity of Eqs. (\ref{constant screening dirac}) and (\ref{constant screening BHZ}).
In the opposite limit where the electron sits on top of the induced charge, $r\ll1/q_0$, it does not feel it at all, resulting in no screening besides $\varepsilon_r$.

In the inset of Fig.~\ref{fig:n_int}, we study the effect of a finite Dirac mass term.
With a finite $\xi_M$, the induced density (as in the case of pure Dirac systems) shows a qualitatively different behavior.
$n_0(r)$ changes sign for sufficiently large $r$, ensuring a vanishing total induced charge.
From a quantitative point of view, a finite negative (positive) $M$ enhances (suppresses) the features of $n_0(r)$,
due to its effect on the interband overlap factor.


\subsection{Long wavelength expansion}  \label{sec: Expansion of Pi}

An analytic discussion of the polarization function is only possible in the limit
 $X\rightarrow0$. Here, we focus on the limit of vanishing mass $\xi_M\rightarrow0$ to extract an analytic formula of the plasmon dispersion.
 An expansion of $\Pi^{R}$ in $X$ gives, for $\Omega>X$
\begin{align}
\Pi^{R}= & -\frac{g_s}{8\left|B\right|}X^{2} \Big( 2\frac{\Omega-\sqrt{1+\Omega^{2}}\mathrm{ArcSinh}\left(\Omega\right)}{\pi\Omega^{3}} \label{eq:Pi undoped expanded}
\\ + & i\frac{1}{\Omega\left(1+\sqrt{1+\Omega^{2}}\right)} \Big)+\mathcal{O}\left(X^{4}  \right)\nonumber  \\
\underset{\Omega\gg1}{=} & -\frac{g_s}{8\left|B\right|}\frac{X^{2}}{\Omega^{2}}\left(2\frac{1-\ln\left(2\right)-\ln\left(\Omega\right)}{\pi}+i\right)\nonumber \\+&
 \mathcal{O}\left(\Omega^{-3}\right)+\mathcal{O}\left(X^{4}\right) \nonumber \end{align}
 where one finds an $\Omega^{-2}$ behaviour with an additional logarithmic
correction for the real part in the high frequency limit.

Calculating the plasmon dispersion by performing an expansion of Eq.~(\ref{eq:Plasmon Equation}) up to second order in $\frac{\Gamma}{\Omega}$, one finds the
linear dispersion
\begin{align}
\Omega_{p}=\frac{1}{8}\pi g\alpha X+\mathcal{O}\left(X^{2}\right) \label{eq:Plasmonfrequency linear}
\end{align}
%
 which is only valid for sufficiently large $\alpha$, such that the conditions $\Re\left[\Pi^{R}\left(X,\Omega_{p}\right)\right]>0$
  and $\Omega>X$ are fulfilled. The linearity of the dispersion follows from Eq.~(\ref{eq:Pi undoped expanded}) only by inclusion of the damping via $\Gamma$. 
  Without the substitution $\Omega\rightarrow\Omega-i\Gamma$, $\Re\left[\varepsilon\left(X,\Omega_p\right)\right]=0$ has no sensible solution for $\Omega_p$.
  The damping ratio is given by
  \begin{equation}
\frac{\Gamma}{\Omega} \underset{\Omega\ll1}=1-\frac{\Omega^2}{8}+\mathcal{O}\left(\Omega^3\right)
 \end{equation}
underlining the importance of damping in this limit. The plasmon is only well defined 
for a finite $\Omega_p>\Omega_{c}$, with $\frac{\Gamma}{\Omega}\mid_{\Omega=\Omega_{c}}\lesssim c$ 
where $0<c<1$ sets the limit for the detectability of the plasmons, for example in the recent experiment~\cite{pietro2013} $c$ was shown to be of the order $0.5$. Eq.~(\ref{eq:Plasmonfrequency linear}) translates this into a finite momentum 
scale $q>\frac{q_0}{g\alpha}\frac{8\Omega_{c}}{\pi}$ with the intrinsic plasmon 
length scale $l_0=\frac{g\alpha}{q_0}$, given by the Coulomb interaction strength times the charge decay length $\frac{1}{q_0}$, see Sec. \ref{sec:screening_int}. 
We interpret $l_0$ as the length scale up to which charge separation due to Coulomb interaction can 
occur and give rise to the interband plasmons, in an undoped and therefore overall neutral system. 

In the opposite limit of high frequencies, the term $\ln\left(\Omega\right)$
spoils a simple $\sqrt{X}$ behaviour of the plasmon dispersion. In this limit, we can extract the analytic form of the damping rate
\begin{align}
\frac{\Gamma}{\Omega} & \underset{\Omega\gg1}{=}\frac{1}{3\pi}\left(-\ln\left(\frac{4\Omega^{2}}{e^{3}}\right)+\sqrt{3\pi^{2}+
\ln\left(\frac{4\Omega^{2}}{e^{3}}\right)^{2}}\right)+\mathcal{O}\left(\frac{1}{\Omega}\right) \label{eq:Damping expansion}\\
 & \underset{\Omega\rightarrow\infty} {=}\frac{\pi}{2\ln\left(\frac{4\Omega^{2}}{e^{3}}\right)} \nonumber
 \end{align}
 with Euler's number $e$, yet the plasmon dispersion can only be calculated numerically.

 In the following discussion of the different excitation spectra, we will use these analytic results to check our numerics in the limits of small momenta and low and high frequencies.


\subsection{Excitation spectrum}

\begin{figure}
\includegraphics[width=4.2cm]{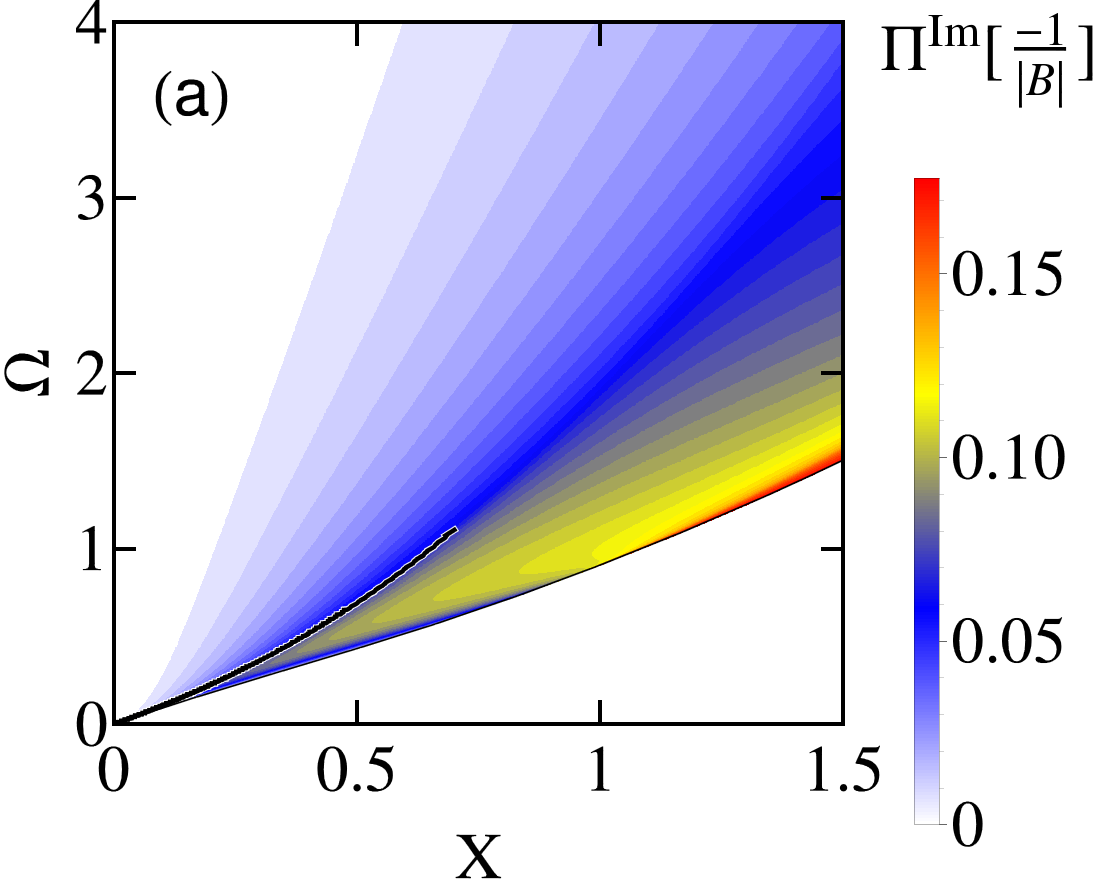}
\includegraphics[width=4.2cm]{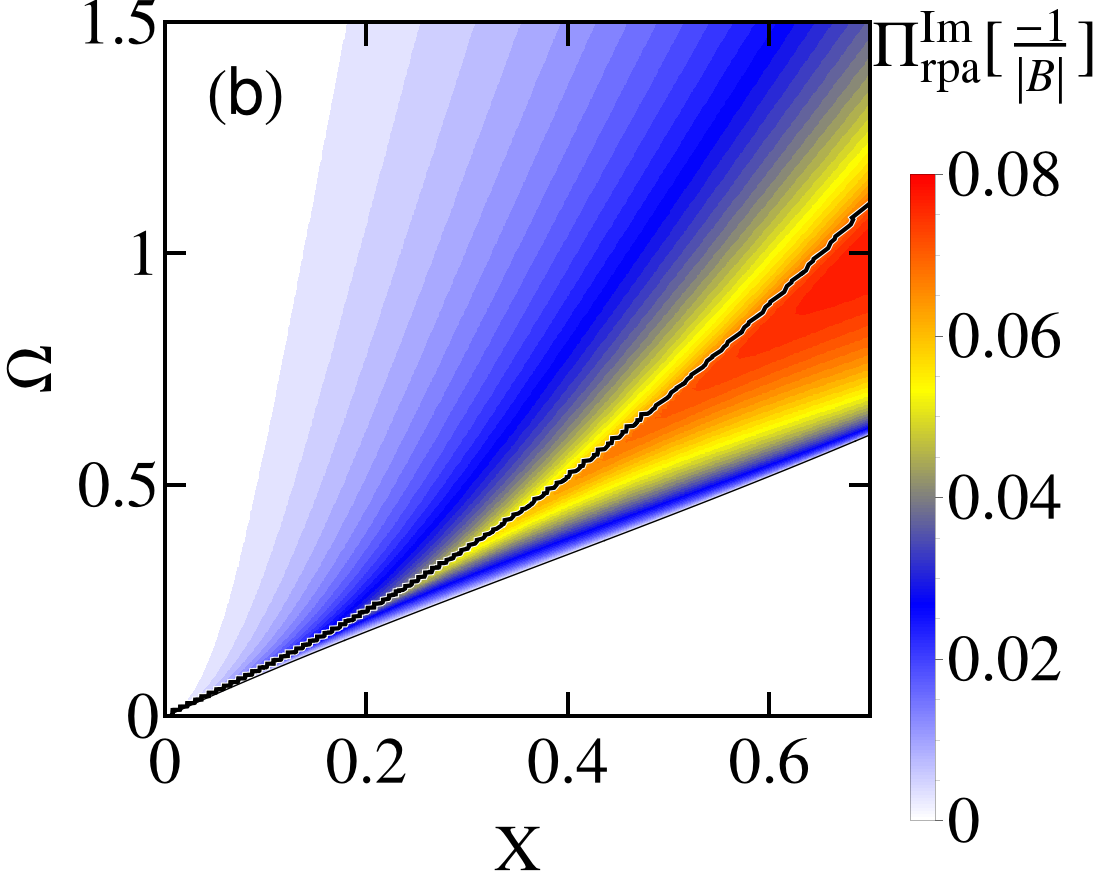}
\includegraphics[width=4.2cm]{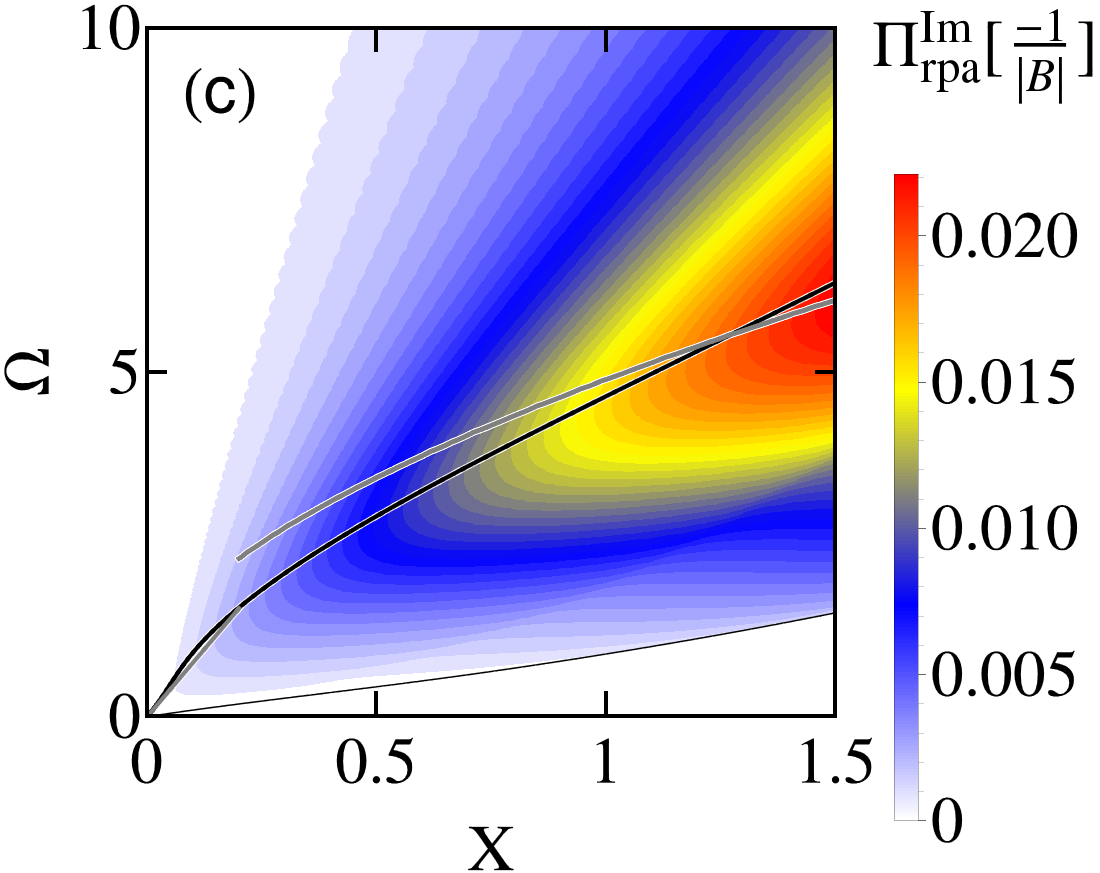}
\includegraphics[width=4.2cm]{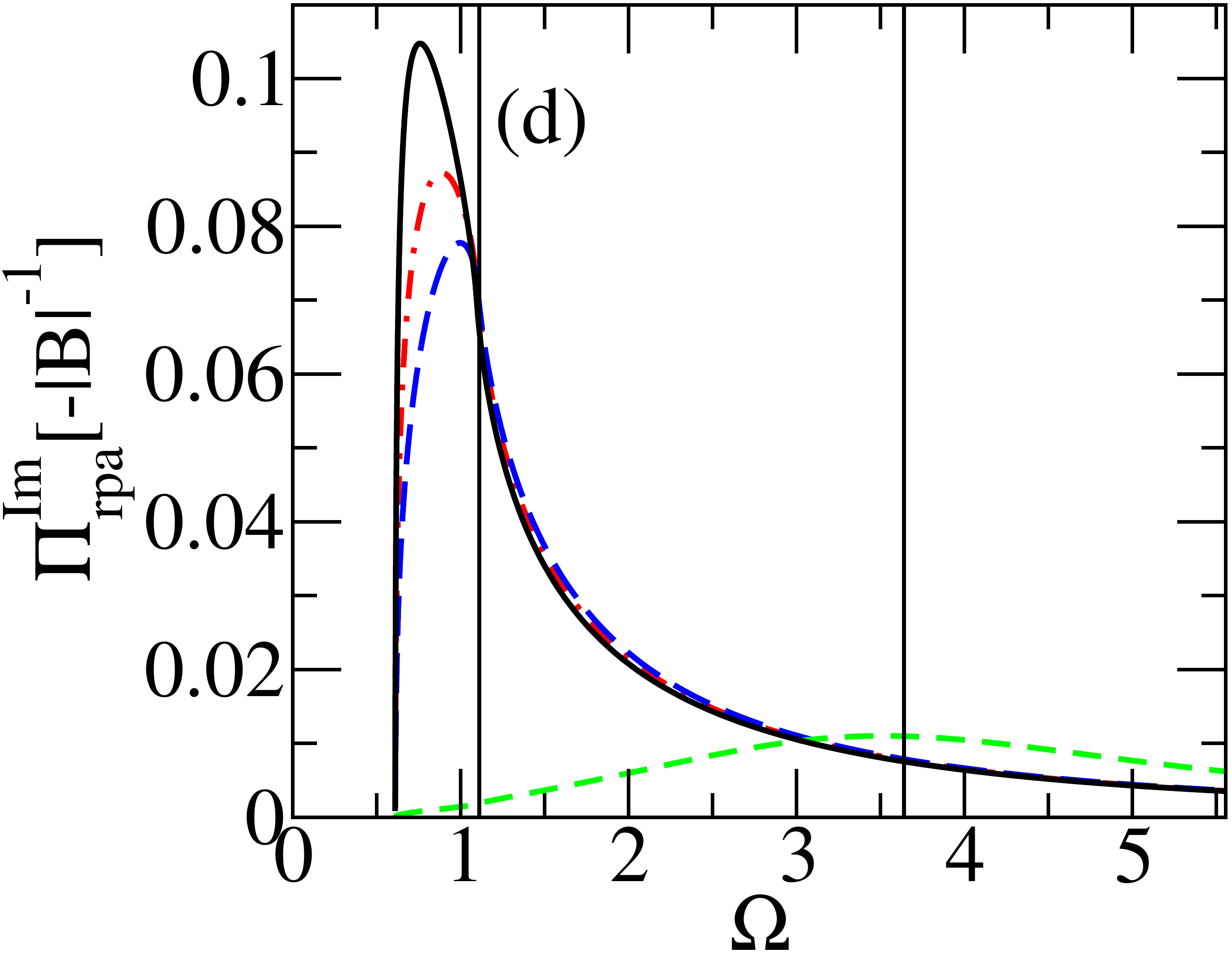}

\caption{(Color online) Plots of $\Pi^{Im}$ (a) and $\Pi^{Im}_{rpa}$
for $\alpha=0.4$ (b) and $\alpha=10$ (c). (d) shows linecuts for fixed $X=0.7$ with $\alpha \in \{0,0.2,0.4,10\}$ in black solid, red dot-dashed, 
blue long dashed and green short dashed lines, respectively. $\xi_{M}=0$ and $\xi_{D}=-0.5$.
\label{fig:intrinsic spectra}}

\end{figure}

The non-interacting single-particle excitation spectrum is given by
$\Pi^{Im}$, which we plot in Fig.~\ref{fig:intrinsic spectra}
(a)
 for $\xi_{M}=0$ and $\xi_{D}=-0.5$. Due to energy conservation,
there are no excitations beneath a frequency $\Omega_{min}$. In contrast
to graphene, where one observes a diverging behaviour of the polarization
at $\Omega_{min}$, here $\Pi^{Im}$ increases continously
from $0$. %
This is due to the broken particle-hole symmetry ($\xi_{D}<0$)
which ensures that the lowest energy excitations correspond to processes exciting particles from the valence
band to the proximity of the Dirac point, where, however, 
the density of states is zero. 
The excitation spectrum shows a maximum for small
momenta $X<1$ which lies beneath the plasmon dispersion given by the black line, perturbatively calculated from
Eq.~(\ref{eq:Plasmon Equation}) up to order $\left(\frac{\Gamma}{\Omega}\right)^2$ for $\alpha=0.4$.

Considering a finite Coulomb interaction, the excitation spectrum is given by
 $\Pi^{Im}_{rpa}$ plotted
in Fig.~\ref{fig:intrinsic spectra} (b) for $\alpha=0.4$ and (c)
for $\alpha=10$. The maximum of the spectrum shifts to higher energies
compared to the non-interacting one, indicating the formation of a
collective excitation in the system, i.e. a plasmon. This is proven by
solving the plasmon equation (Eq.~(\ref{eq:Plasmon Equation})) perturbatively
up to order $\left(\frac{\Gamma}{\Omega}\right)^{2}$, with the dispersion
plotted as a black line on top of the spectrum. Additionally,
the dispersion based on the expansion of $\Pi^{R}$ in the limit $X\rightarrow0$
for $\Omega\ll1$ and $\Omega\gg1$ are plotted as gray lines in Fig.~\ref{fig:intrinsic spectra}(c).

The plasmon dispersion relation starts linearly for small $q$, as one would expect
for a neutral system without doping. At high energies on the other
hand, a free-particle behaviour could be expected, leading to the
usual $\sqrt{q}$ dispersion known from doped systems. 
Although Eq.~(\ref{eq:Pi undoped expanded}) shows that this picture is only partly
true due to the logarithmic correction of $\Pi^{Re}$,
Fig.~\ref{fig:intrinsic spectra} (c) indicates a qualitative agreement.

Fig.~\ref{fig:intrinsic spectra} (d) shows linecuts of $\Pi^{Im}_{rpa}$ for fixed $X=0.7$ with $\alpha \in \{0,0.2,0.4,10\}$. 
Additionally, the black vertical lines indicate the plasmon frequency for $\alpha\leq0.4$ (left line) and $\alpha=10$ (right line).
For $\alpha=0.2$ the maximum of the interacting spectrum lies between the maximum of the non-interacting spectrum and the plasmon frequency, 
indicating that single-particle and collective excitations are equally strong. Increasing the interaction to $\alpha=0.4$, the maximum of the 
interacting spectrum and the plasmon frequency almost coincide, therefore the plasmon dominates over the single-particle excitation.  
At very large interactions $\alpha=10$, the plasmon is the only relevant excitation in the system. 

Increasing the Coulomb interaction broadens the plasmon peak and reduces its height as shown in Fig.~\ref{fig:intrinsic spectra} (d). 
This seems contrary to the picture of a plasmon as a sharp interaction-induced charge resonance, 
suggesting that these interband plasmons may not be well-defined for high energies. 
Yet this is a false conclusion. In Sec.~\ref{F-sum rule interpretation exc spectrum} we discussed that the contribution of the resonance to 
the f-sum rule is the actual measure of importance of a resonance. It can be estimated by multiplying the peak 
height in $\Pi^{Im}_{rpa}$ by $\Omega_p$, while the relevant peak width is given by $\frac{\Gamma}{\Omega_p}$. The latter 
is decreasing with $\Omega_p$ according to Eq.~(\ref{eq:Damping expansion}). 
From this normalization of the peak we conclude that the discussed interband plasmons fulfill the interpretation as
sharp interaction-induced charge resonances, with the width $\frac{\Gamma}{\Omega_p}$ decreasing with 
increasing plasmon frequency, above the critical frequency $\Omega_c$ as defined in Sec.~\ref{sec: Expansion of Pi}. 


\subsection{F-sum rule}

\begin{figure}
\includegraphics[width=6cm]{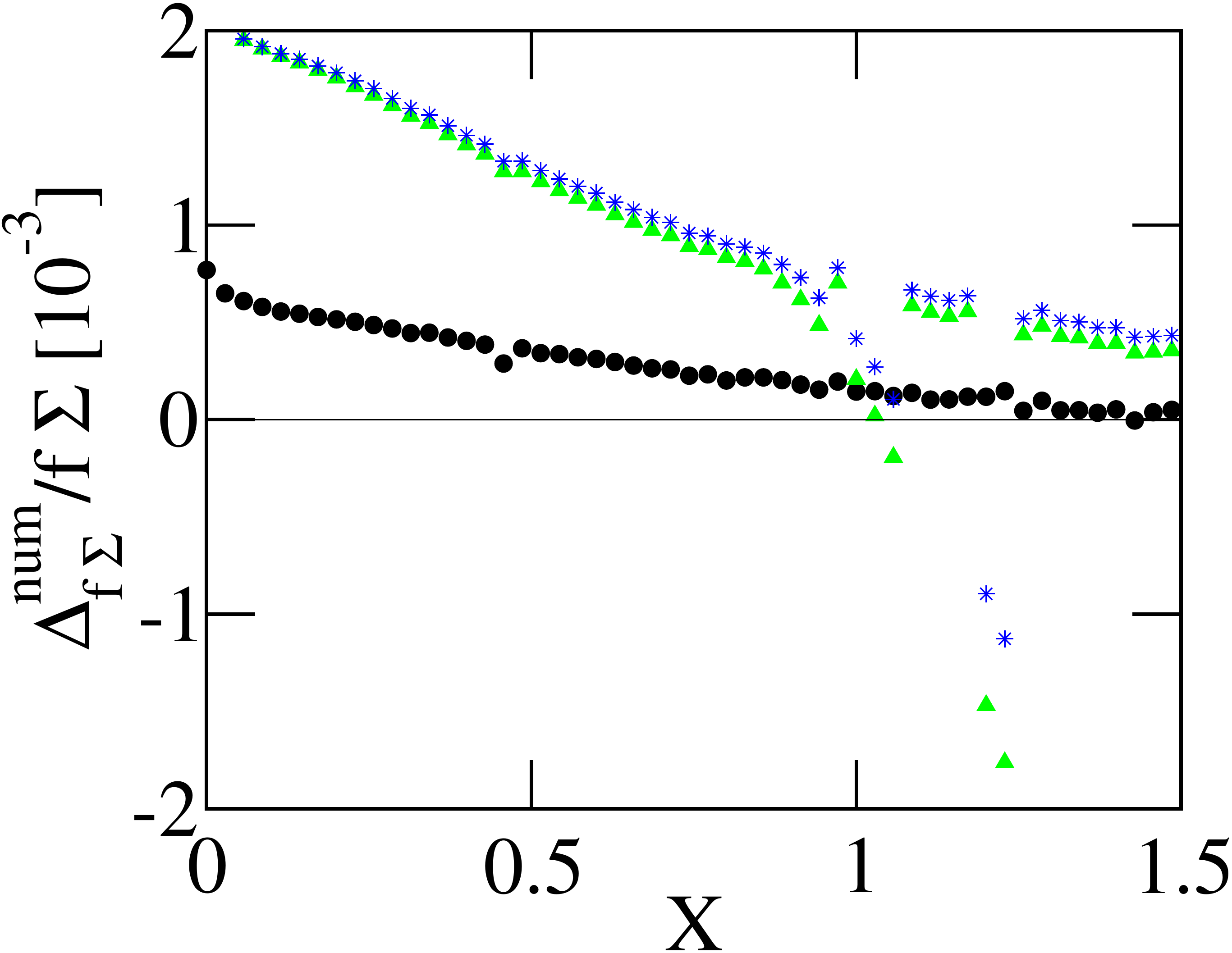}

\caption{(Color online) The ratio $\frac{\Delta_{f\sum}^{num}}{f\sum}$, with $\Delta_{f\sum}^{num}$
being the difference between the numercial and analytical f-sum rule.
Black dots are for the non-interacting spectrum, while blue stars
stand for $\alpha=0.4$ and green triangles for $\alpha=10$. $\beta=3$, $X_{max}=6$, $\xi_{M}=0$ and $\xi_{D}=-0.5$. 
The deviations around $X\gtrsim1$ stem from numerical instabilities, which are however negligibly small. \label{fig:f-sum numerical undoped}}

\end{figure}

The f-sum rule provides a check for our numerics. In Fig.~\ref{fig:f-sum numerical undoped}
 we plot the ratio $\frac{\Delta_{f\sum}^{num}}{f\sum}$, with $\Delta_{f\sum}^{num}=f\sum^{num}-f\sum$
where $f\sum^{num}$ is the numerical calculated f-sum rule and $f\sum$
the analytic one. The deviation are of the order $10^{-3}$, comparable to the analytical uncertainty, see Sec.~\ref{F-sum rule introduction}, and thus
negligible. The f-sum has to be the same for interacting and not interacting
systems. We find a slight dependence on the interaction strength $\alpha$,
which could be a numerical artifact, depending on $\Lambda$, or a real $\alpha$ dependence
like in graphene, where spectral weight is missing for small frequencies, cf. Eq.~(14) in Ref.~\onlinecite{sabio2008} ($\Pi^{RPA}<\Pi^{R} \ \forall q,\omega$ 
for the undoped Dirac model). As the effect declines with increasing cutoff $\Lambda$,
we conclude that the RPA approximation in the BHZ model misses no spectral weight compared
to the full Coulomb interaction, even in the undoped limit.

\section{Doped system} \label{sec_doped}
In this section, we extend our analysis to finite doping $\mu>0$, where a net charge density is present in the system.
Doping the system has two effects: one is the Fermi blocking of interband
excitations (red arrow in Fig.~\ref{fig:AB bandstructure} (a)) for small $X$ and $\Omega$. 
The other is the appearance of intraband
excitations (green arrow in Fig.~\ref{fig:AB bandstructure} (a)), which are absent in the intrinsic limit.
Again, first we study the polarization and screening properties of the system in the static limit, where we also study Friedel 
oscillations due to the scattering on a charged impurity.
Then, we study the dynamical polarization function in the long-wavelength limit, where we obtain an analytical expression for the collective plasmonic modes
of the system.
Finally we numerically compute the dynamical polarization function in the full range of momenta and frequencies in the full parameter space of the BHZ model, 
analyzing the effect of each of the model parameters.
Particular emphasis is put on the coexistence of interband and intraband plasmons and on how the BHZ model interpolates between the Dirac and 2DEG behavior.


\subsection{Static limit} \label{Static limit}

\begin{figure}
 \centering
 \includegraphics[width=8cm]{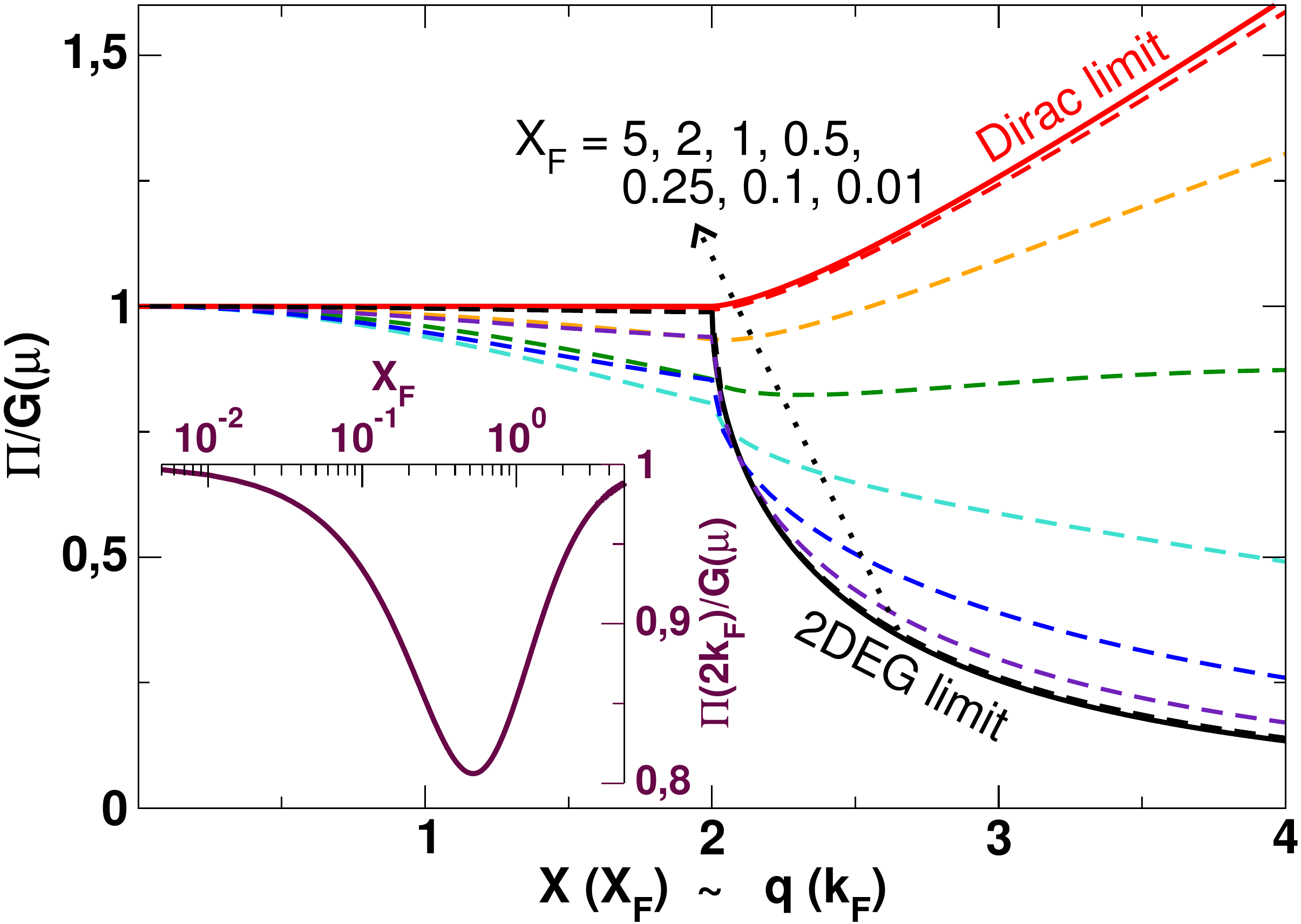}
 \caption{(Color online) Static polarization function $\tilde\Pi(X)$ of the BHZ model for $\xi_M=\xi_D=0$ at finite doping for different value of $X_F$, normalized by the DOS $G(\mu)$.
 In the inset, details on the value of $\Pi(2 X_F)$ as a function of $X_F$ are given.
 }
 \label{fig:normalized}
\end{figure}

In Fig.~\ref{fig:normalized},
we present the static polarization function $\tilde\Pi(X)=\Pi(X)/G(\mu)$ at finite doping,
conveniently normalized by the density of states at the Fermi level $G(\mu)$.
This normalization stands out naturally from the long wavelength property of the polarization function
\begin{equation}
\lim_{q\rightarrow0} \Pi(q)=\Pi_\mu(0)=G(\mu). \label{eq:lwl_Pi}
\end{equation}
For the BHZ model at finite doping, $\tilde\Pi(X)$ has a pronounced dependence on the extrinsic parameter $X_F=k_F/q_0$.
For $X_F\ll1$ ($X_F\gg1$) the Fermi level falls in a region where locally the dispersion curve has predominant Dirac (2DEG) character.
In a 2DEG system, the static polarization assumes the following analytic form~\cite{giuliani}
\begin{eqnarray}
\tilde\Pi(q) =  1-\Theta(q-2 k_F) \frac{\sqrt{q^2-4 k_F^2}}{q},
\end{eqnarray}
while in the Dirac limit we have~\cite{gorbar2002,ando2006}
\begin{eqnarray}
\tilde\Pi(q)\hspace{-0.05cm} =\hspace{-0.05cm} 1\hspace{-0.05cm}-\hspace{-0.05cm}\Theta(\hspace{-0.02cm}q\hspace{-0.05cm}-\hspace{-0.05cm}2 k_F\hspace{-0.05cm})\hspace{-0.05cm} \left[\hspace{-0.05cm}
\frac{\sqrt{\hspace{-0.05cm}q^2\hspace{-0.05cm}-\hspace{-0.05cm}4 k_F^2}}{2 q} \hspace{-0.05cm}- \hspace{-0.05cm}\frac{q}{4k_f}\hspace{-0.05cm}
\arctan{\frac{\hspace{-0.05cm}\sqrt{\hspace{-0.05cm}q^2\hspace{-0.05cm}-\hspace{-0.05cm}4 k_F^2}}{2k_F}}\hspace{-0.05cm}\right].
\end{eqnarray}
Our calculations for the BHZ model with $\xi_M=\xi_D=0$ correctly reproduce the Dirac and 2DEG limits for $X_F\ll1$ and $X_F\gg1$, respectively.
We note that with a finite $B$ term and nonzero $X_F$ the polarization will always have a decay behavior for $q>q_0$.
In the 2DEG and Dirac limit one finds $ \tilde\Pi(q)=1$ for $q<2k_F$, coincidence due to the balancing effect of dispersion curve and overlap factor.
Interestingly, in the BHZ model we observe instead a deviation from unity, shown in details in the inset of Fig.~\ref{fig:normalized}, 
which has a maximum for $X_F\approx 0.5$.
In the 2DEG limit, $X_F\gg 1$, $\tilde \Pi(X)$ has a strong discontinuity in its first derivative at $X=2X_F$, 
while for decreasing $X_F$ this discontinuity decreases and finally
vanishes in the Dirac limit, where the discontinuity affects only the second derivative.


\subsection{Screening}
We already analyzed in section~\ref{sec:screening_int} the intrinsic response of a BHZ system to a test charge, 
when no net charge density is present in the system.
While the intrinsic response is realized on intrinsic scales of the model $1/q_0$, the 'metallic' response (at finite electronic density) 
is characterized by the Fermi wave length $\pi/k_F$.
Therefore it is convenient to express $n_\mu(X)$ as a function of dimensionless units $\tilde X = X/X_F=k/k_F$, due to the presence of a discontinuity at $\Pi_\mu(2 X_F)$.
The induced charge density $n_\mu$ is given by
\begin{eqnarray}
  n_\mu(r) = \eta_\mu \int d\tilde X \frac{J_0(r k_F \tilde X)}{1-\alpha g_0(\tilde X X_F)}~\frac{\tilde \Pi_\mu(\tilde XX_F)}{1 - \alpha g(\tilde XX_F)},
  \label{eq:n_mu_Y}
\end{eqnarray}
where, using the property Eq.~(\ref{eq:lwl_Pi}), we have emphasized the dependence of the induced density on the DOS at the Fermi level, which now appears in the
scaling factor $\eta_\mu=\eta_0 |B| G(\mu) X_F= \frac{e^2 k_F}{4\pi \epsilon_0\epsilon_r^2} G(\mu)$.
We note that the integral also depends on the parameters $\alpha$ and $X_F$
(and naturally on $\xi_M$ and $\xi_D$, when finite).

\begin{figure}
 \centering
 \includegraphics[width=8.5cm]{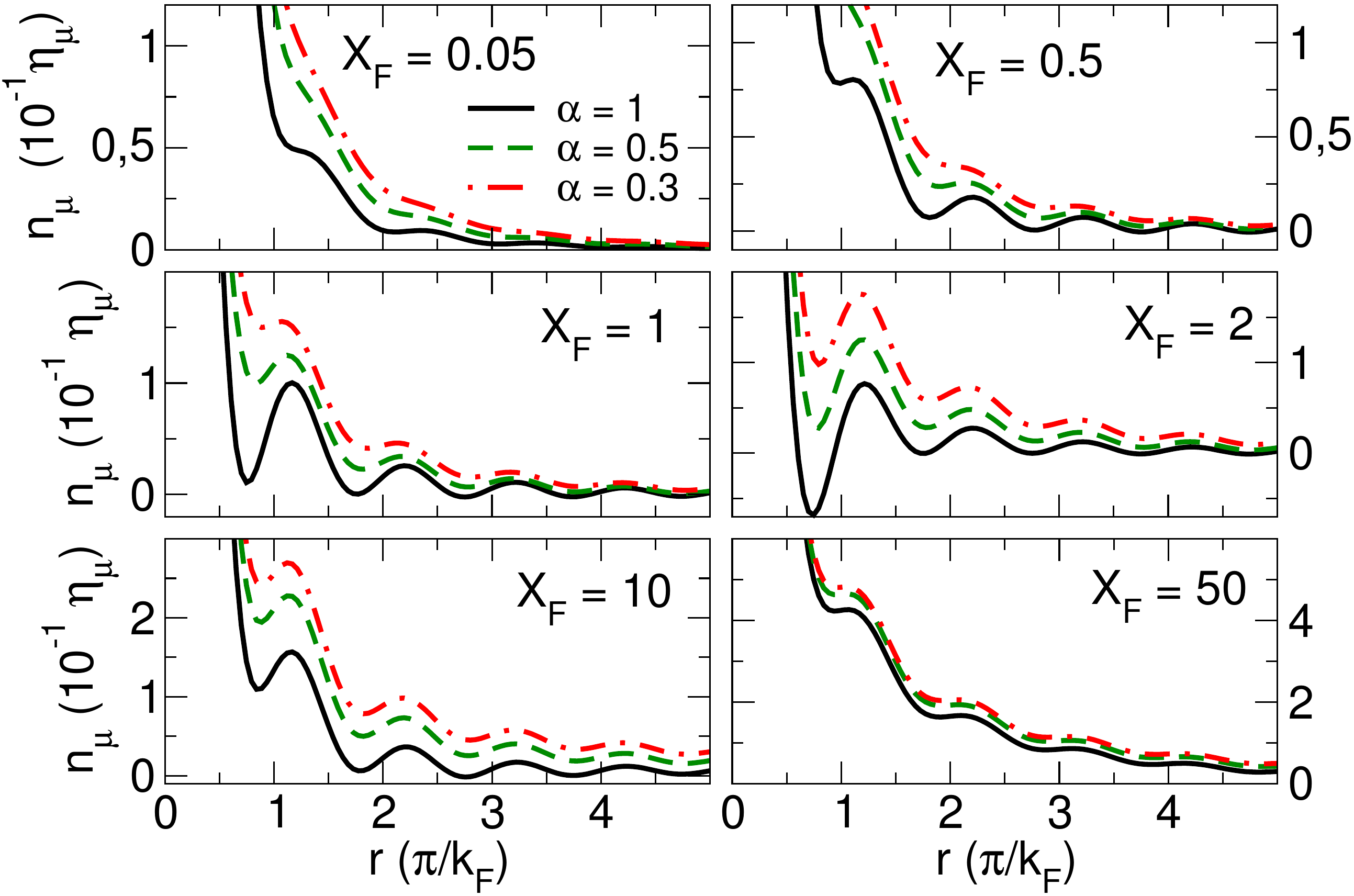}
 \caption{(Color online) Induced charge density in real space for the BHZ model for $\xi_M=\xi_D=0$ and $\alpha=0.3$, $0.5$ and $1$.
 Data in different panels belong to systems with $X_F=0.05$, $0.5$, $1$, $2$, $10$ and $50$.
 All calculation are obtained by keeping $\epsilon_r=10$ and $k_F=0.1$~nm$^{-1}$, and varying $A\approx 0.143$, $0.29$ and $0.47$~eV~nm,
 for $\alpha=0.3$, $0.5$ and $1$, respectively, while varying the parameter $B$ accordingly to $X_F$.
 In the panel $X_F=50$ (2DEG limit), the three curves with $\alpha=0.3$, $0.5$ and $1$ are quite close and correspond to similar $\beta$ parameter ($\beta=0.01$, $0.008$ and $0.007$, respectively).
 }
 \label{fig:n_dop}
\end{figure}
In Fig.~\ref{fig:n_dop}, we present the induced screening electronic radial density for the BHZ model for $\xi_M=\xi_D=0$ due to a point-like test charge.
Each panel corresponds to a different value of the ratio $X_F=k_F/q_0$, and
within each panel curves differing by the Dirac fine structure constant $\alpha$ are presented.
Friedel oscillations appear of period $\pi/k_F$, which become more defined for larger $\alpha$.
We also note that density oscillations are more prominent for $k_F\approx q_0$ than in the Dirac ($X_F\ll 1$) and 2DEG limits ($X_F \gg 1$).
In the 2DEG limit the $\alpha$ parameter is ill defined and should be replaced by the more general parameter $\beta=\frac{e^2 G(\mu)}{2 \epsilon_o \epsilon_r k_F}$
, characterizing the dielectric response of the system.
%

The presence of Friedel oscillations and their asymptotic behavior are related through the Lighthill theorem~\cite{lighthill} to discontinuities in the static polarization function and its derivatives (see for example Ref.~\onlinecite{badalyan2010} for a detailed discussion). A discontinuity like $|q-2k_F|^{\nu}\Theta\left(q-2k_f\right)$ in $\Pi\left(q\right)$, with $\Theta$ the Heaviside step function and $\nu\in\mathbb R$, translates into a decay of the oscillations in $n\left(r\right)$ with leading order $r^{-\nu-3/2}$. One finds $\nu=1/2$ ($\nu=3/2$) for the leading order discontinuity of a 2DEG (Dirac) system, such that the first (second) and all higher derivatives of the static polarization function are discontinuous at $q=2k_F$.  
Analyzing the Friedel oscillations for the BHZ model, one finds a composition of two different contributions with an asymptotic decay at large distances as $r^{-2}$ (2DEG contribution) and $r^{-3}$ (Dirac contribution), respectively.
As a consequence,  the discontinuity in the 
RPA polarization function of the BHZ model at $q = 2k_F$ can be very well approximated by a combination of 2DEG ($\nu=1/2$) and Dirac ($\nu=3/2$) contribution.
In the Dirac (2DEG) limit, the effect of the discontinuity in the second (first) derivative becomes predominant and oscillations purely decay in leading order as $r^{-3}$ ($r^{-2}$).


\subsection{Long wavelength expansion, plasmon dispersion}

At finite doping, for small momenta $X$, the polarization function is governed by intraband
excitations, as the interband excitations are Fermi-blocked. 
We perform an expansion in this limit, for $\Omega>X$, to gain an analytical insight into the physics at finite doping and 
derive an analytical formula for the plasmon dispersion.
In particular, intraband plasmons are expected to be the dominant excitation for small momenta, similarly to the 2DEG and Dirac case. 
We expand the polarization function up to order $X^4$
\begin{align*}
\left|B\right|\Re\left[\Pi^{R}\left(X,\Omega\right)\right]= \Pi_{44}\frac{X^{4}}{\Omega^{4}}+\Pi_{42}\frac{X^{4}}{\Omega^{2}}+\Pi_{40}X^{4}\\ + \Pi_{22}\frac{X^{2}} {\Omega^{2}} + \Pi_{20}X^{2}+\mathcal{O}\left(\Omega^{2}\right)
\end{align*}
and use it to solve Eq.~(\ref{eq:Plasmon Equation}). 
We obtain a plasmon dispersion of
 \begin{equation}
\Omega=\sqrt{2\pi\alpha\Pi_{22}}\sqrt{X}+\left(\frac{\Pi_{44}}{\Pi_{22}^{\frac{3}{2}}}\frac{1}{\sqrt{8\pi\alpha}}+\sqrt{2\pi^{3}\alpha^{3}\Pi_{22}}\Pi_{20}\right)X^{\frac{3}{2}}\label{eq:Plas_Freq_BHZ}
\end{equation}
with the leading coefficient
  \begin{align}
\Pi_{22}= & g_s\Theta\left(X_f\right)\left(\frac{X_{f}^{2}\left(1+2X_{f}^{2}+2\xi_{M}\right)}{4\pi\sqrt{X_{f}^{2}+\left(\xi_{M}+X_{f}^{2}\right)^{2}}}-\gamma\xi_{D}\frac{X_{f}^{2}}{2\pi}\right) \nonumber \\
= & g_s\Theta\left(X_f\right)\frac{1}{4\pi}\left(\left|\Omega_{f}\right|+\underset{-\left|\xi_M\right|<\ldots<\left|\Omega_{f}\right|}{\underbrace{\Pi_{inter}\left(X_{f}\right)}}\right) \nonumber \\
\underset{\underset{X_{f}\rightarrow0}{\xi_{M}=0}}{=} & g_s\frac{X_{f}}{4\pi}+\mathcal{O}\left(X_{f}^{2}\right)=g_s\frac{\left|\Omega_{f}\right|}{4\pi}+\mathcal{O}\left(X_{f}^{2}\right) \label{eq:Pi Doped expanded Dirac limit}\\
\underset{\underset{X_{f}\rightarrow\infty}{\xi_{M}=0}}{=} & g_s\frac{X_{f}^{2}}{2\pi}\left(1-\gamma\xi_{D}\right)+\mathcal{O}
\left(\frac{1}{X_{f}^{2}}\right)=g_s\frac{\left|\Omega_{f}\right|}{2\pi}+\mathcal{O}\left(\frac{1}{X_{f}^{2}}\right)  \label{eq:Pi Doped expanded 2DEG limit}
\end{align}
with $\gamma=sgn\left[\Omega_{f}\right]$ and $\Pi_{inter}\left(X_{f}\right)=\frac{X_{f}^{4}-\xi_{M}^{2}}{\sqrt{X_{f}^{2}+\left(\xi_{M}+X_{f}^{2}\right)^{2}}}-\gamma\xi_{D}X_{f}^{2}$.
In the limit of zero mass, $\Pi_{inter}\left(X_{f}\right)$ interpolates smoothly between $0$ for $X_{f}\rightarrow0$
and $\left|\Omega_{f}\right|$ for $X_{f}\rightarrow\infty$. The former case corresponds to the Dirac limit,
where one finds the plasmon frequency
\[
\omega=A\sqrt{\frac{g\alpha k_{f}}{2}}\sqrt{q}=\sqrt{\frac{ge^{2}\mu}{8\pi\varepsilon_{0}\varepsilon_{r}}}\sqrt{q}\]
 in the literature~\cite{wunsch2006}, being identical to Eqs.~(\ref{eq:Plas_Freq_BHZ},\ref{eq:Pi Doped expanded Dirac limit}). The latter case is the 2DEG limit, where
 one finds the plasmon dispersion
\begin{align*}
\omega= & \sqrt{\frac{e^{2}N}{2m\varepsilon_{0}\varepsilon_{r}}}\sqrt{q}=\sqrt{\frac{ge^{2}\mu}{4\pi\varepsilon_{0}\varepsilon_{r}}}\sqrt{q}
\end{align*}
in the literature~\cite{giuliani}, with $N=\frac{g}{4\pi}k_{f}^{2}$ the carrier density and $m=\frac{1}{2\left|B+D\right|}$. This in in agreement with Eqs.~(\ref{eq:Plas_Freq_BHZ},\ref{eq:Pi Doped expanded 2DEG limit}).

Thus the BHZ model as a function of its parameters reproduces the plasmon dispersion in the Dirac and 2DEG limits and interpolates between
them. 
We note that for $k_{f}\rightarrow0$ the term $\Pi_{22}$ is zero and the intraband plasmon disappears. In this limit, 
the leading order contribution $\mathcal{O}\left(\frac{X^{2}} {\Omega^{2}}\right)$ of the intrinsic polarization, Eq.~\ref{eq:Pi undoped expanded}, takes the place of $\Pi_{22}$. The crucial difference between the extrinsic and the intrinsic polarization is that the latter has a finite imaginary part of order $\mathcal{O}\left(\frac{X^{2}} {\Omega^{2}}\right)$, 
leading to the linear dispersion of the interband plasmons. Yet for finite $k_{f}>0$, 
these interband plasmons are supressed due to the Fermi blockade of the interband excitations and only exist if their plasmon frequency 
exceeds both the chemical potential $\Omega_f$ and the critical frequency 
$\Omega_c$ as defined in Sec.~\ref{sec: Expansion of Pi}, see for example Fig.~\ref{fig:Im epsi BHZ Xf01}.

Besides the different scaling with momenta in the limit $q\rightarrow0$, also the scaling with $\alpha$ is different for
the inter- and intraband plasmons, Eqs.~(\ref{eq:Plasmonfrequency linear}) and (\ref{eq:Plas_Freq_BHZ}): 
linear vs. square root. This will have important consequences in the following when we will
discuss how to separate the two different collective excitations.


\subsection{Excitation spectrum: Interpolation between Dirac and 2DEG regime}

We begin the discussion of the doped spectrum by looking at the limiting
results of 2DEG and Dirac system. From this, we then find that we can interpolate
between them by changing the Fermi momentum. Interestingly, by considering
the cases of broken particle-hole symmetry and large masses, we also find regimes which are distinct from the Dirac and 2DEG limit. As an example,
these regimes support both inter- and intraband plasmons at parameters which are realistic for HgTe QWs.

In all the following plots, the boundaries of the single-particle spectrum will
be indicated by faint black lines, the isolines $\Pi^{Re}=0$ by red lines.
The plasmon dispersions are plotted as black curves (full result from perturbation theory) and gray curves (expanded result in limit $X\rightarrow0$).


\subsubsection{Limiting case: 2DEG}

\begin{figure}
\includegraphics[width=4.2cm]{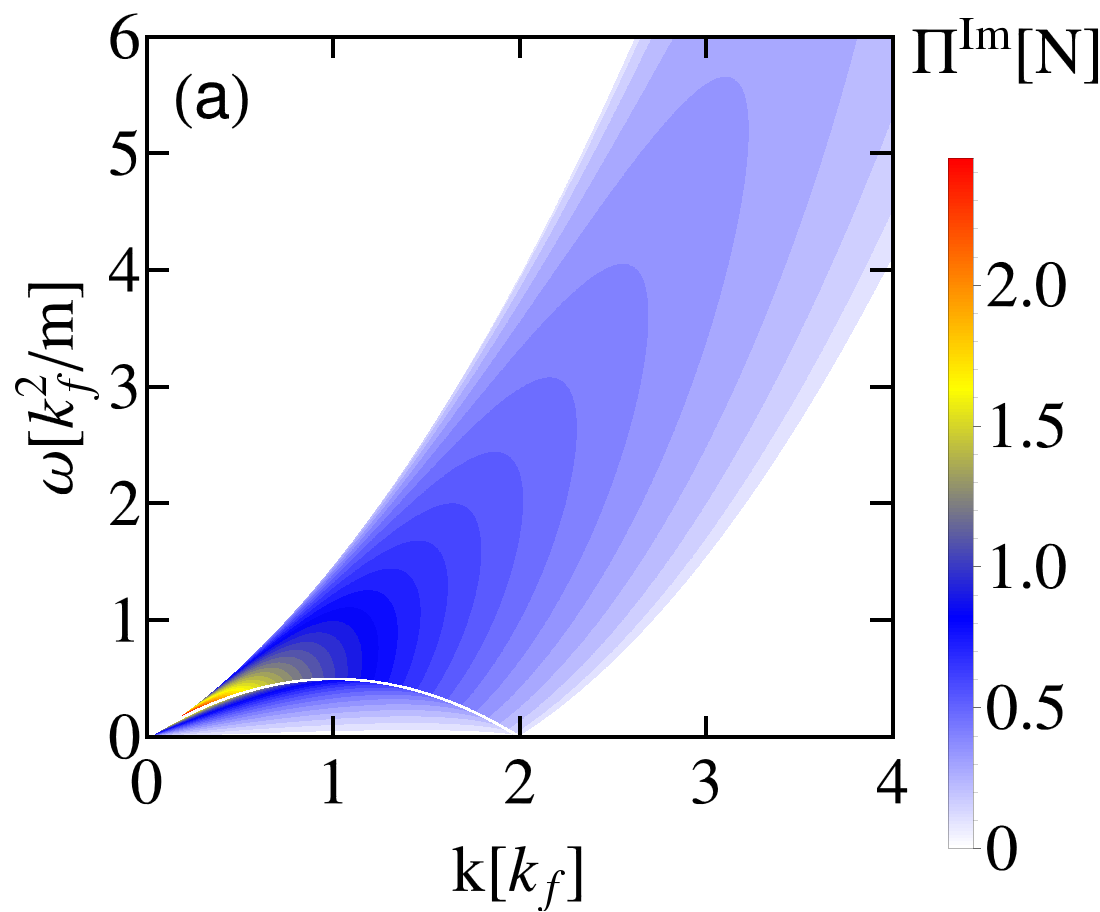}
\includegraphics[width=4.2cm]{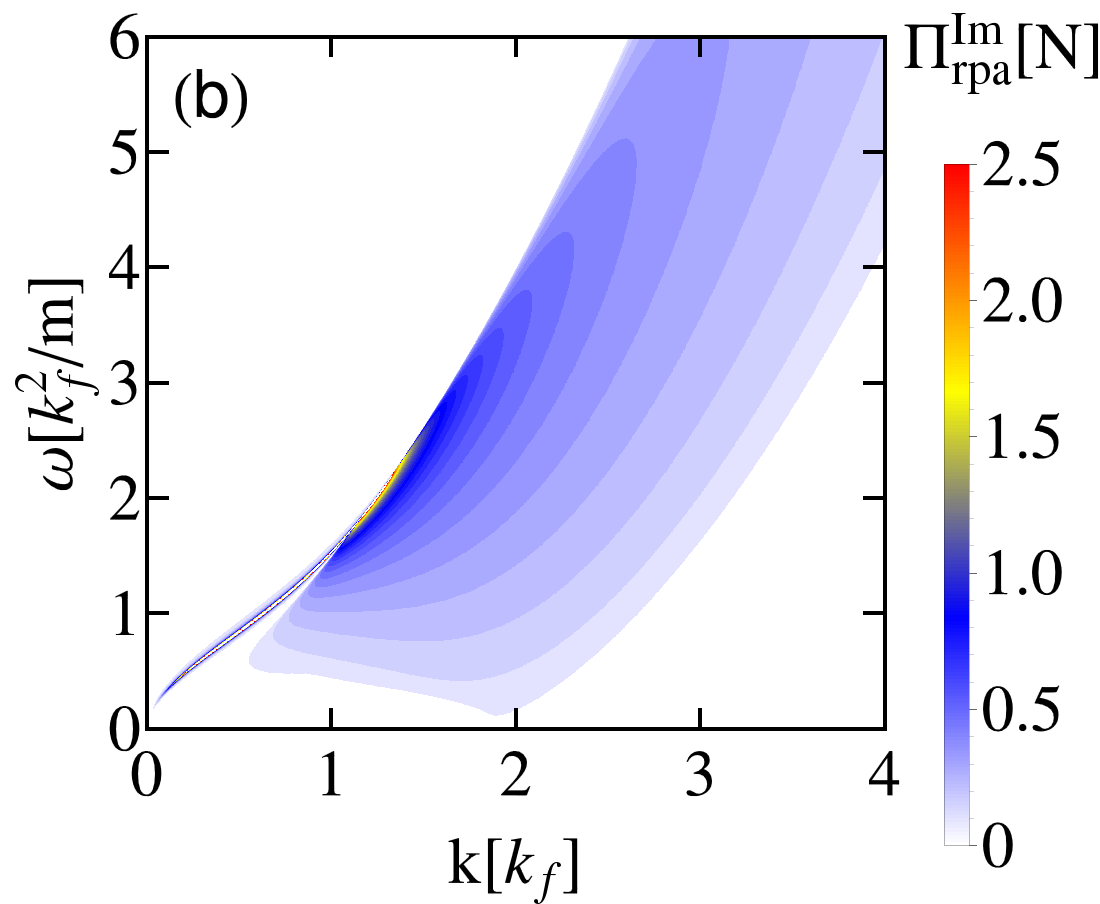}

\caption{(Color online) Spectrum of a 2DEG. (a) Imaginary part of $\Pi^{R}$
with $N=\frac{g_s m}{2\pi\hbar^{2}}$ and $g_s$ the degeneracy factor. (b) $\Pi^{Im}_{rpa}$
for $r_s=2$, with $v_q=\frac{r_s k_f}{Nq}$ the Coulomb interaction. We add an artificial
damping in the region of $\Pi^{Im}=0$
to make the plasmons visible.  \label{fig:Pi 2DEG}}

\end{figure}

In the 2DEG limit, only intraband excitations are possible. 
The polarization function has a well-known analytical form~\cite{giuliani}, therefore we can easily 
plot the non-interacting spectrum in Fig.~\ref{fig:Pi 2DEG} (a).
$\Pi^{Im}$ is peaked for $q,\omega\rightarrow0$ closely to the upper
boundary of the spectrum. 
It decays to zero instead for large momenta and frequencies like $\Pi^{Im}\propto q^{-1}$, 
if one considers a fixed ratio $\omega\propto q^2$ within the SPE region.

The interacting spectrum is shown in Fig.~\ref{fig:Pi 2DEG} (b).
An intraband plasmon appears with the usual $\sqrt{q}$ dispersion
for $q\rightarrow0$. 
It absorbes all of the spectral weight in this limit, thus
$\Pi^{Im}_{rpa}$ is suppressed in the SPE region. 
For intermediate momenta, the plasmon dispersion lies in the SPE region and the plasmon decays and broadens.
For larger momenta and frequencies, the interacting
and non-interacting spectra agree qualitatively.

\subsubsection{Limiting case: Dirac}

\begin{figure}
\includegraphics[width=4.2cm]{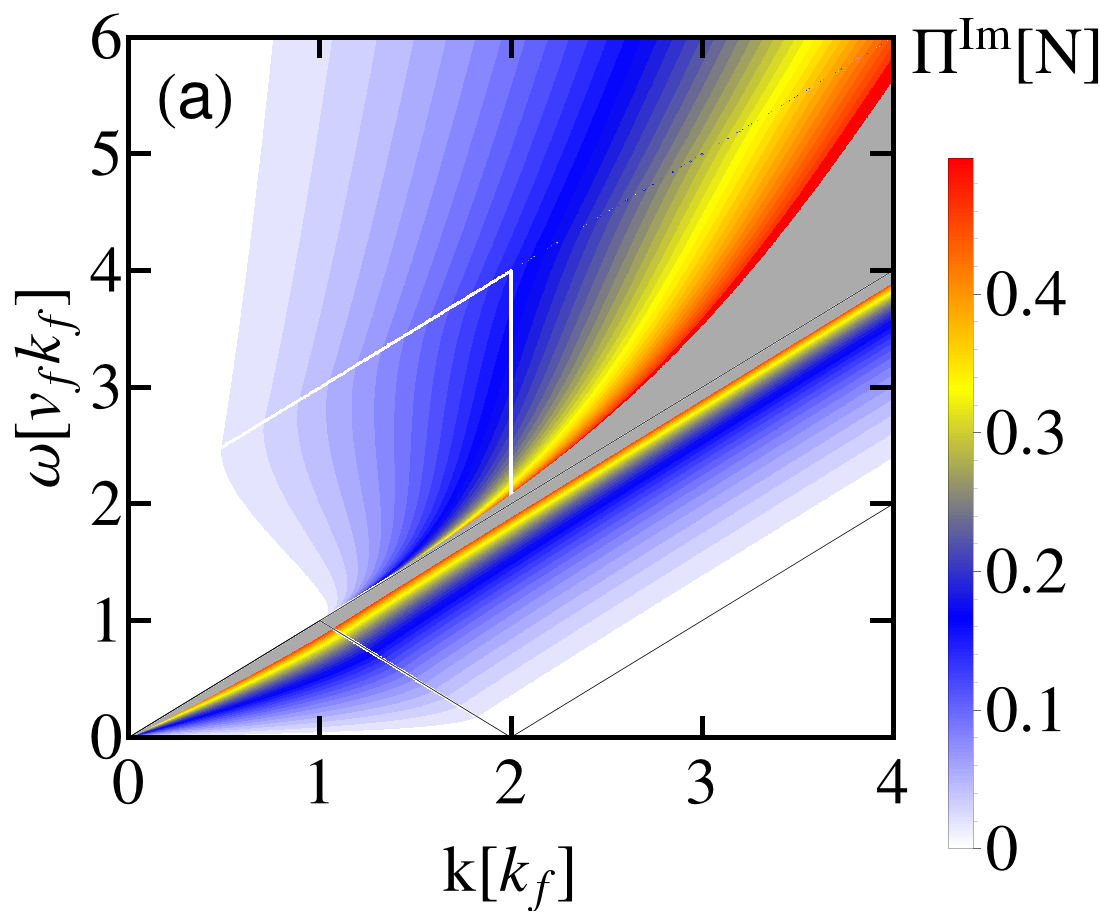}
\includegraphics[width=4.2cm]{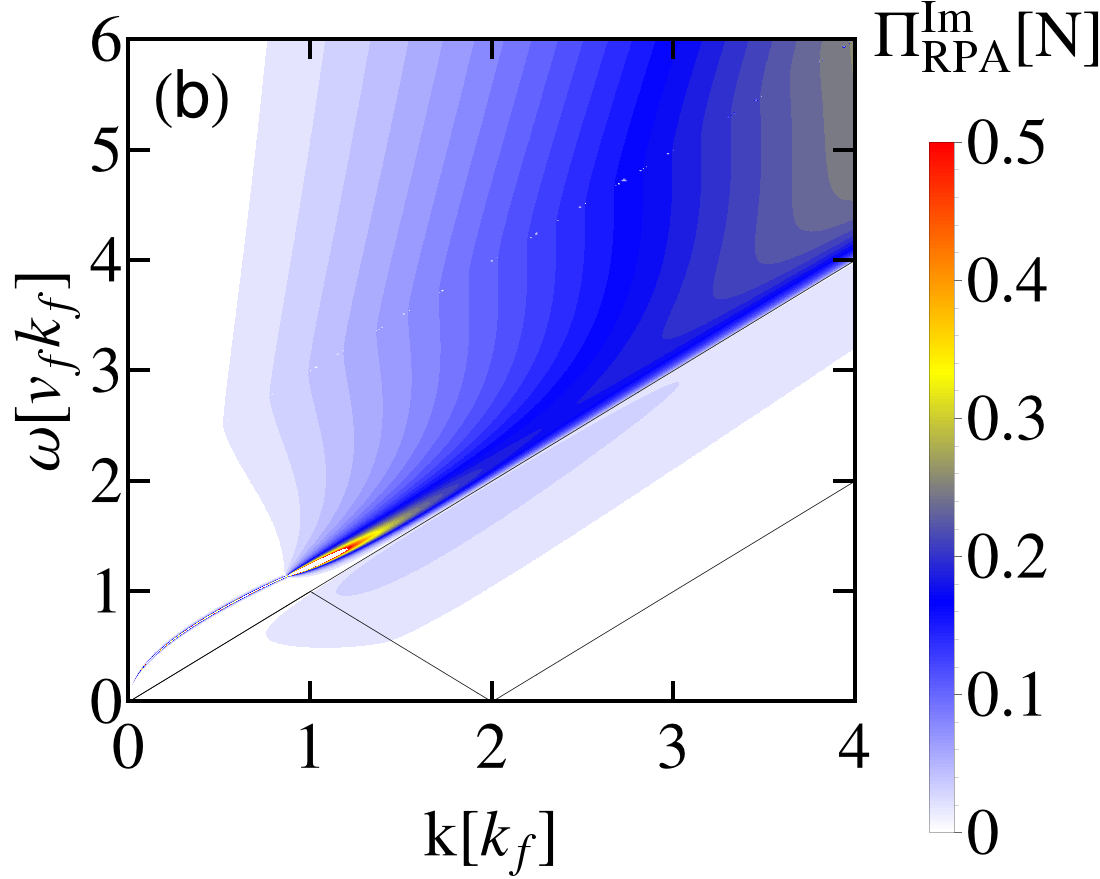}

\caption{(Color online) Plots for Dirac case. (a) $\Pi^{Im}$ with $N=\frac{g_s k_f}{\hbar v_f}$ and $g_s$ the degeneracy factor. In the gray area, the colorscale is exceeded due to the divergency of $\Pi^{Im}$. (b) $\Pi^{Im}_{rpa}$
for $r_s=2\pi g_s\alpha=4\pi\cdot0.6$, with $v_q=\frac{r_s k_f}{Nq}$ the Coulomb interaction. We add an artificial damping in the regions of
$\Pi^{Im}=0$ to make the plasmons
visible. \label{fig:Pi Graphene}}

\end{figure}

The Dirac spectrum comprises both inter- and intraband excitations.
The polarization function still has a well-known analytical expression~\cite{wunsch2006,hwang2007}, of which we 
plot the non-interacting spectrum $\Pi^{Im}$ in Fig.~\ref{fig:Pi Graphene} (a).
The intraband excitations occur for higher energies $\omega>v_f q$, while for intraband excitations less energy is needed, $\omega<v_f q$. 
Both excitation spectra touch at $v_f q=\omega$, where they diverge. 
Only the Fermi-blockade suppresses the interband transitions in $\Pi^{Im}$
for $q<2k_{f}$ and cures the divergency, see Fig.~\ref{fig:Pi Graphene} (a) for $\omega>v_f q$.
One finds a $\omega^{-1}$ decay for high frequencies.

The interacting spectrum $\Pi^{Im}_{rpa}$
is plotted in Fig.~\ref{fig:Pi Graphene} (b) for $\alpha=0.6$. 
Similar to the 2DEG, all of the intraband spectral weight is absorbed by
a plasmon in the limit $q\rightarrow0$ and the divergence at $v_f q=\omega$ is cured. 
Interestingly, for sufficient large interaction strength $\alpha$ the plasmon decays in the interband spectrum. 
For larger momenta and frequencies, we note that the intraband polarization does not recover the non-interacting value,
as it does for the 2DEG, but remains much smaller. 
Therefore single-particle intraband excitations are blocked altogether for all momenta and frequencies in this limit.
The missing spectral weight goes into a charge resonance at higher frequencies in the interband spectrum~\cite{sabio2008}. 
Yet, this resonance is not a solution of the plasmon equation and therefore not a plasmon~\cite{stauber2014}.


\subsubsection{BHZ model, $\xi_{M}=\xi_{D}=0$} \label{sec: BHZ model no mass no D}

\begin{figure}
\includegraphics[width=4.2cm]{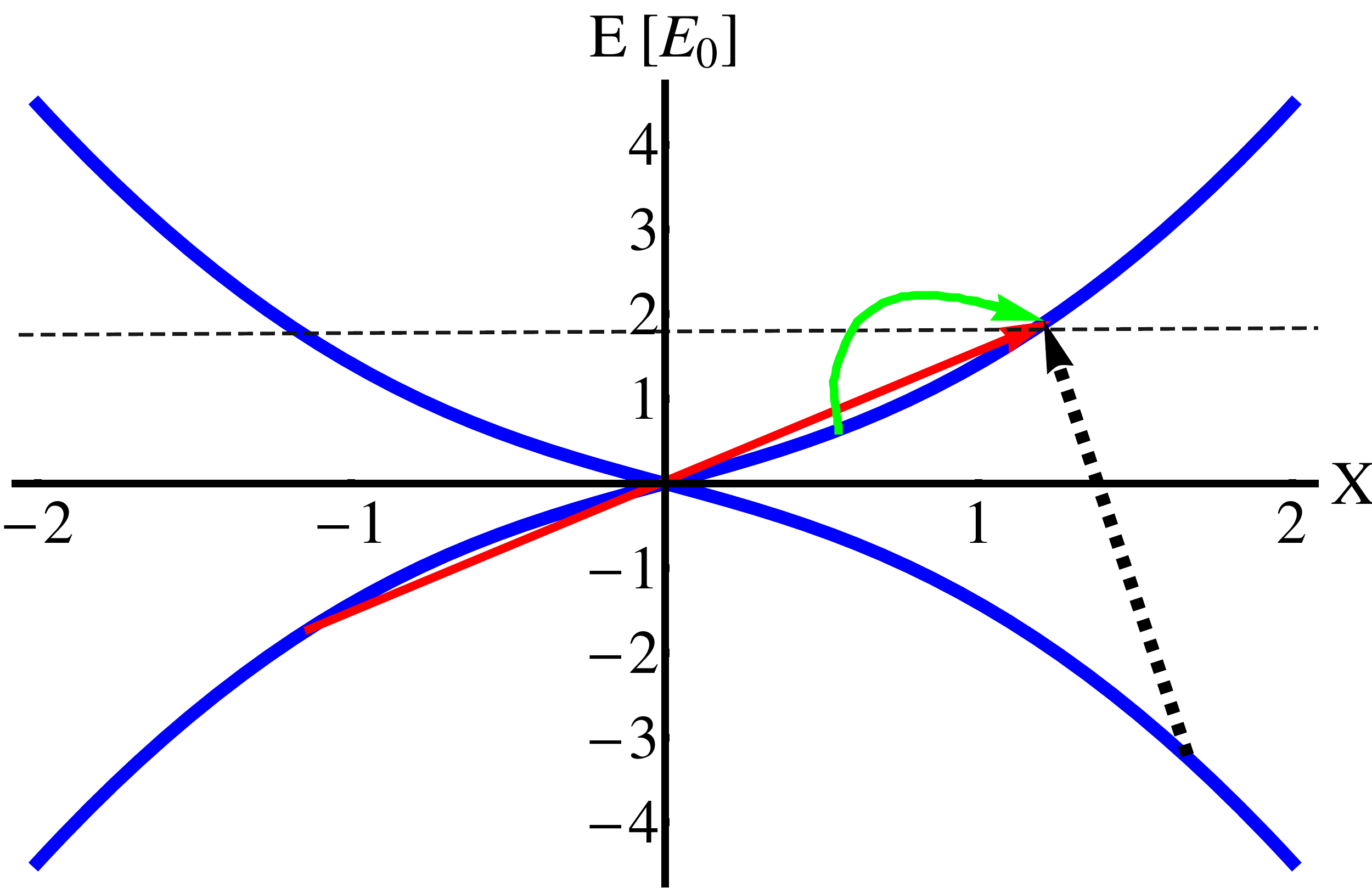}
\includegraphics[width=4.2cm]{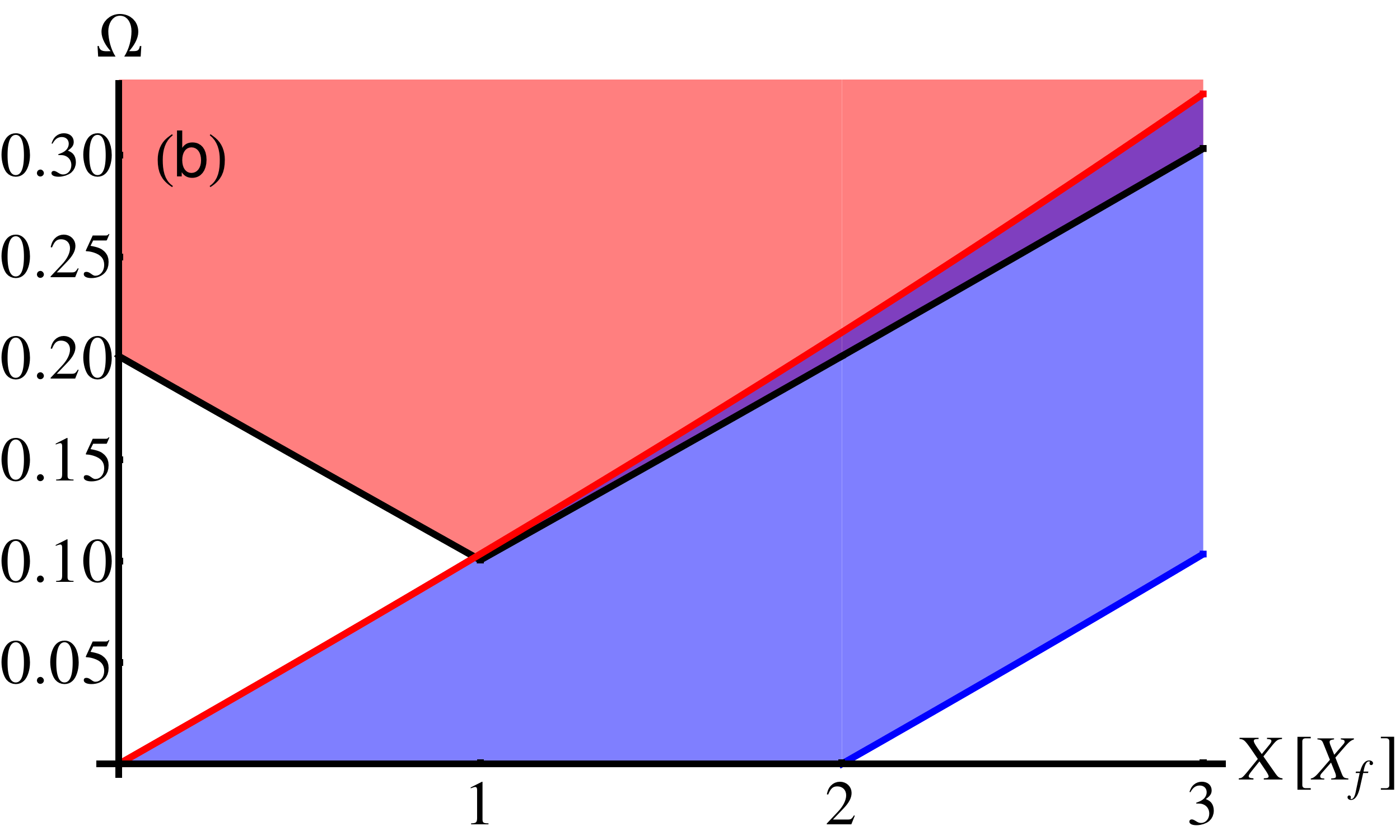}
\includegraphics[width=4.2cm]{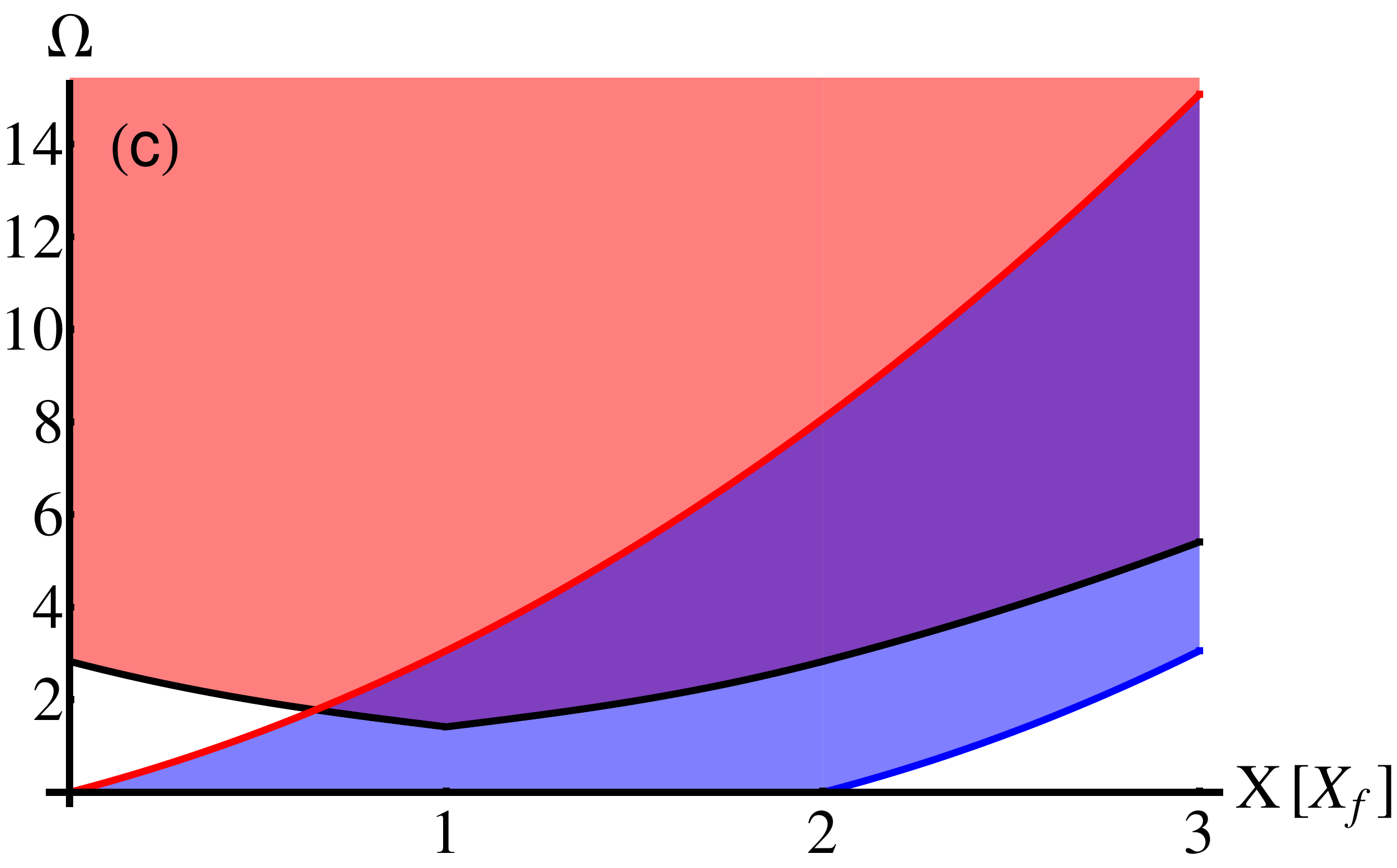}
\includegraphics[width=4.2cm]{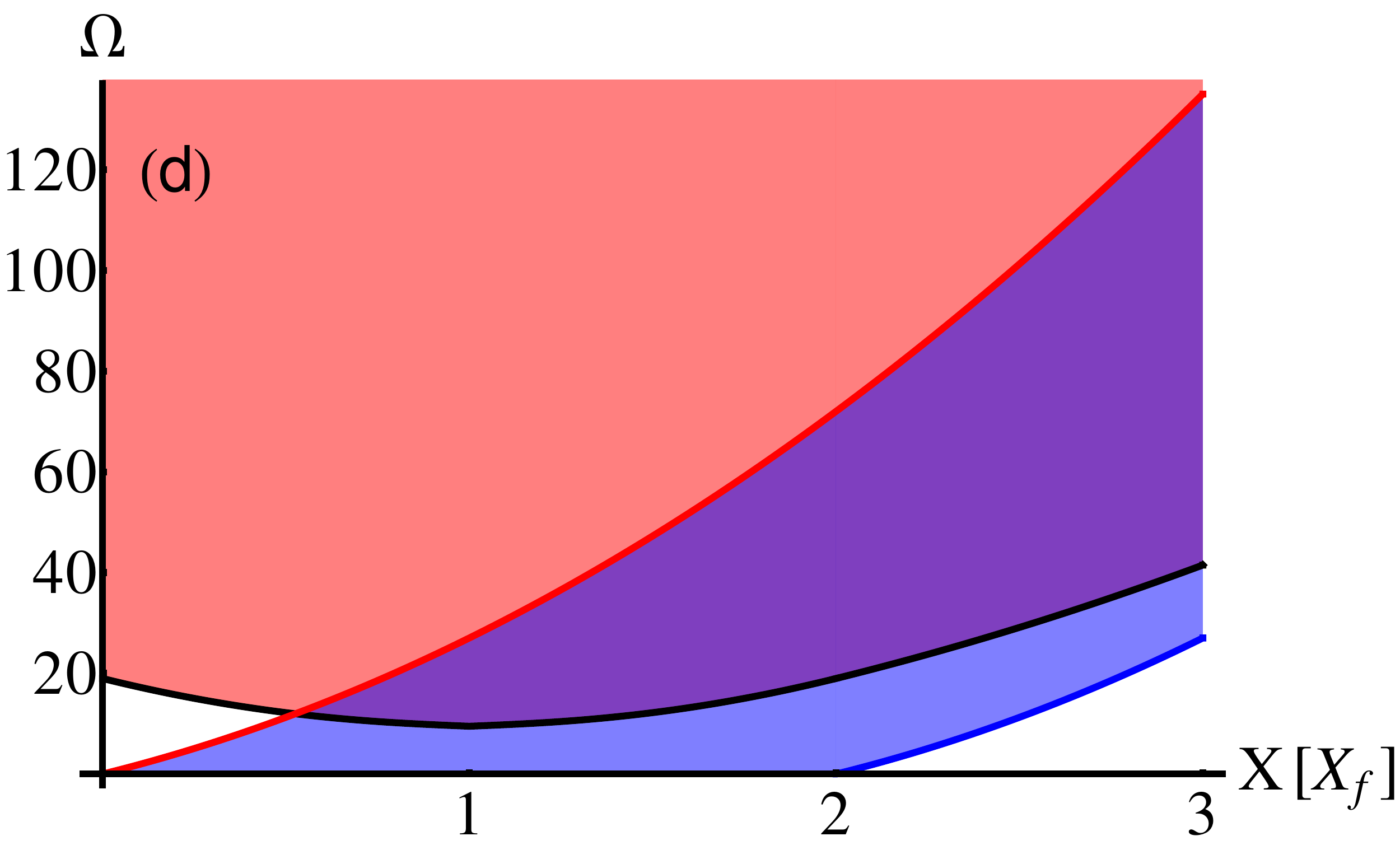}

\caption{(Color online) (a) Bandstructure of the BHZ model with indicated intraband (interband) transitions, green arrow (red and black, dashed arrow),
and finite chemical potential. (b)-(d) Boundaries of the spectrum for $X_{f}\in\left\{ 0.1,1,3\right\} $. Interband
spectrum in red, intraband spectrum in blue and mixed area in purple.
\label{fig:AB bandstructure}}

\end{figure}

The bandstructure of the BHZ model without mass and particle-hole
symmetry breaking is shown in Fig.~\ref{fig:AB bandstructure} (a).
 The  interband single-particle excitations lying lowest in energy are symmetric in momentum as shown by the red arrow in Fig.
 \ref{fig:AB bandstructure} (a), going from $-\boldsymbol{X}$ to $\boldsymbol{X}$. Due to particle-hole symmetry, this leads to nesting
 and thus one expects these excitations to dominate the interband spectrum.
 Interband excitations as indicated by the dashed, black arrow on the other hand,
 going from momentum $\boldsymbol{X}+\boldsymbol{X_f}$ to $\boldsymbol{X_f}$ with $\boldsymbol{X}\Vert\boldsymbol{X_f}$, are suppressed due to
 imperfect nesting of the different sized electron and hole cones, as well as by a small overlap factor. The latter can
 be cured by introducing a large negative mass, as will be shown in Sec. \ref{sec_BHZ_mass}. Then these excitations have a considerable
 influence onto the polarization for small energies, helping with the
  formation of interband plasmons, following the ideas presented in Sec. \ref{(Anti-)Screening and intrinsic plasmons}.
  In the pure Dirac system, these processes are forbidden by helicity.

By varying the doping level we can modify the excitation spectrum of the system [see Figs. \ref{fig:AB bandstructure} (b)-(d)] to resemble that 
of a Dirac system ($X_{f}\ll1$) or of a 2DEG ($X_{f}\gg1$), or to obtain an intermediate behavior ($X_{f}\sim1$). 
In the pictures we highlight the boundaries of the excitation spectra, with the red area corresponding to the interband spectrum and 
the blue area to the intraband spectrum. 
The overlap between the two is indicated by the purple area.
The boundaries of the spectra vary from the linear graphene behavior to the $q^{2}$ dependence of the 2DEG.
In general, the mixing of linear and quadratic dispersion leads to an overlap of the inter- and intraband spectrum. 
This affects the visibility of the interband plasmons, which can be hidden due to strong single-particle damping.


\paragraph{Weak doping of $X_{f}=0.1$: }

The extrinsic ($k_{f}$) and intrinsic ($q_0$) scales of the system are separated
by one order of magnitude. 
As the Fermi surface lies in the (almost) linear part of the spectrum, we expect that on the $k_{f}$
scale we resemble graphene. 
The physics on the $q_0$ scale on the other hand should be
more or less untouched by the doping, and the system should behave as in the intrinsic limit.

\begin{figure}
\includegraphics[width=4.2cm]{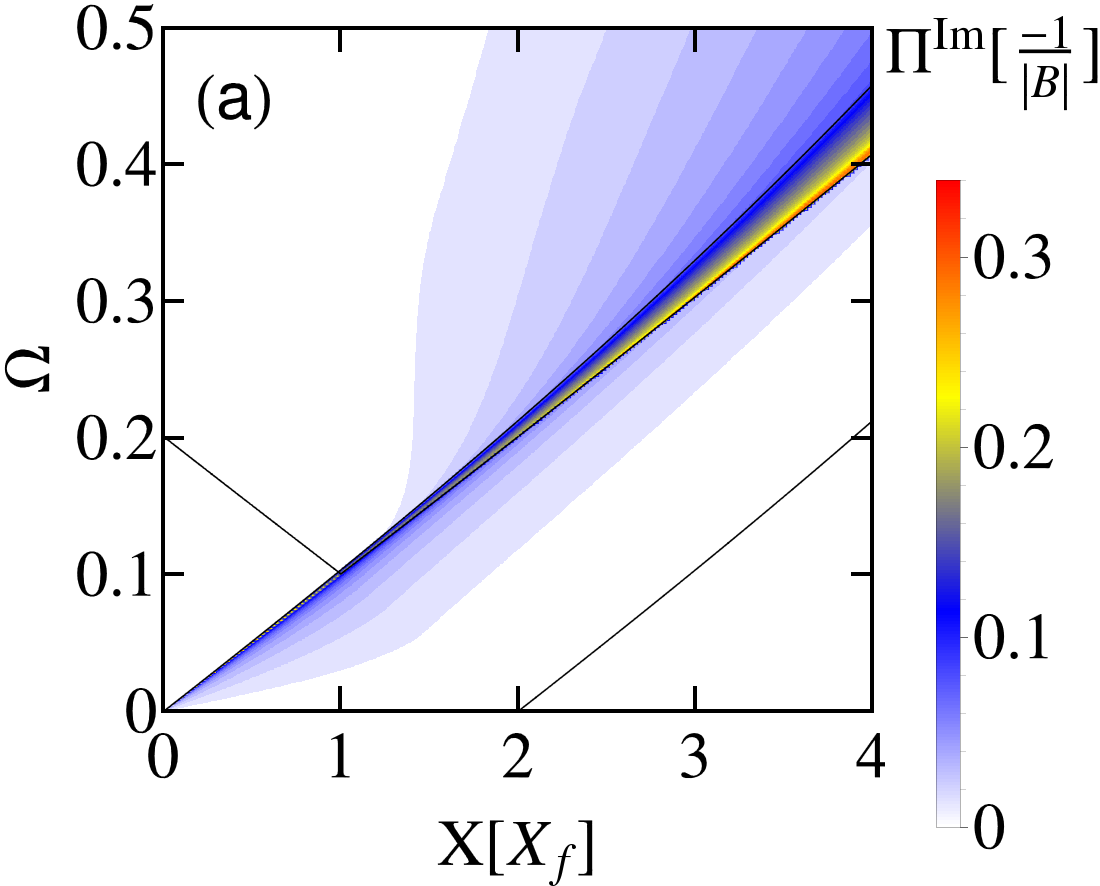}
\includegraphics[width=4.2cm]{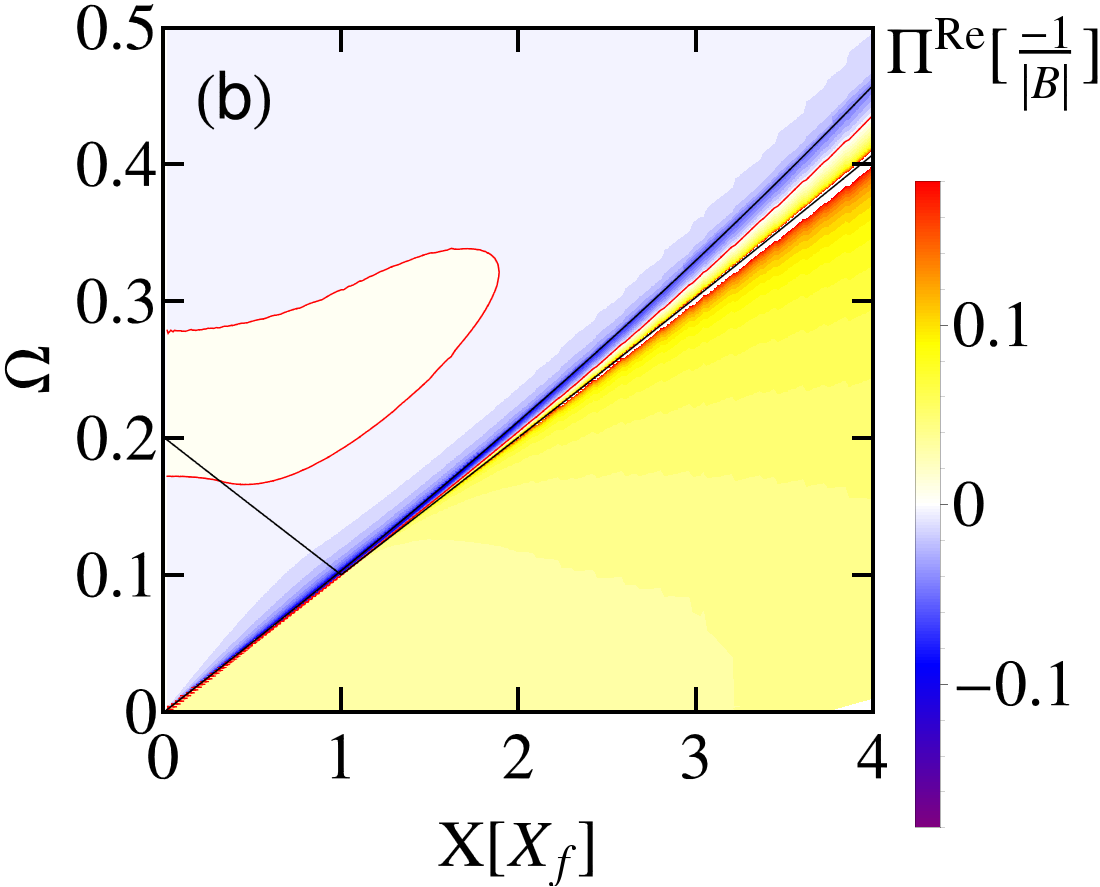}

\caption{(Color online) Imaginary (a) and real part (b) of the polarization function for $X_{f}=0.1$. The red line indicates 
$\Pi^{Re}=0$. 
\label{fig:ImP BHZ Xf01}}

\end{figure}

We plot $\Pi^{R}$ in Fig.~\ref{fig:ImP BHZ Xf01}.
 Comparing panel (a) to Fig. \ref{fig:Pi Graphene},
one finds good agreement with the Dirac case. The biggest deviation
is found in the peak of $\Pi^{Im}$ at $\Omega_{min}^{inter}$,
which is not symmetric as for a Dirac system due to the overlap of inter- and intraband spectrum [Fig.~\ref{fig:AB bandstructure} (b)].
The finite quadratic part in the spectrum cures the divergency formerly occuring in the Dirac limit.
The real part of $-\Pi^{R}$ is strongly negative only at the upper boundary of the intraband spectrum. 
This indicates that for small interactions, only one plasmon will dominate the excitation spectrum on the Fermi scale.

\begin{figure}
\includegraphics[width=4.2cm]{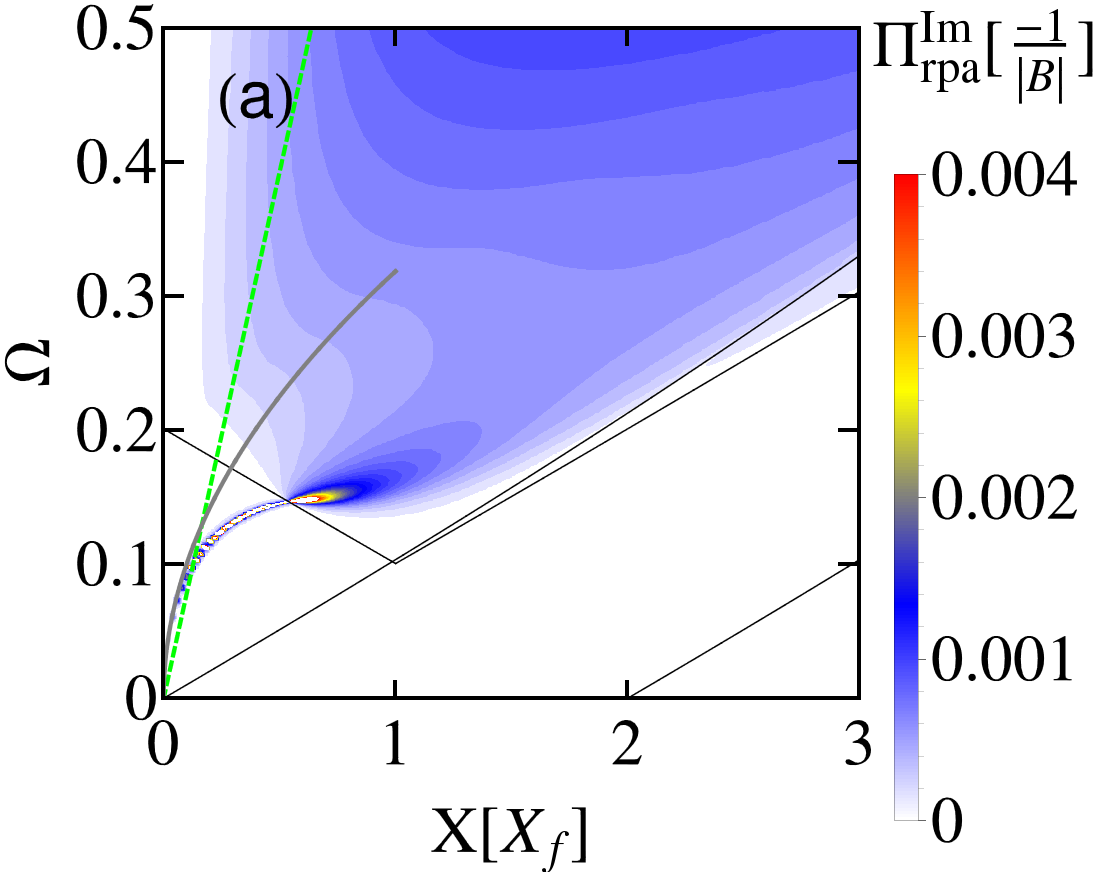}
\includegraphics[width=4.2cm]{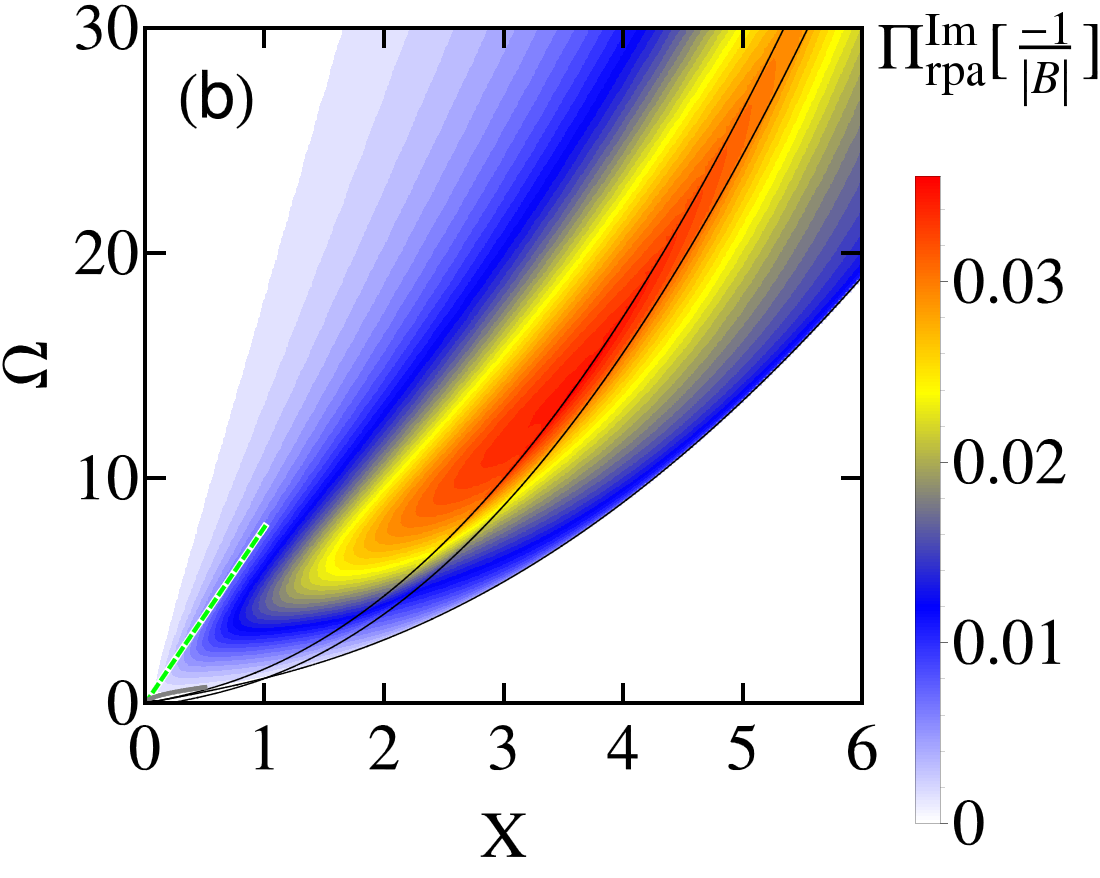}

\caption{(Color online) $\Pi^{Im}_{rpa}$ for $X_{f}=0.1$ with $\alpha=10$ on the $k_f$ scale (a) and the $q_0$ scale (b).
We add an artificial damping in the regions of $\Pi^{Im}=0$
to make the plasmons visible. \label{fig:Im epsi BHZ Xf01}}

\end{figure}

As we are interested in the regime where both inter- and intraband plasmons are visible, 
we look at the interacting spectrum, given in Fig.~\ref{fig:Im epsi BHZ Xf01}
by plotting $\Pi^{Im}_{rpa}$, for a strong interaction $\alpha=10$.
On the Fermi scale [panel (a)], the intraband plasmon absorbs all spectral weight from the intraband spectrum.
The dispersion agrees with the perturbative dispersion from the expansion in Eq.~(\ref{eq:Plas_Freq_BHZ}) in the limit $X\rightarrow0$, plotted as a gray curve. 
The green, dashed line shows the linear dispersion of the interband plasmon in the undoped limit, based on Eq.~(\ref{eq:Plasmonfrequency linear}). 
On the Fermi scale, it is not obvious
that there is an interband plasmon, although the interacting polarization function develops a smeared resonance around 
the perturbative interband plasmon dispersion for high momenta. 
Switching to the intrinsic scale, Fig.~\ref{fig:Im epsi BHZ Xf01} (b), one finds the interband plasmon, corresponding to the 
single peak in $-\Pi^{Im}_{rpa}$,
unperturbed by doping for momenta much larger than $k_f$. 
The dispersion is the same as for a plasmon in the undoped limit~\cite{juergens2014}. The two black lines near the peak are just the
 boundaries of the intraband excitation spectrum, which does not play a role here. 

As in the limit of $X\rightarrow0$ the interband plasmon dispersion scales linearly with $\alpha$, $\Omega_p\propto \alpha$ see Eq.~(\ref{eq:Plasmonfrequency linear}), 
while the intraband plasmon frequency is proportional to $\sqrt{\alpha}$,
lowering the interaction strength will lead to an overlap of the two resonances below some critical $\alpha$. 


\paragraph{Strong doping of $X_{f}=3$: }


Strong doping of the system significantly increases the spectral weight, as shown in Fig.~\ref{fig: f-sum rule} (b), 
with the increase of intraband excitations, while most of the interband excitations are Fermi-blocked, 
leading therefore to an effective decoupling of the two bands.
We expect the overall spectrum to be governed by intraband excitations and to resemble the spectrum of a 2DEG, as the Fermi surface lies in the (almost)
quadratic part of the spectrum.

\begin{figure}
\includegraphics[width=4.2cm]{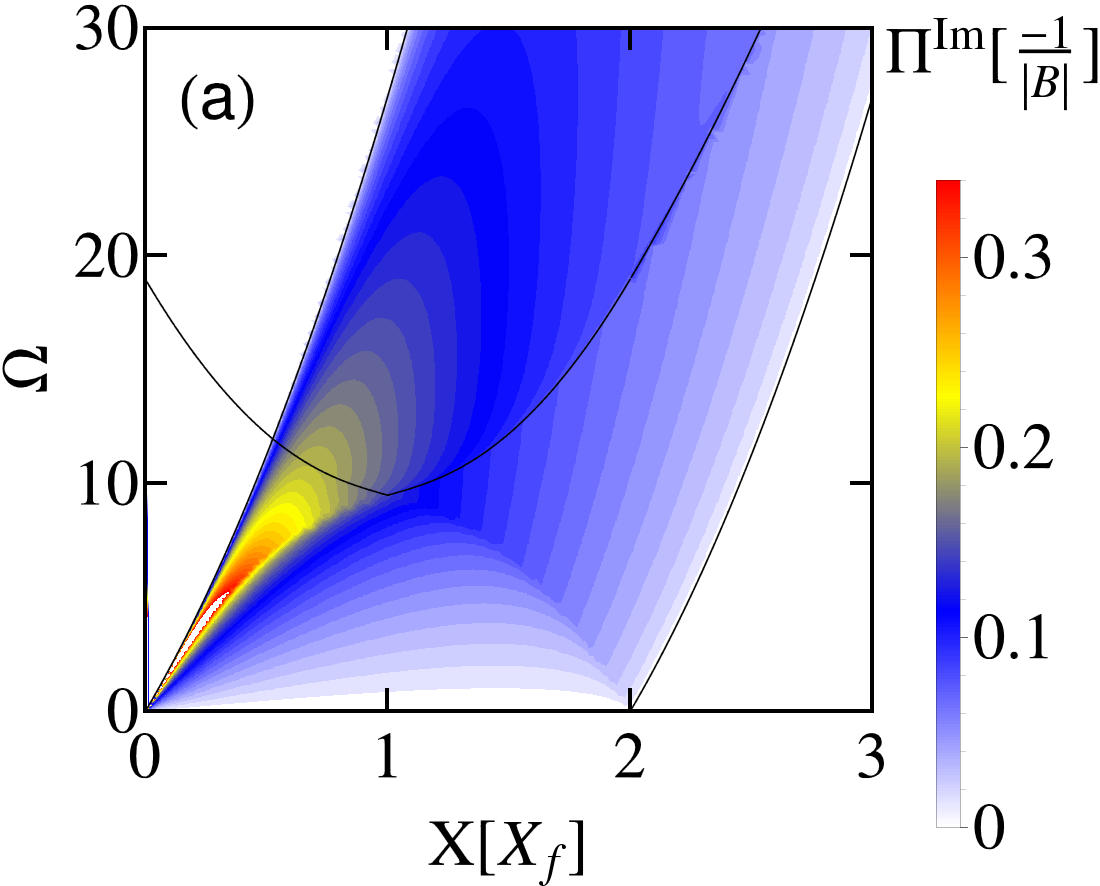}
\includegraphics[width=4.2cm]{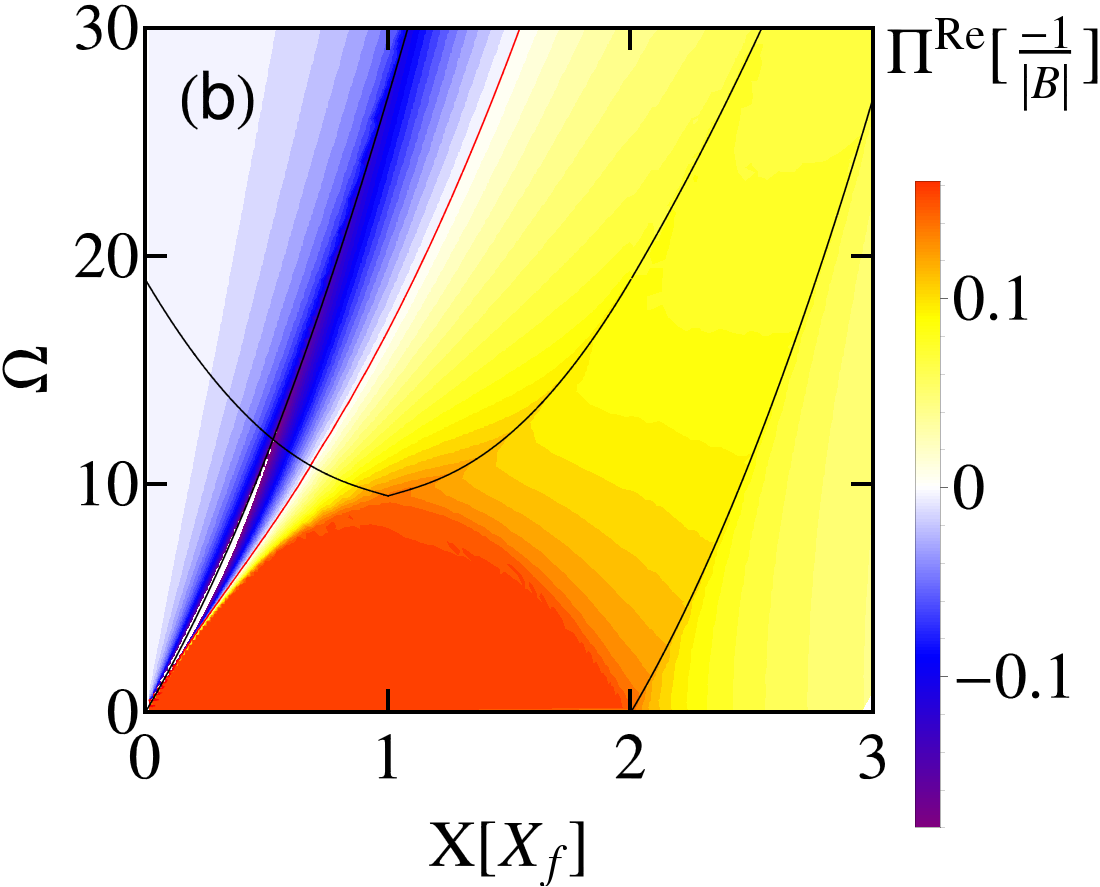}

\caption{(Color online) The imaginary (a) and real part (b) of the polarization function for $X_{f}=3$. The red line indicates 
$\Pi^{Re}=0$. \label{fig:ImP BHZ Xf3}}

\end{figure}

The corresponding $\Pi^{R}$ is plotted in Fig.~\ref{fig:ImP BHZ Xf3}.
 The single-particle spectrum in panel (a) is peaked at small momenta and at energies close to the upper
 bound of the intraband spectrum. 
 The interband part of the spectrum leads only to minor deviations from the 2DEG case
[compare with Fig.~\ref{fig:Pi 2DEG} (a)].
The real part of $-\Pi^{R}$ in panel (b) is strongly negative at the upper boundary of the intraband spectrum,
indicating that only a single intraband plasmon will dominate the interacting spectrum. 
We additionally note that the static limit property for which the polarization is a constant $ \tilde\Pi(X)=1$ for $X<2X_F$, discussed in Sec. \ref{Static limit}, 
extends also to an area of finite $\Omega$.

\begin{figure}
\includegraphics[width=4.2cm]{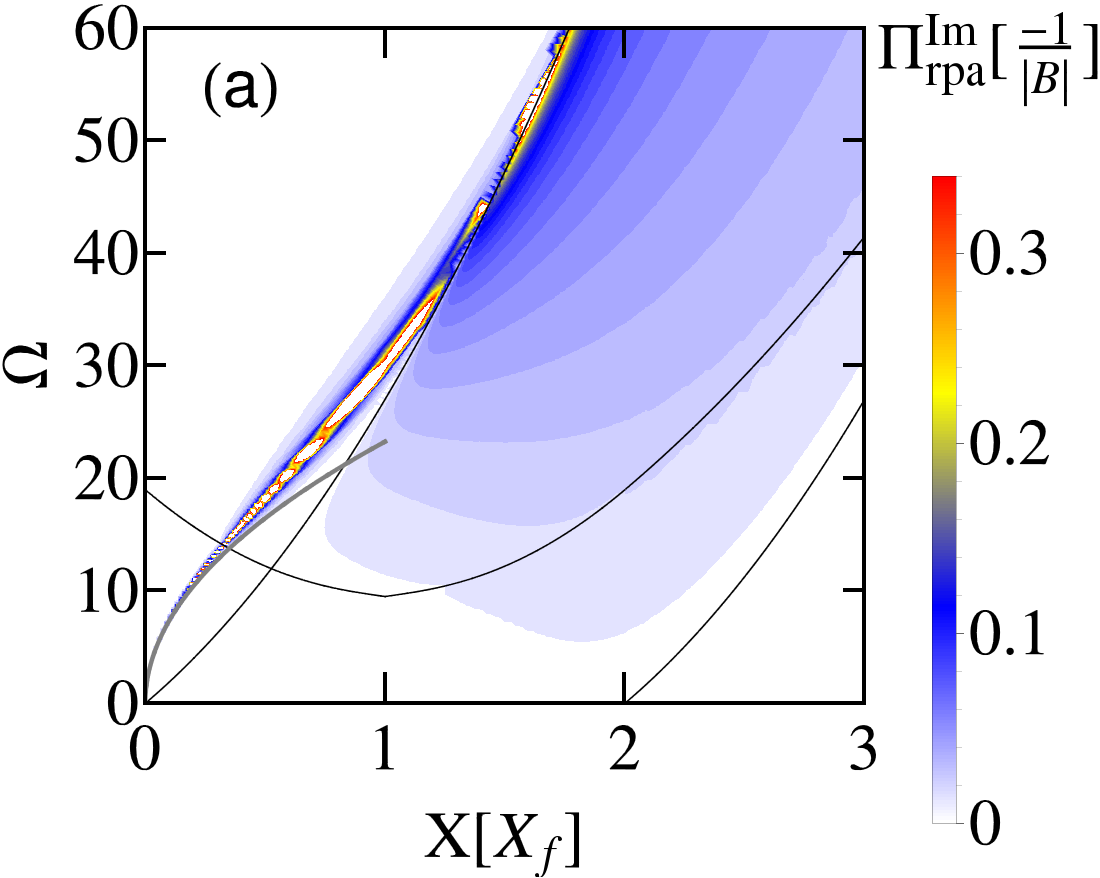}
\caption{(Color online) $\Pi^{Im}_{rpa}$ for $X_{f}=3$ and $\alpha=10$.
We add an artificial damping in the regions of $\Pi^{Im}=0$
to make the plasmons visible. \label{fig:Im epsi BHZ Xf3}}

\end{figure}

The interacting spectrum is shown in Fig.~\ref{fig:Im epsi BHZ Xf3} by
plotting $\Pi^{Im}_{rpa}$ for the interaction
strength  $\alpha=10$.
Even for this large Coulomb interaction, we only find the intraband plasmon. 
This is as expected due to the combined effects of Fermi blocking of interband excitations and increased spectral weight for intraband transitions. 
The interband plasmon lies in the large overlap of inter- and intraband spectrum, cf. Fig.~\ref{fig:Im epsi BHZ Xf01} (b), and it is therefore 
heavily damped and not visible in the overall spectrum.


\paragraph{Intermediate doping of $X_{f}=1$:}

For intermediate doping levels like $X_{f}=1$, a mixture of Dirac and 2DEG behaviour is expected, due to the similar importance of 
inter- and intraband excitations.

\begin{figure}
\includegraphics[width=4.2cm]{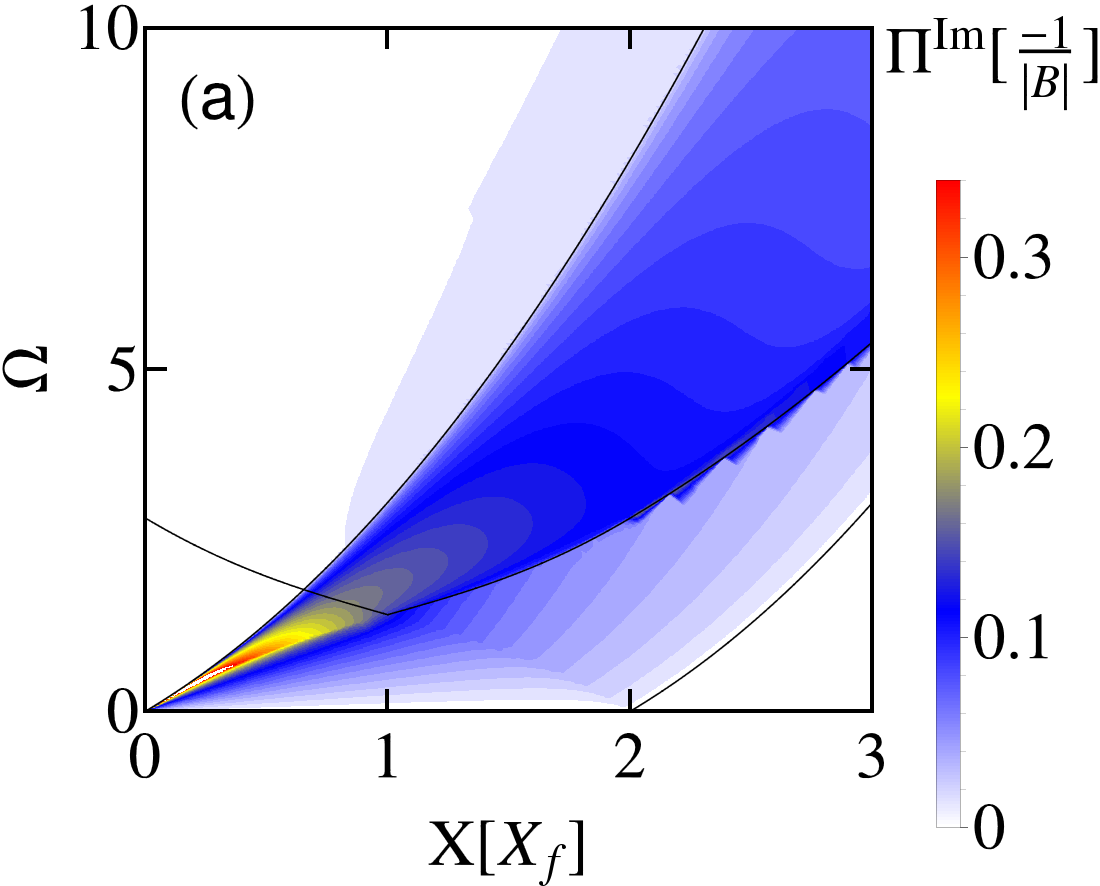}
\includegraphics[width=4.2cm]{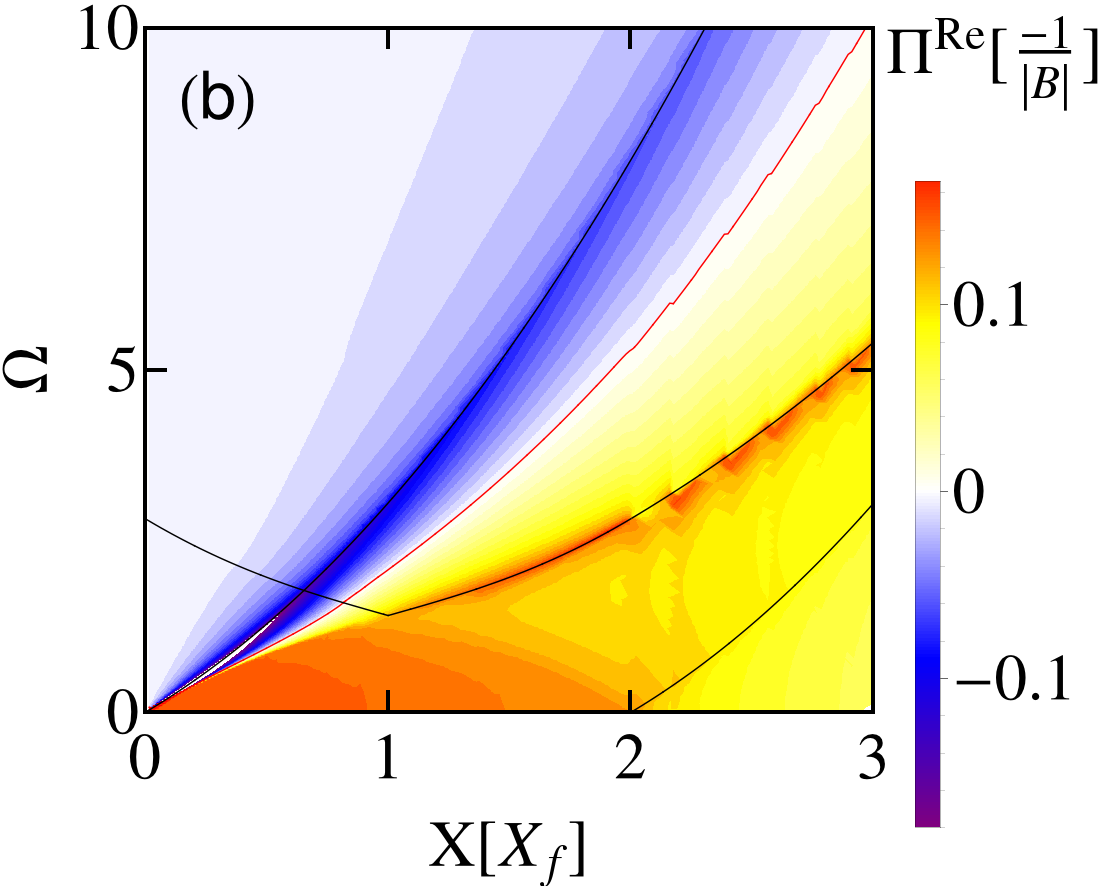}

\caption{(Color online) Imaginary (a) and real part (b) of the polarization function for $X_{f}=1$. The red line indicates 
$\Pi^{Re}=0$.  \label{fig:ImP BHZ Xf1}}

\end{figure}

We plot the polarization function $\Pi^{R}$ in Fig.~\ref{fig:ImP BHZ Xf1}.
 Indeed, the single-particle spectrum in panel (a) looks like a combination of Figs. \ref{fig:ImP BHZ Xf01} (a) and
\ref{fig:ImP BHZ Xf3} (a). While the the shape of the polarization resembles
the one of the 2DEG, the interband spectrum is now more pronounced
and even dominating for $X>2X_{f}$. Therefore we could expect both kinds of excitations giving rise to a plasmon mode.
The real part of $-\Pi^{R}$ in panel (b) shows again just a single minimum, following the upper boundary of the
intraband spectrum. The deviations from the constant behaviour
$ \tilde\Pi(X)=1$ for $X<2X_F$ in the case of intermediate doping, see Sec. \ref{Static limit}, are also found for finite $\Omega$.

\begin{figure}
\includegraphics[width=4.2cm]{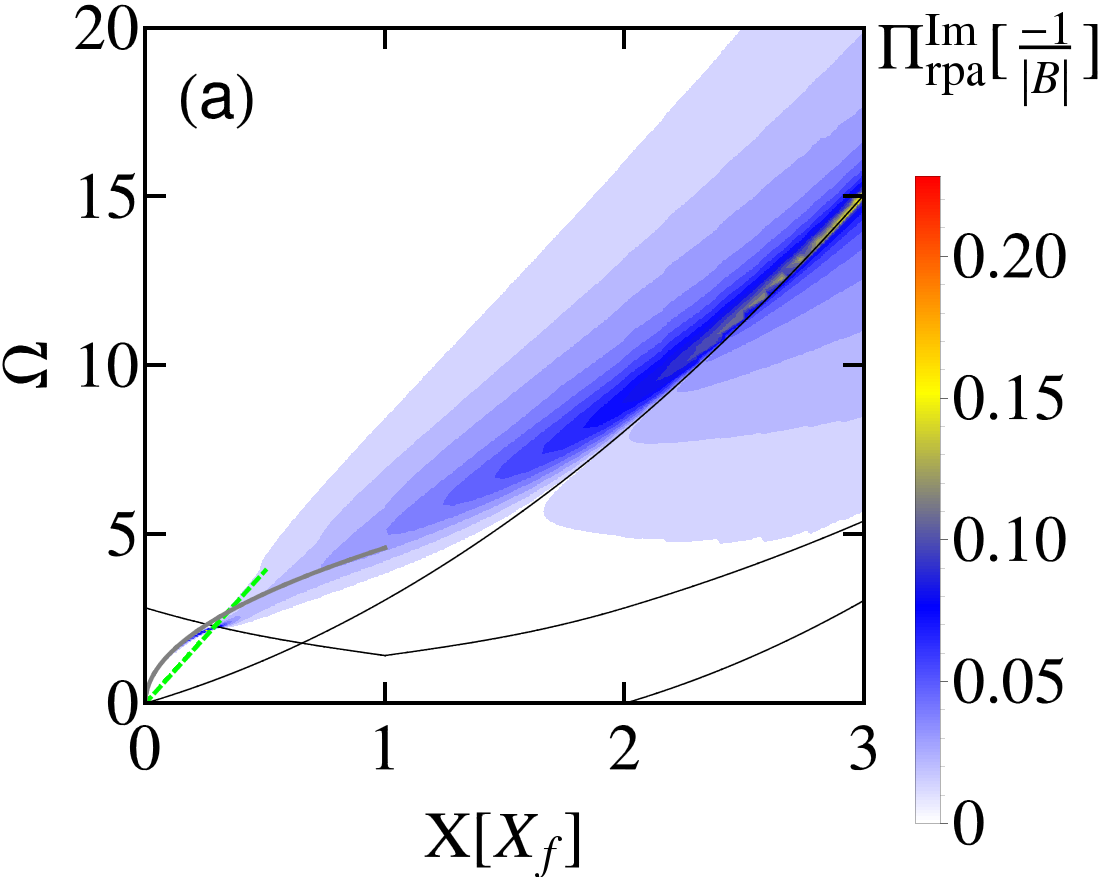}

\caption{(Color online) $\Pi^{Im}_{rpa}$ for $X_{f}=1$ with $\alpha=10$.
We add an artificial damping in the regions of $\Pi^{Im}=0$ to
make the plasmons visible. \label{fig:Im epsi BHZ Xf1}}

\end{figure}

The interacting spectrum is shown in Fig.~\ref{fig:Im epsi BHZ Xf1} by
plotting $\Pi^{Im}_{rpa}$ for an interaction
strength $\alpha=10$. %
It is dominated by a single resonance, lying above the intraband part of the single-particle spectrum.
For small momenta, this resonance corresponds to the intraband plasmon. Yet for intermediate momenta, a
comparison with the interband plasmon dispersion in Fig.~\ref{fig:Im epsi BHZ Xf01} (b) indicates that also the interband plasmon
contributes to the resonance. A clear distinction between the two is then not possible anymore.

In summary, doping the system offers the possibility to change the excitations
spectrum on the Fermi scale from a Dirac to a 2DEG type. The interacting excitation spectrum is
usually governed by a single intraband plasmon, while the interband plasmon is hidden in the
single-particle background. Only large interaction strengths offer a possibility to see both plasmons in the spectrum.
In the following, we will now analyze the influence of both broken p-h symmetry and finite masses, which both offer a way to
separate the two plasmons and make them visible in the total spectrum.


\subsection{Hg(Cd)Te quantum wells: BHZ model with finite $\xi_{D}$} \label{sec: Hg(Cd)Te quantum wells}

A broken particle-hole symmetry, $\xi_{D}\neq0$, with small or vanishing mass is the experimental relevant case
for HgTe QWs. 
It also offers the possibility of blocking the interband SPE
spectrum close to the minimal excitation energy $\Omega_{min}^{inter}$, resulting in less damped interband plasmons~\cite{juergens2014}.
\subsubsection{Spectrum}

\begin{figure}
\includegraphics[width=5.5cm]{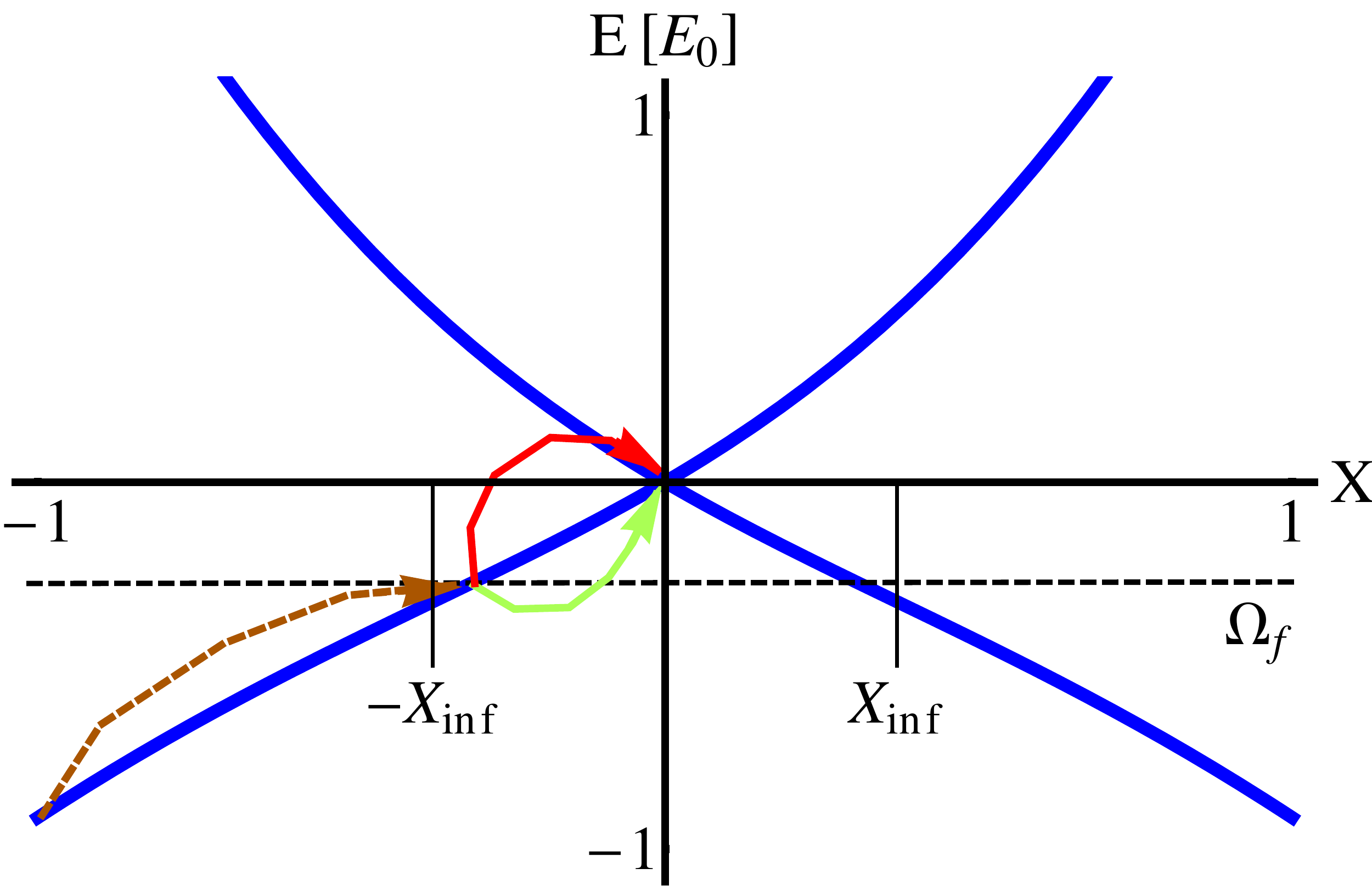}

\caption{(Color online) Bandstructure for $\xi_{D}=-0.5$. Both low energy interband excitations
(red arrow) and high energy intraband excitations (green arrow) involve
the Dirac point. \label{fig:BHZ spectrum finite D}}

\end{figure}

\begin{figure}
\includegraphics[width=4.2cm]{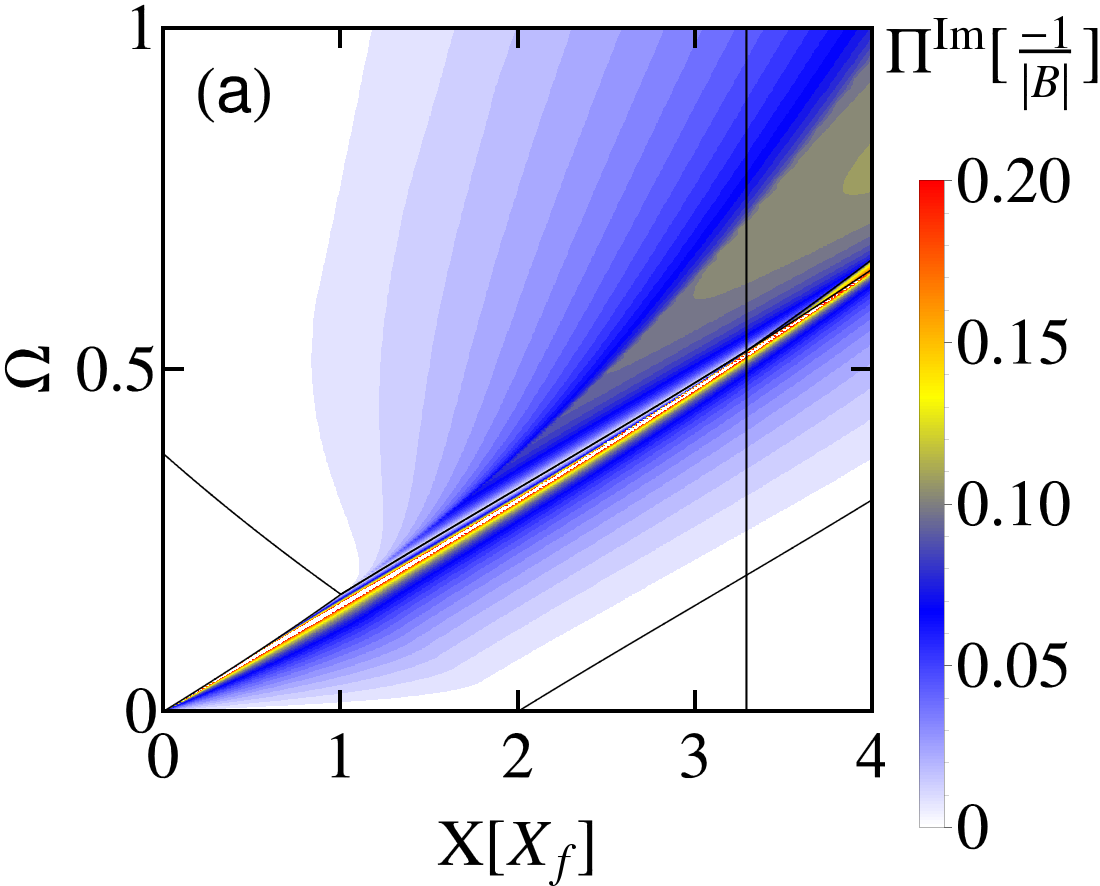}
\includegraphics[width=4.2cm]{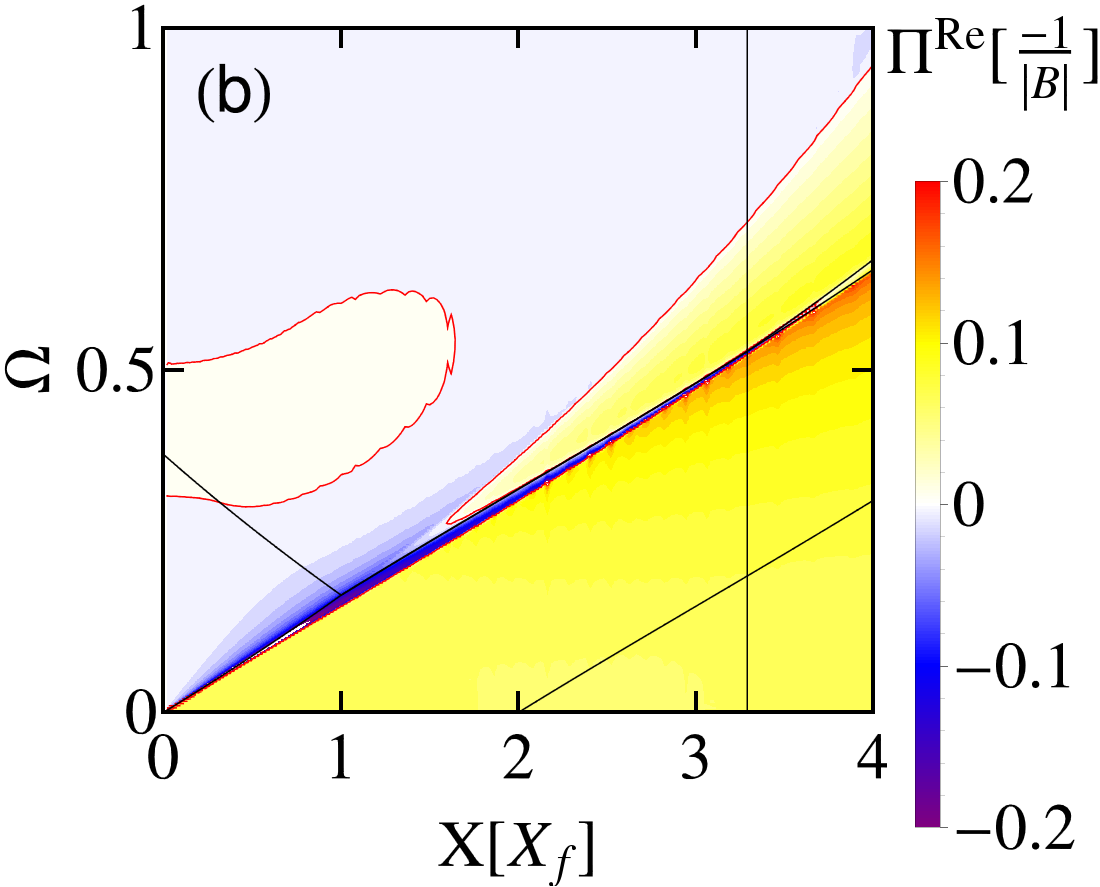}

\caption{(Color online) $\Pi^{Im}$ (a) and $\Pi^{Re}$ (b)
for $X_{f}=\frac{1}{2}X_{inf}=0.185$ and $\xi_{D}=-0.5$. The red line indicates 
$\Pi^{Re}=0$. \label{fig:Polarization BHZ ABkinf12}}

\end{figure}

\begin{figure}
\includegraphics[width=4.2cm]{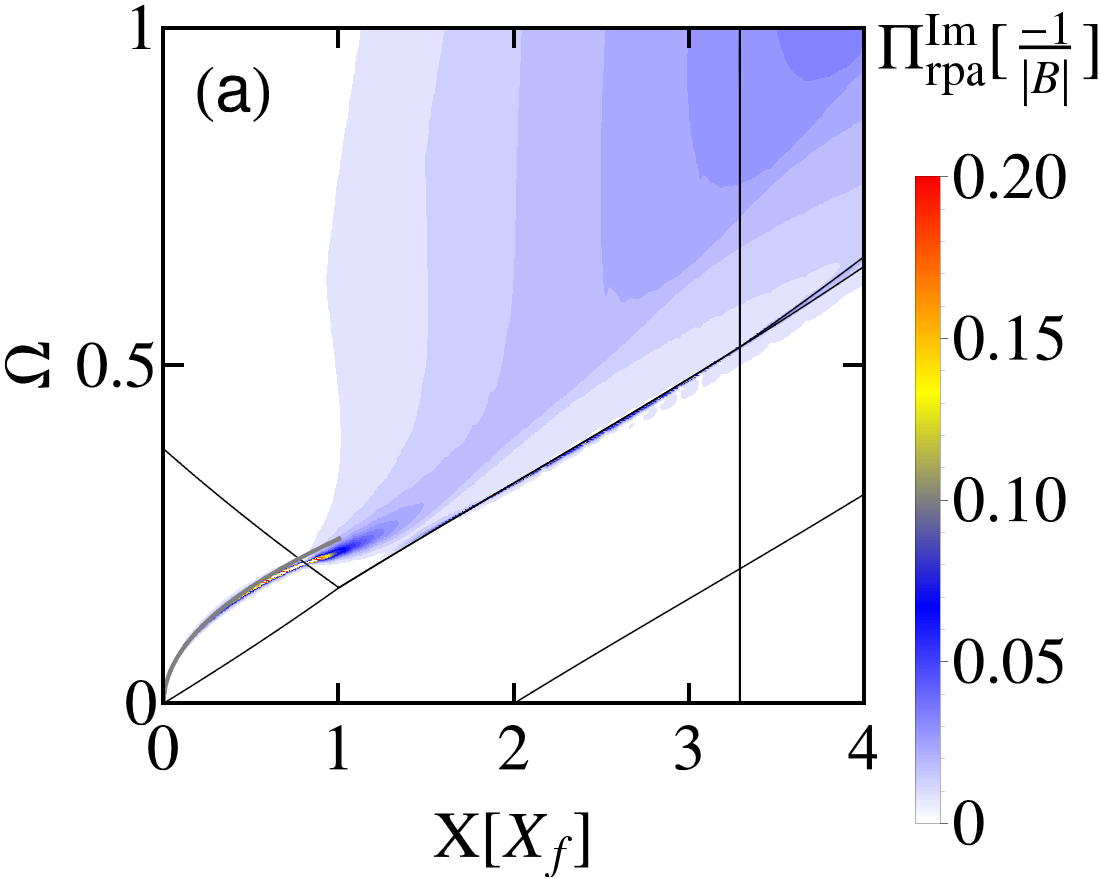}
\includegraphics[width=4.2cm]{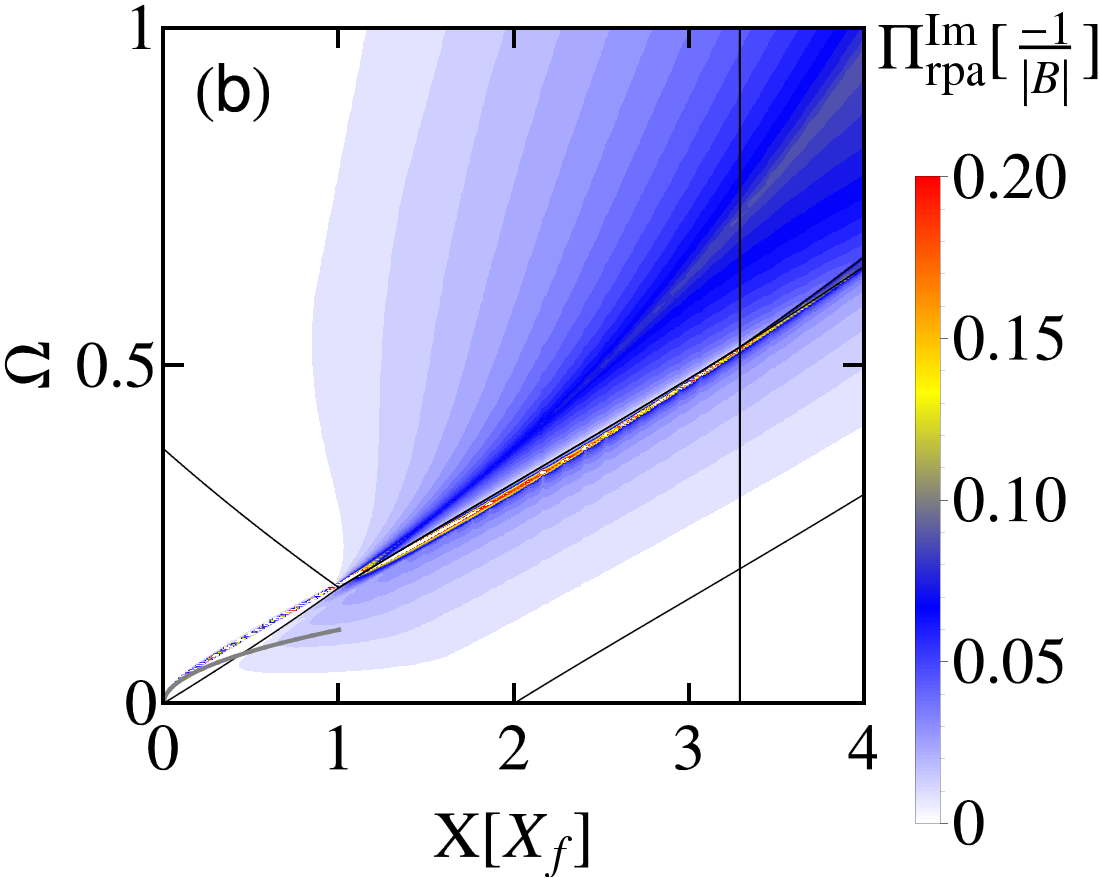}
\includegraphics[width=4.2cm]{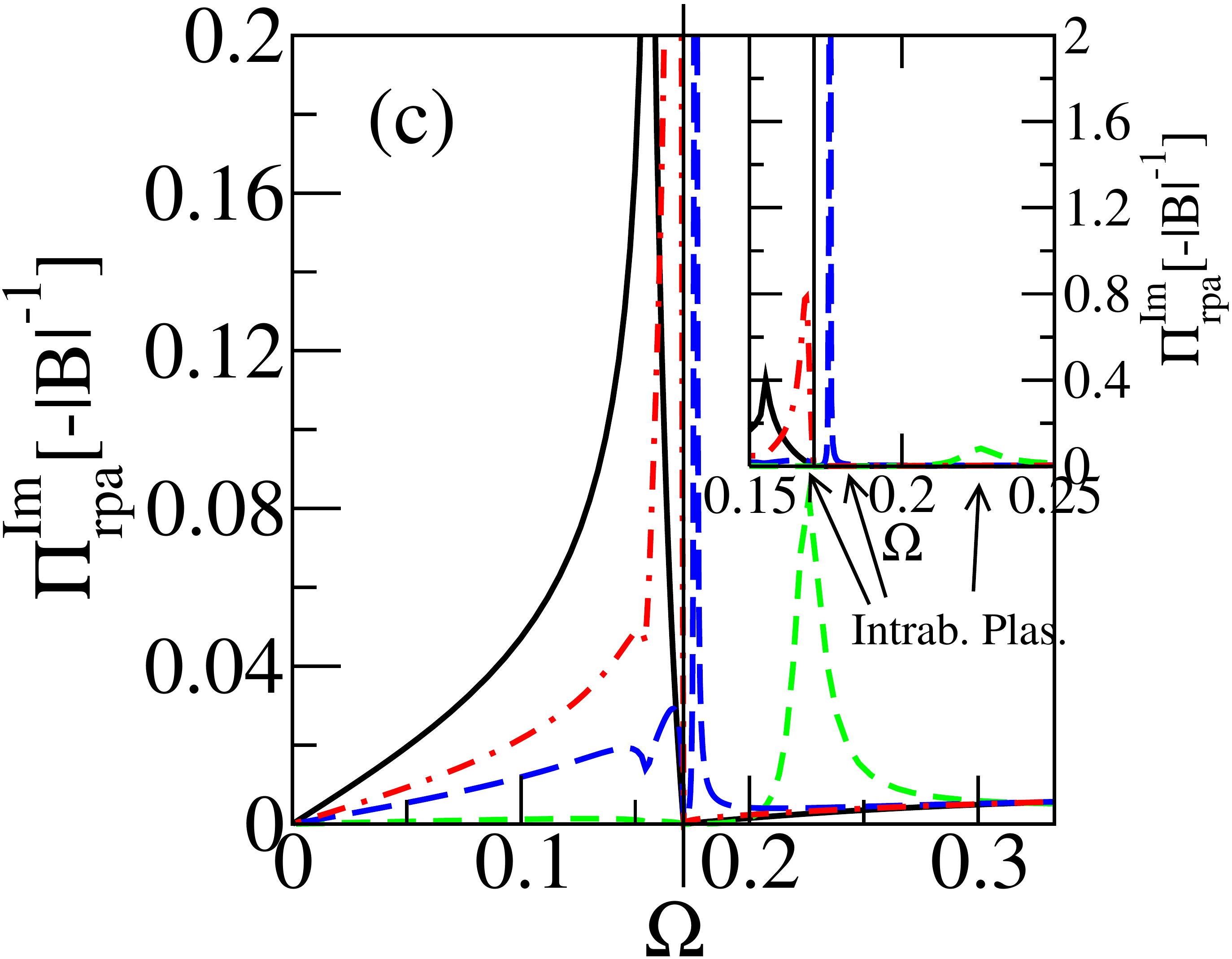}
\includegraphics[width=4.2cm]{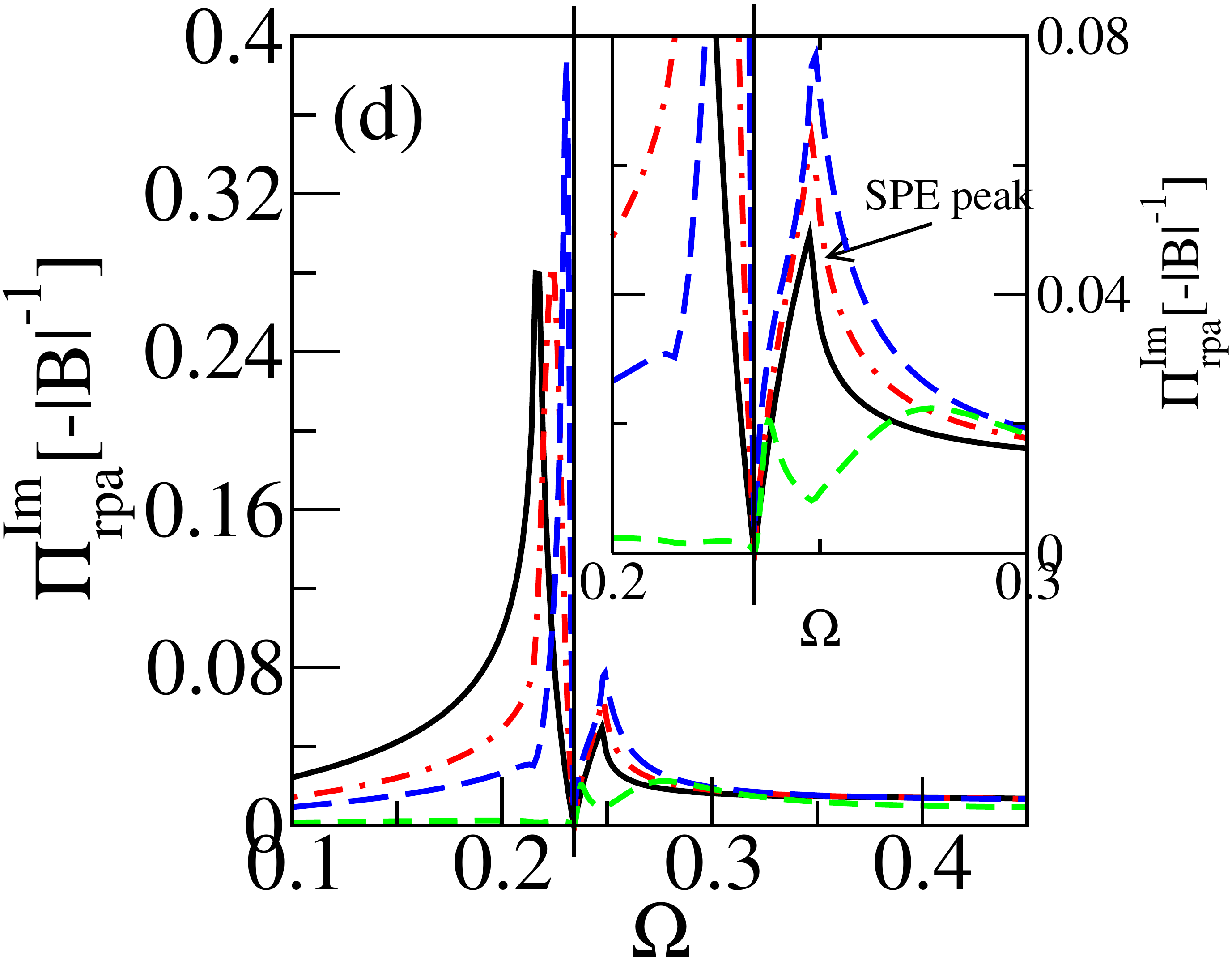}
\includegraphics[width=4.2cm]{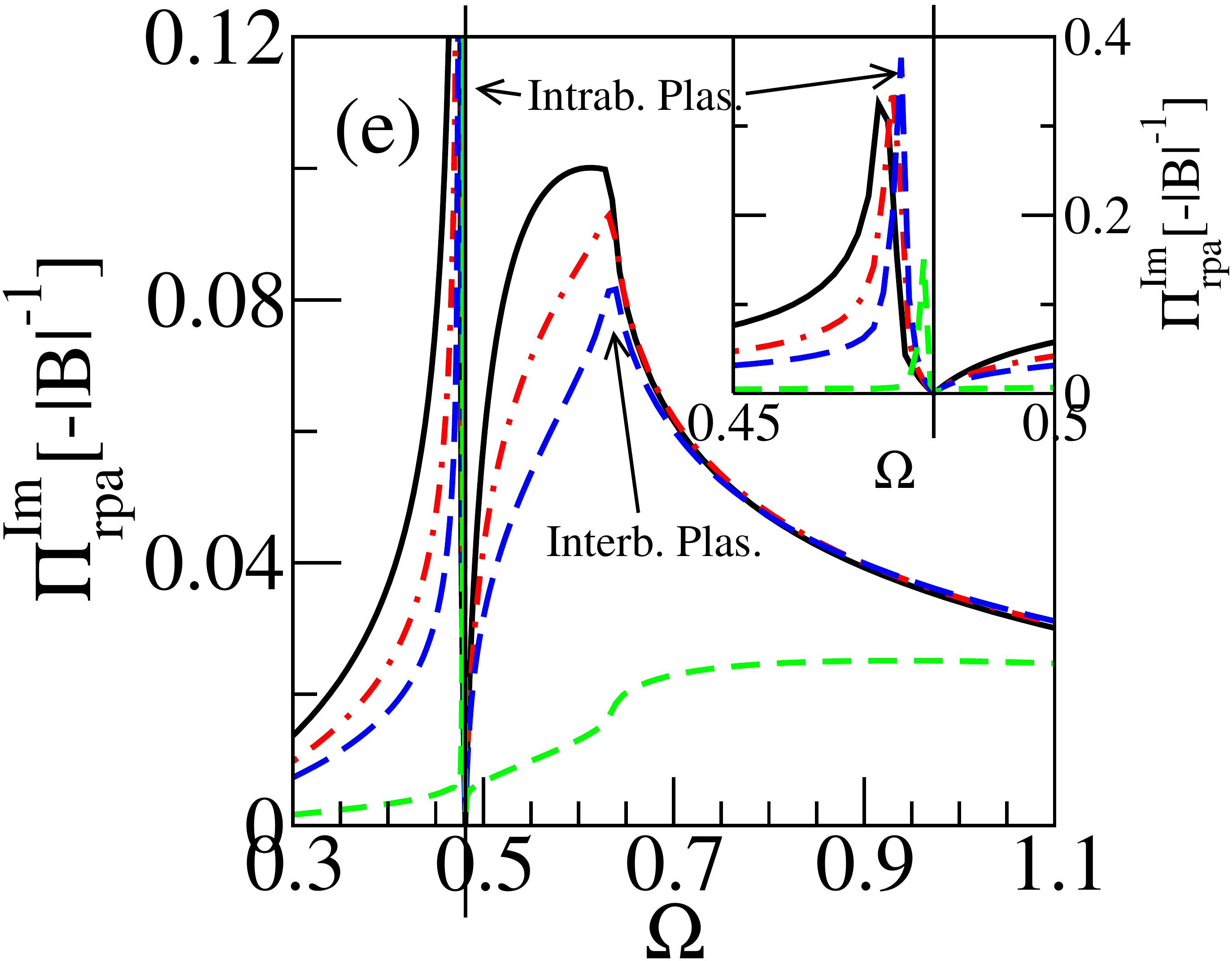}

\caption{(Color online) Interacting polarization function $\Pi^{Im}_{rpa}$ for $X_{f}=\frac{1}{2}X_{inf}=0.185$,
 $\xi_{D}=-0.5$, $\xi_{M}=0$ and $\alpha=2$ (a) and $\alpha=0.4$ (b). (c)-(e) show linecuts for fixed $X=X_f$, $X=1.4X_f$ and $X=3X_f$, respectively,
 with $\alpha \in \{0,0.2,0.4,2\}$ in black solid line, red dot-dashed line, blue long dashed line and green short dashed line, respectively.
 The black, vertical line separates the inter- and intraband SPE region. \label{fig:Im Eps BHZ ABkinf12}}

\end{figure}

Here, we want to use a similar effect for the intraband excitations in order to separate
the inter- and intraband spectrum as well as the two plasmon modes. 
The broken p-h symmetry introduces an inflection point into the spectrum, $\frac{\partial^2}{\partial X^2}\epsilon_{X,\lambda}=0|_{X=X_{inf}}$, with momentum $X_{inf}$ and energy $\Omega_{inf}$.
For $\xi_D<0$, it lies in the hole part [$\lambda=-1$] of the spectrum. 
With a sufficiently small Fermi momentum, $X_f\lesssim X_{int}$, the highest
energy intraband excitations involve the Dirac point for momenta on the order of the Fermi momentum, see Fig.~\ref{fig:BHZ spectrum finite D}.
The same is true for the lowest energy interband excitations. 
Due to the vanishing density of states at the Dirac point, both kind of excitations are suppressed, 
and therefore inter- and intraband SPE spectrum are effectively separated in energy and momentum.
This situation is shown in Fig.~\ref{fig:Polarization BHZ ABkinf12} (a)
and Fig.~\ref{fig:Im Eps BHZ ABkinf12} (c)-(e), where the imaginary part of the polarization goes to zero between inter- and intraband parts of the spectrum, fully
separating them.
There, we choose $X_{f}=\frac{1}{2}X_{inf}$, $\xi_{D}=-0.5$ and $\xi_{M}=0$. 
This blocking effect holds for small momenta up to roughly $2X_{inf}$, indicated by the black vertical line in Fig.~\ref{fig:Polarization BHZ ABkinf12}
at $X\approx3.2$. 
For larger momenta the high energy intraband excitation go from deep in the valence band directly to the Fermi surface - the blocking
effect of the Dirac point is gone, see dashed arrow in Fig. \ref{fig:BHZ spectrum finite D}.

In Fig.~\ref{fig:Polarization BHZ ABkinf12} (b), $\Pi^{Re}$ shows one major difference in comparison to the p-h symmetric case of weak doping 
in Fig.~\ref{fig:ImP BHZ Xf01} (b). At the border of intra- and interband spectrum a strong antiscreening region is formed. For sufficiently low $\alpha$ a 
plasmon should exist there, clearly separated from the second antiscreening region at higher $\Omega$, giving rise to the possibility of observing both intra- and interband plasmons.
This can be seen in Fig.~\ref{fig:Im Eps BHZ ABkinf12}, where we plot $\Pi^{Im}_{rpa}$
for $\alpha=2$ (a) and $\alpha=0.4$ (b). Panel (c) - (e) show line cuts for fixed momenta $X \in \{1,1.4,3\}X_f$ and different interaction strengths
$\alpha \in \{0,0.2,0.4,2\}$. 

For large interaction strength $\alpha=2$, the intraband plasmon decays into the interband SPE spectrum, see panel (a) for $X\approx X_f$ and the green short dashed line in panel (c). 
Most of the spectral weight stays there also for larger momenta, as $\Pi^{Im}_{rpa}$ is close to 
0 in the intraband SPE region and the resonance between 
inter- and intraband SPE spectrum is weak. The latter can be best seen in the insets of panel (d) and (e), represented by 
the green short dashed line peaked slightly above [(d)] or below [(e)] the black vertical line separating intra- and interband SPE region. 
Yet even with the peak being small, it indicates the formation of a slighly damped plasmon, but with small spectral weight. 
The missing spectral weight is transferred to higher energies into the interband SPE region. For intermediate momenta, a second plasmon branch forms, see panel (a) for $X_f<X<2X_f$ and the second peak of the green short dashed line in the inset of panel (d). For even higher momenta, $X>2X_f$, it overlaps with the forming interband plasmon leading to a broad charge resonance without clear peak, see green short dashed line in panel (e) for $\Omega>0.7$. 

The picture changes for smaller interaction strength. 
For $\alpha=0.4$, the intraband plasmon decays in the region between inter- and intraband SPE spectrum, indicated by the strong peak of the blue long dashed line in 
panel (c). As the single-particle excitations in this region are suppressed 
due to the Dirac point, the plasmon leads to a high and narrow peak of $\Pi^{Im}_{rpa}$. 
Considering larger momenta $X>X_f$, the resonance is split: one part forms an intraband plasmon in the intraband SPE region, 
see blue long dashed line peaked slightly below the black vertical line in panels (d) and (e). The second part stays in the interband SPE region, 
where it enhances the SPE peak [black line in the inset of panel (d)] for intermediate momenta $X_f<X<2X_f$. 
For momenta $X\gtrsim2X_f$, an interband plasmon forms, as shown in panel (e). 
There, the broad single-particle peak [black line] around $\Omega=0.6$ gets reshaped into a clear peaked resonance [blue long dashed line] - the interband plasmon.


\subsubsection{Experimental parameters}
 Taking the experimental parameters from Sec. \ref{sec: experimental_parameters}, $q_{0}\approx0.4\ \frac{1}{\mathrm{nm}}$
 and $E_{0}\approx 140\ \mathrm{meV}$,
 one finds for the plots in Fig.~\ref{fig:Im Eps BHZ ABkinf12} the Fermi momentum
 $k_f\approx0.07\ \frac{1}{\mathrm{nm}}$ and chemical potential $\mu\approx -24\ \mathrm{meV}=-\hbar\cdot36\ \mathrm{THz}$.
 The plot range is therefore $q\in\left[0,0.74\right]q_{0}=\left[0,0.3\right]\frac{1}{\mathrm{nm}}$
 and $\omega\in\left[0,1\right]\mathrm{\frac{E_{0}}{\hbar}}=\left[0,210\right]\mathrm{THz}$ and
 thus of the right order of magnitude for experimental techniques like Raman spectroscopy or
 electron loss spectroscopy.


\subsubsection{Spectral weight and the f-sum rule}

Both plasmonic resonances in Fig.~\ref{fig:Im Eps BHZ ABkinf12} (b) overlap 
for $X\approx X_f$, before they separate for higher momenta. Therefore the question arises whether one can really speak
of a clear distinction between inter- and intraband plasmons for larger momenta. 
Here, we want to study the f-sum rule and thus the spectral weight of the different
excitations.

\begin{figure}
\includegraphics[width=4.2cm]{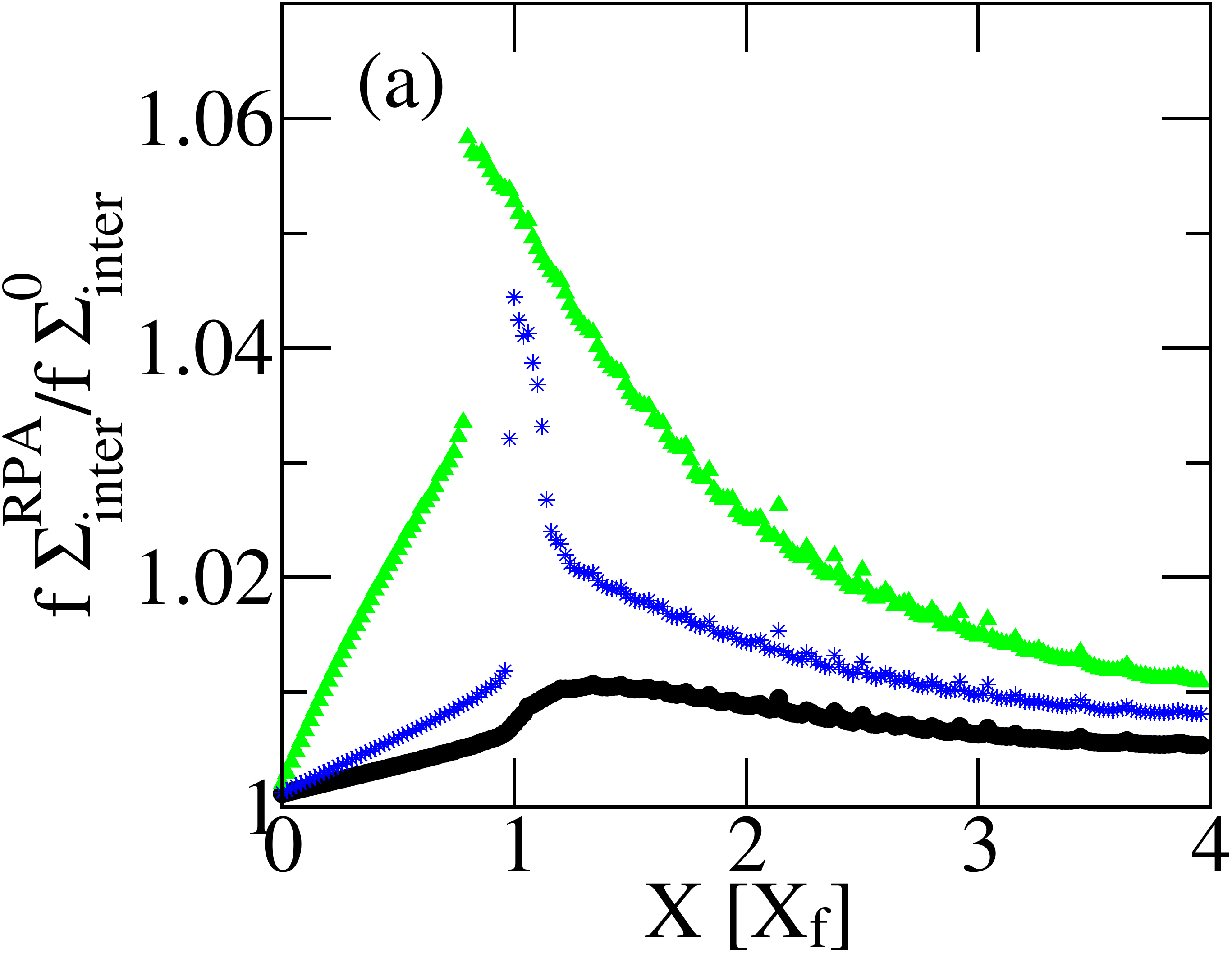}
\includegraphics[width=4.2cm]{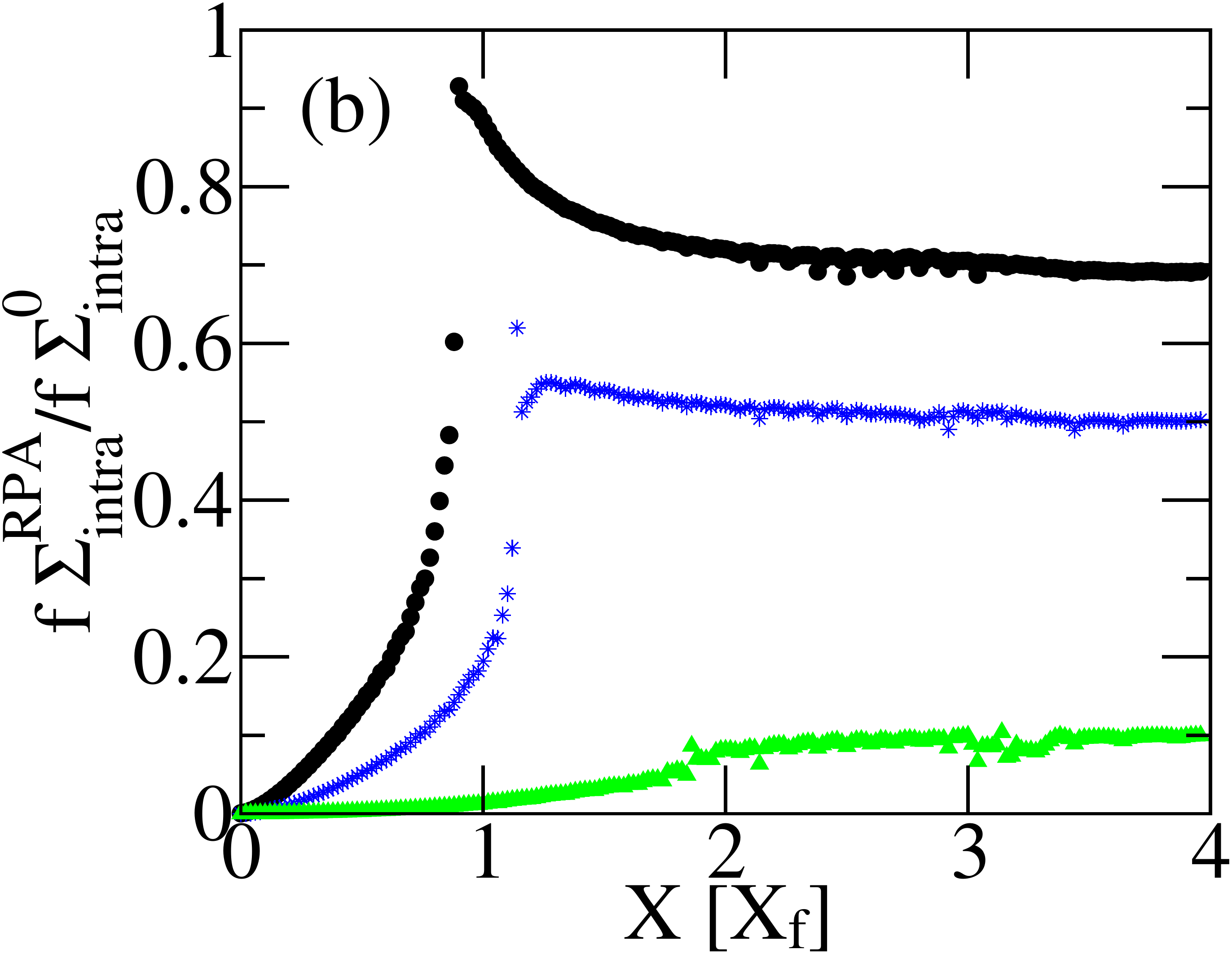}

\caption{(Color online) (a) The ratio $\frac{f\sum_{inter}^{RPA}}{f\sum_{inter}^{0}}$ of
interacting over non-interacting f-sum rule of the interband excitations.
Black dots are for $\alpha=0.2$, blue stars stand for $\alpha=0.4$
and green triangles for $\alpha=2$. (b) The same for the ratio $\frac{f\sum_{intra}^{RPA}}{f\sum_{intra}^{0}}$
of the intraband excitations. For all plots: $X_{max}=4X_{f}$ and $\beta=4$. \label{fig:f-sum numerical doped}}

\end{figure}

The relative deviations of numerical to analytical f-sum rule are again of the order $10^{-3}$ and thus negligible. Fig.~\ref{fig:f-sum numerical doped} (a)
shows the ratio of spectral weight in the interband SPE region for the interacting over the non-interacting case, $\frac{f\sum_{inter}^{RPA}}{f\sum_{inter}^{0}}$, 
and panel (b) the same for the intraband SPE region. The intraband plasmon lying between these two regions for $X<X_f$ is excluded. As for 
cutoffs $\beta>1$ one usually has $f\sum_{inter}^{0}\gg f\sum_{intra}^{0}$, transfer of spectral weight from one region 
to the other can lead to quantitatively different relative changes of spectral weight in panels (a) and (b). 
As a key result, we find that there is always spectral weight missing in the intraband SPE region. 
For small momenta, $X<X_f$, the weight goes into the undamped intraband plasmon [this follows directly from the conservation of the f-sum rule for interacting and non-interacting systems], 
while at larger momenta it is transferred to higher energies into the
interband SPE region. 
Yet, the increase is only about $2\%$ at $X\gtrsim2X_f$, such that we can conclude that the plasmon between the inter- and intraband SPE region is
a pure intraband plasmon with a reduced spectral weight. 
The plasmon in the interband region is the interband plasmon we know already from the undoped 
system, see Fig.~\ref{fig:intrinsic spectra}, with a slight increased spectral weight from the intraband SPE region.


\subsubsection{Small gap $\xi_M\neq0$}

Deviations in the thickness of a Hg(Cd)Te QW lead to the opening of a small gap in the bandstructure, 
resulting in a topological trivial $\xi_M>0$ or non-trivial $\xi_M<0$ system.  
Apart from the possible  appearance of edge states, which is beyond the scope of this paper, a small mass 
works in opposition to the blocking effect of
finite $\xi_{D}$, as it generates a finite density of states for $X=0$.
In the following, we therefore show that the blocking effect of a finite $\xi_{D}$ is robust against the opening of small gaps. 

\begin{figure}
\includegraphics[width=4.2cm]{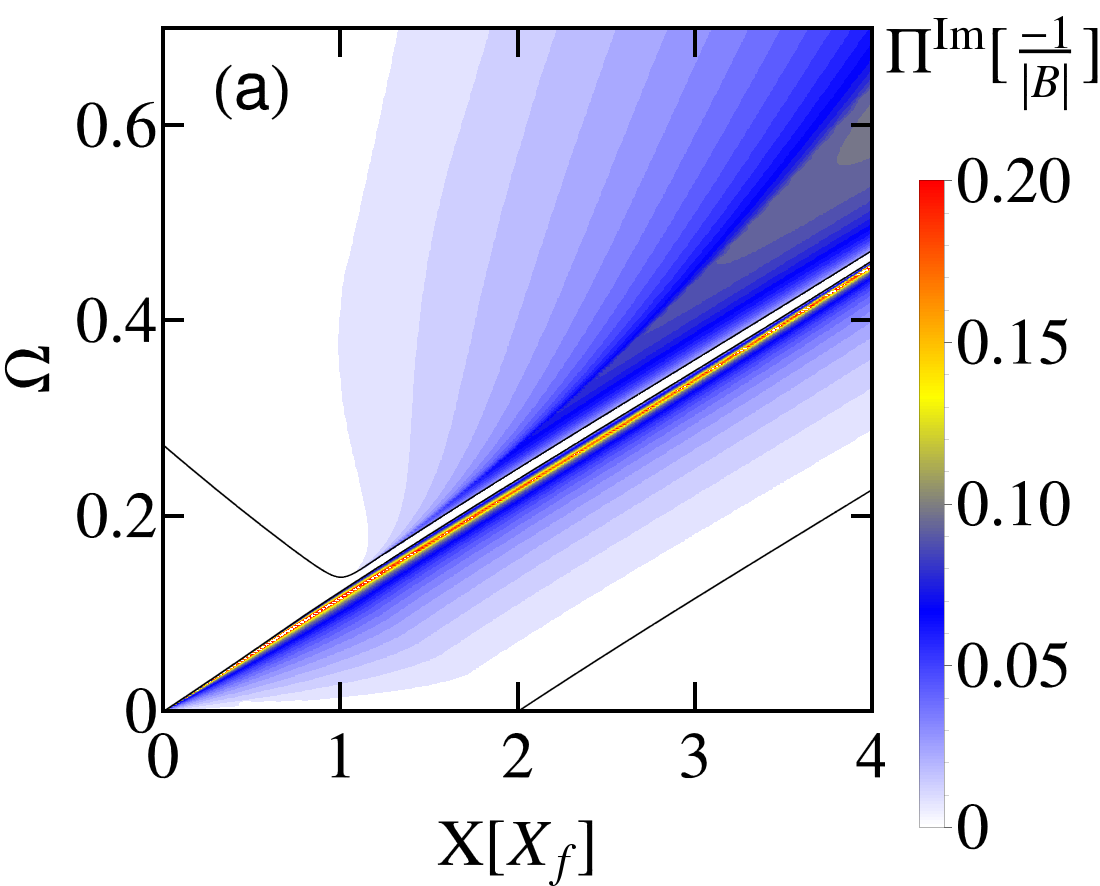}
\includegraphics[width=4.2cm]{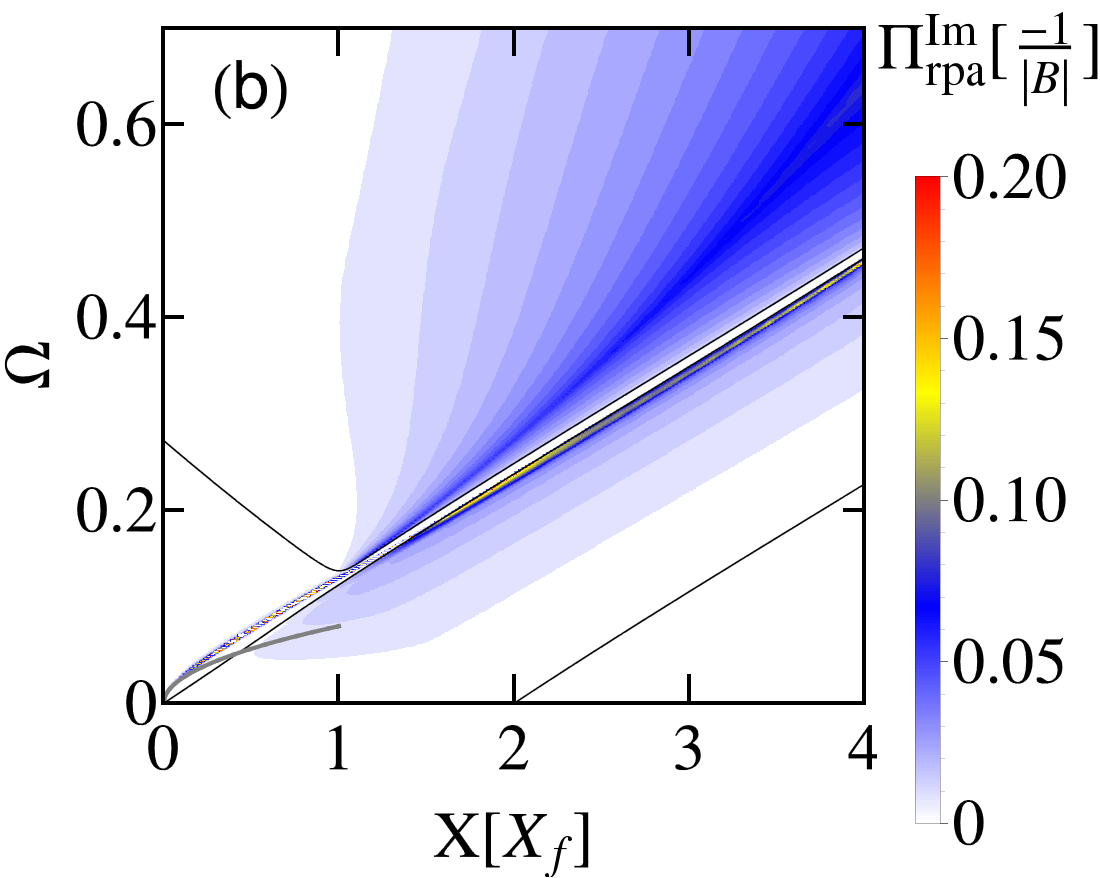}

\caption{(Color online) The imaginary part of the non-interacting polarization $\Pi^{Im}$ (a) and the interacting
 one $\Pi^{Im}_{rpa}$ (b) with $\alpha=0.4$. $X_{f}=0.133<\frac{1}{2}X_{inf}$, $\xi_{M}=0.01$ and
$\xi_{D}=-0.5$. \label{fig:Polarization BHZ MABkinf004}}

\end{figure}

In Fig.~\ref{fig:Polarization BHZ MABkinf004}
we plot the non-interacting and interacting spectrum for $\xi_{D}=-0.5$ and a small mass $\xi_{M}=0.01\approx1.4\ \frac{\mathrm{meV}}{E_{0}}$.
A comparison with Figs. \ref {fig:Polarization BHZ ABkinf12} and
\ref{fig:Im Eps BHZ ABkinf12} shows that the small mass has just the effect of separating the inter- and intraband SPE region additionally.
Thus we conclude that the idea of observing both plasmons in experiments is robust against slight deviations in the mass and therefore 
the thickness of the Hg(Cd)Te QW.


\subsection{Topology: BHZ model with large $\xi_{M}\neq0$ \label{sec_BHZ_mass}}

A finite Dirac mass opens a gap in the bandstructure and changes the pseudospin, and therefore the overlap factor,
in a non-trivial fashion. Thus we can expect in general a quite different behavior for positive and negative mass. Yet,
for these differences to occur on the intrinsic scale and thus influence the interband plasmons, $\left|\xi_M\right|$ should be
of the order of 1. 
In the following, we study such large masses, both negative and positive, with p-h symmetry. 
While not experimentally relevant for HgTe QWs, it offers the possibility to study the effect of a topological bandstructure
on the electronic excitations, including plasmons.
We also note here that the dispersion of the BHZ model becomes
purely parabolic for the mass $\xi_{M}=-\frac{1}{4}$: $\epsilon_{X,\lambda}= \frac{\lambda}{4} +X^2\left(\lambda-\xi_{D}\right)$. 
In this limit, the polarization function, Eq.~(\ref{eq:Pi_par}), 
can be calculated analytically.


\subsubsection{Large, negative mass}

\begin{figure}
\includegraphics[width=4.2cm]{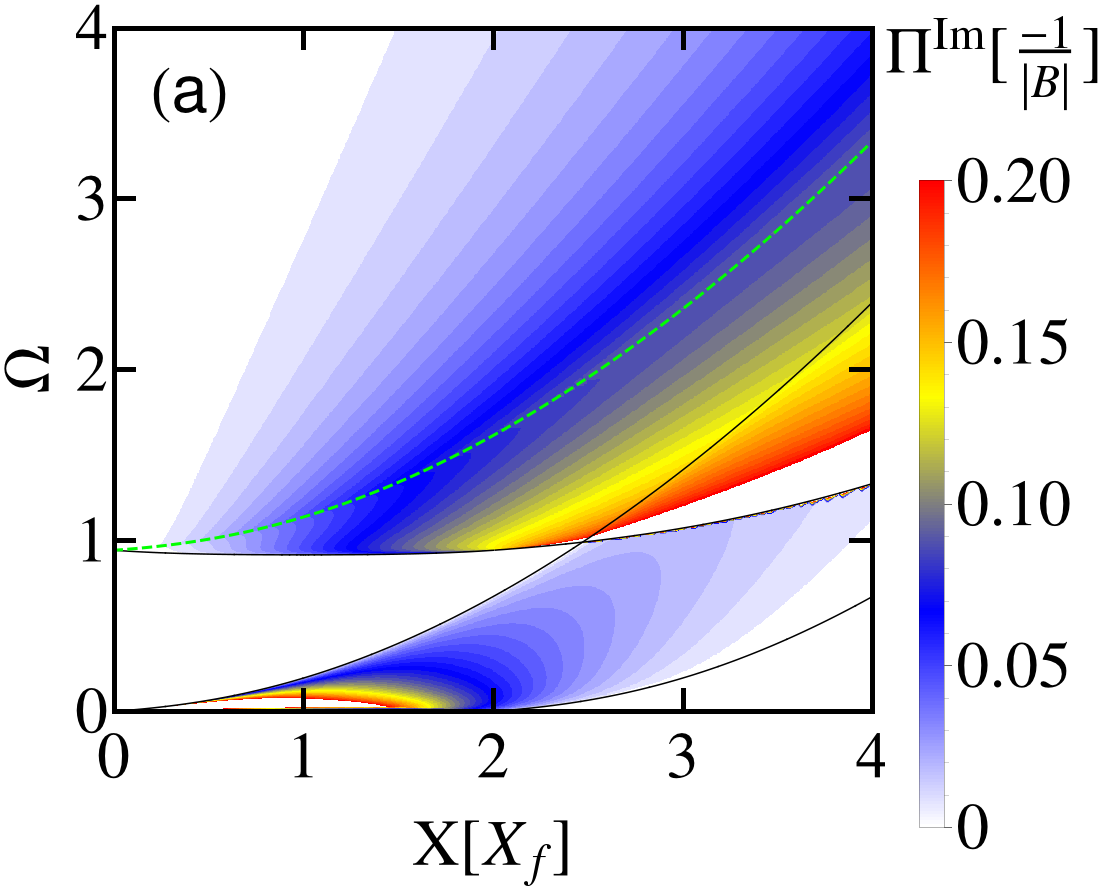}
\includegraphics[width=4.2cm]{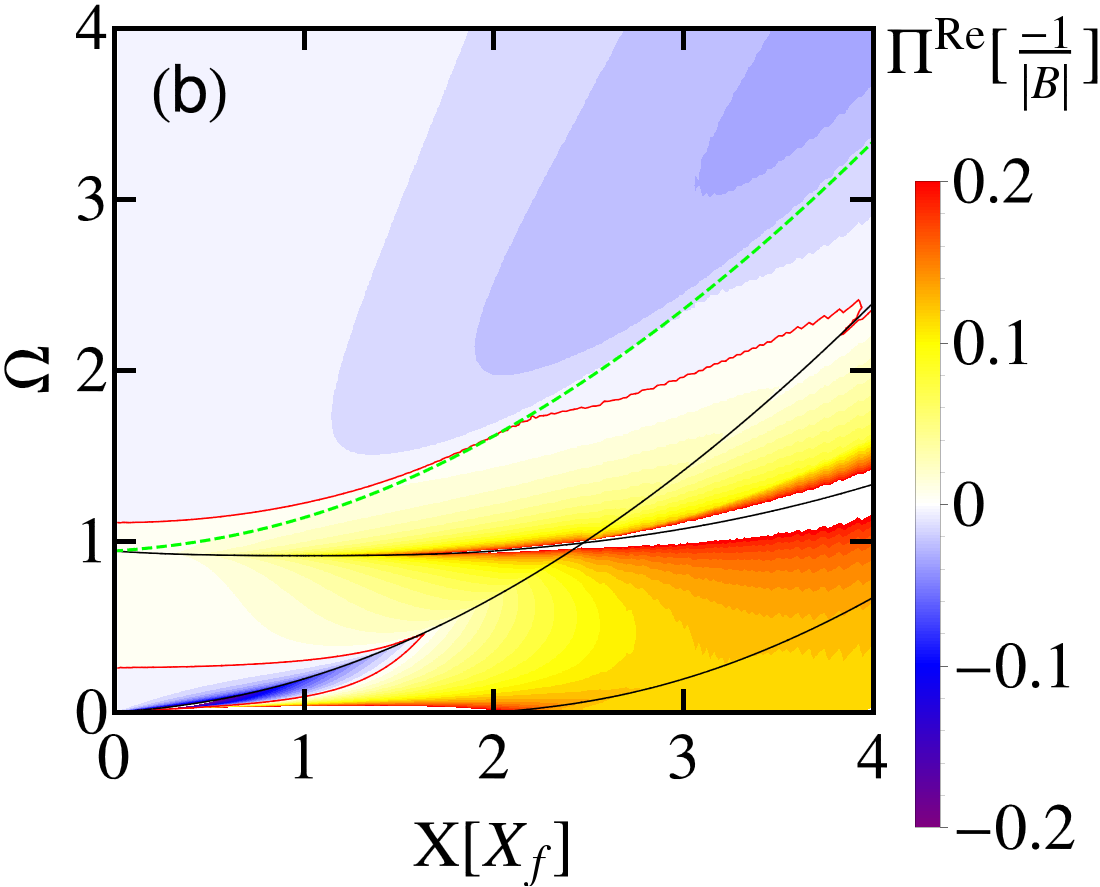}
\includegraphics[width=4.2cm]{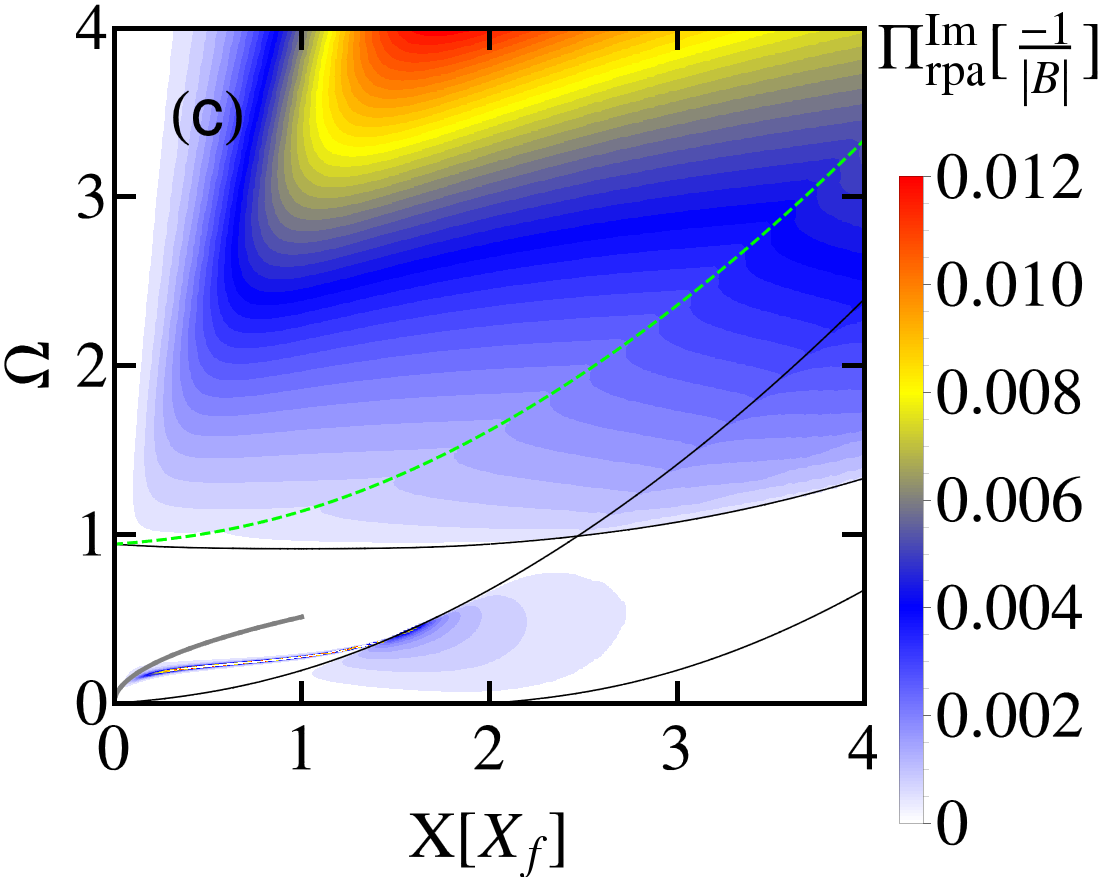}
\includegraphics[width=4.2cm]{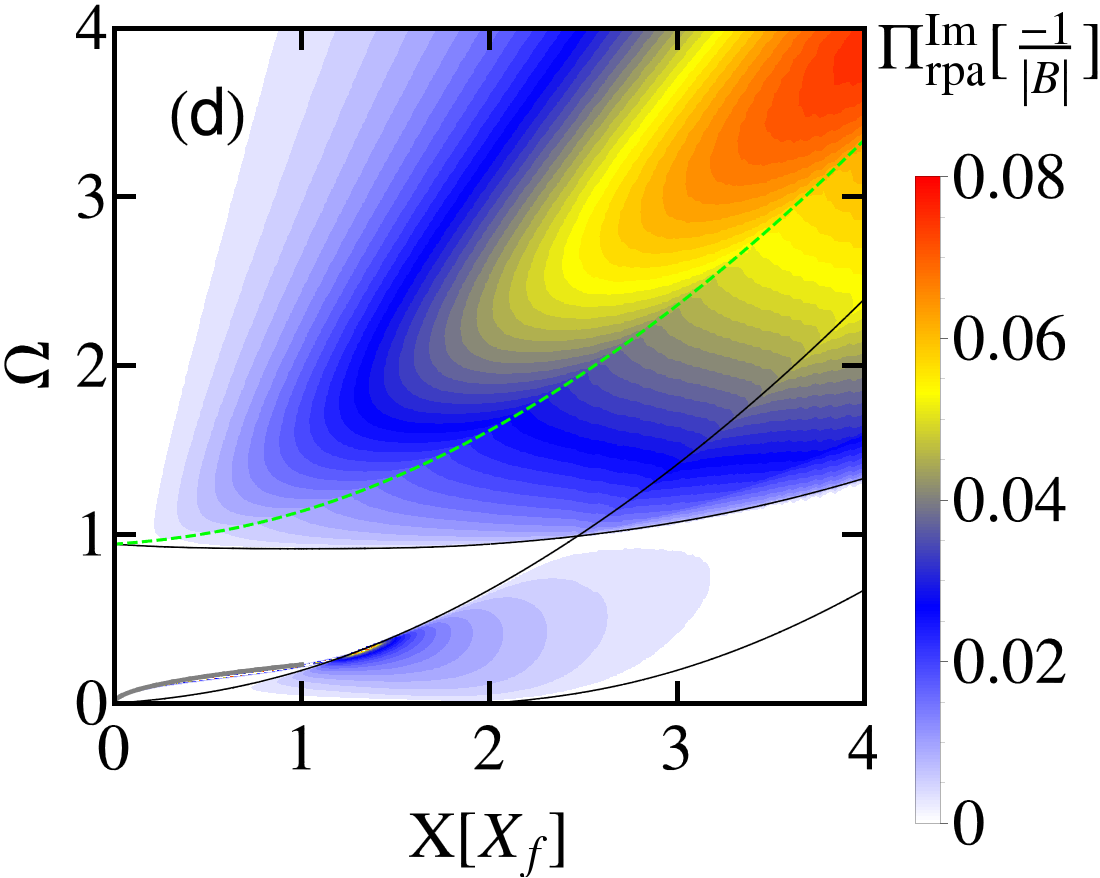}

\caption{(Color online) (a) Imaginary and (b) real part of the polarization $\Pi^{R}$. (c) and (d)
show $\Pi^{Im}_{rpa}$ with $\alpha=10$ and $\alpha=2$, respectively.
$\xi_{M}=-\frac{4}{9}$ and $X_{f}=0.33$ in all plots.
\label{fig:Im epsi BHZ MnAB Xf033}}

\end{figure}

For the parameters $X_{f}=0.33$ and $\xi_{M}=-\frac{4}{9}$ we plot the polarization $\Pi^{R}$ in
 Fig.~\ref{fig:Im epsi BHZ MnAB Xf033} (a) and (b).
The mass separates intra- and interband SPE regions for $X\lesssim2X_f$.
Compared to the massless cases of $X_f=0.1$, Fig.~\ref{fig:ImP BHZ Xf01},
and $X_f=1$, Fig.~\ref{fig:ImP BHZ Xf1}, the interband SPE spectrum is
enhanced due to the combination of enhanced overlap factor and low doping, thus small Fermi
blockade. Due to the flat bandstructure, even for $X_f=0.33$ the chemical potential is
just barely above the gap. An interesting consequence of this strong interband transition can be seen in panel (b),
where we find two distinct areas where $-\Pi^{Re}$ becomes negative. As a consequence,
inter- and intraband plasmons will always be separated, with the intraband plasmon being confined to
low energies. This stems from the fact that the electrons in the conduction band are pseudo-spin polarized, such 
that intraband excitations to much higher momenta and energies, where the pseudo-spin shows in the opposite direction, 
are not possible. 

This is confirmed in panels (c) and (d), where we plot $\Pi^{Im}_{rpa}$
with $\alpha=10$ and $\alpha=2$, respectively. All the spectral weight of the
intraband SPE region goes into the plasmon, which at least for $\alpha=2$
follows very well the $\sqrt{X}$ law. The interband spectrum is dominated
by the interband plasmon, having of course a much broader peak due to damping (finite $\Im\left[\Pi^R\right]$).

 The dashed, green line in the interband spectrum in Fig.~\ref{fig:Im epsi BHZ MnAB Xf033}
 indicates the energy, at which excitation processes going from momentum
 $\boldsymbol{X}+\boldsymbol{X_f}$ to $\boldsymbol{X_f}$ with $\boldsymbol{X}\Vert\boldsymbol{X_f}$ are possible, see
 black, dashed arrow in Fig.~\ref{fig:AB bandstructure} (a).
 Usually suppressed by the overlap factor, a large negative mass enhances the overlap of these
 excitations to near unity for small Fermi momenta. Fig.~\ref{fig:Im epsi BHZ MnAB Xf033} (b) and (d) show
 that the interband plasmons mainly occur above this line, indicating that the described
 excitation process is important for the collective excitation. As the process is forbidden by helicity
 in the pure Dirac system, it is one reason why the BHZ model supports intrinsic plasmons while the Dirac model does not.

\subsubsection{Large, positive mass}

\begin{figure}
\includegraphics[width=4.2cm]{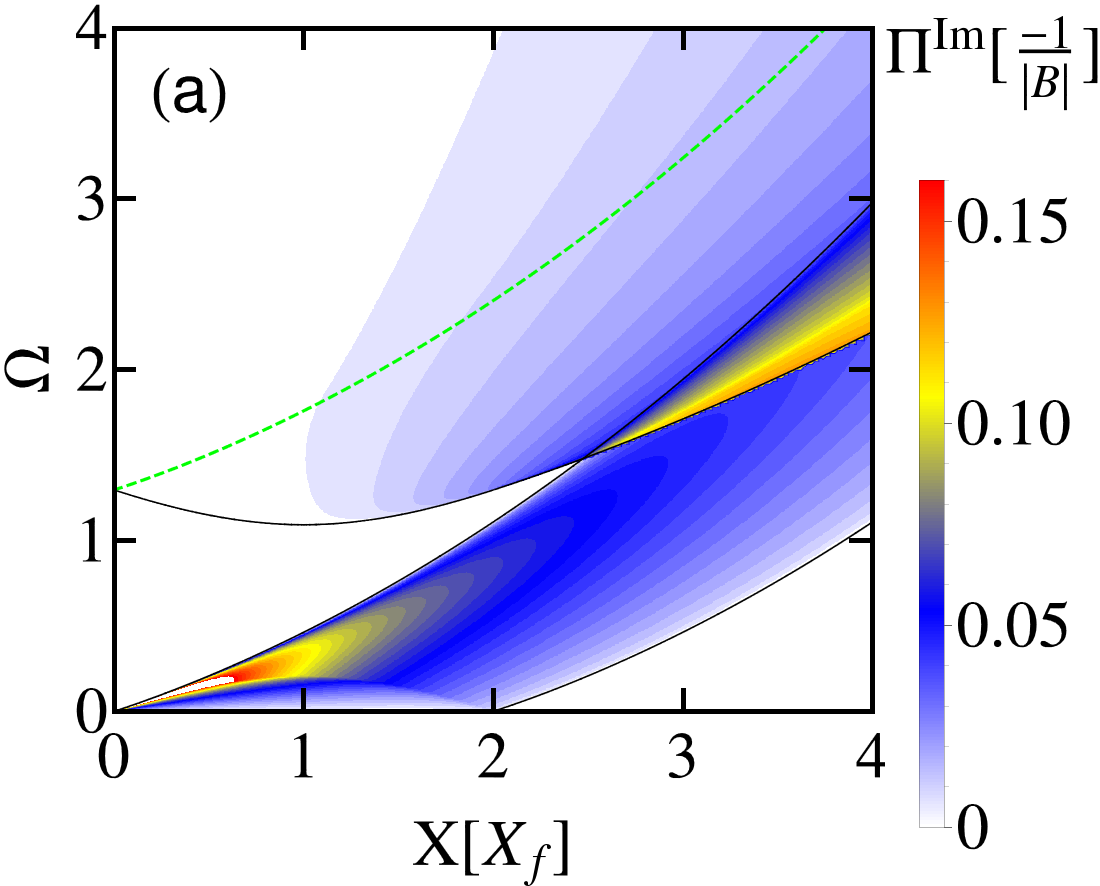}
\includegraphics[width=4.2cm]{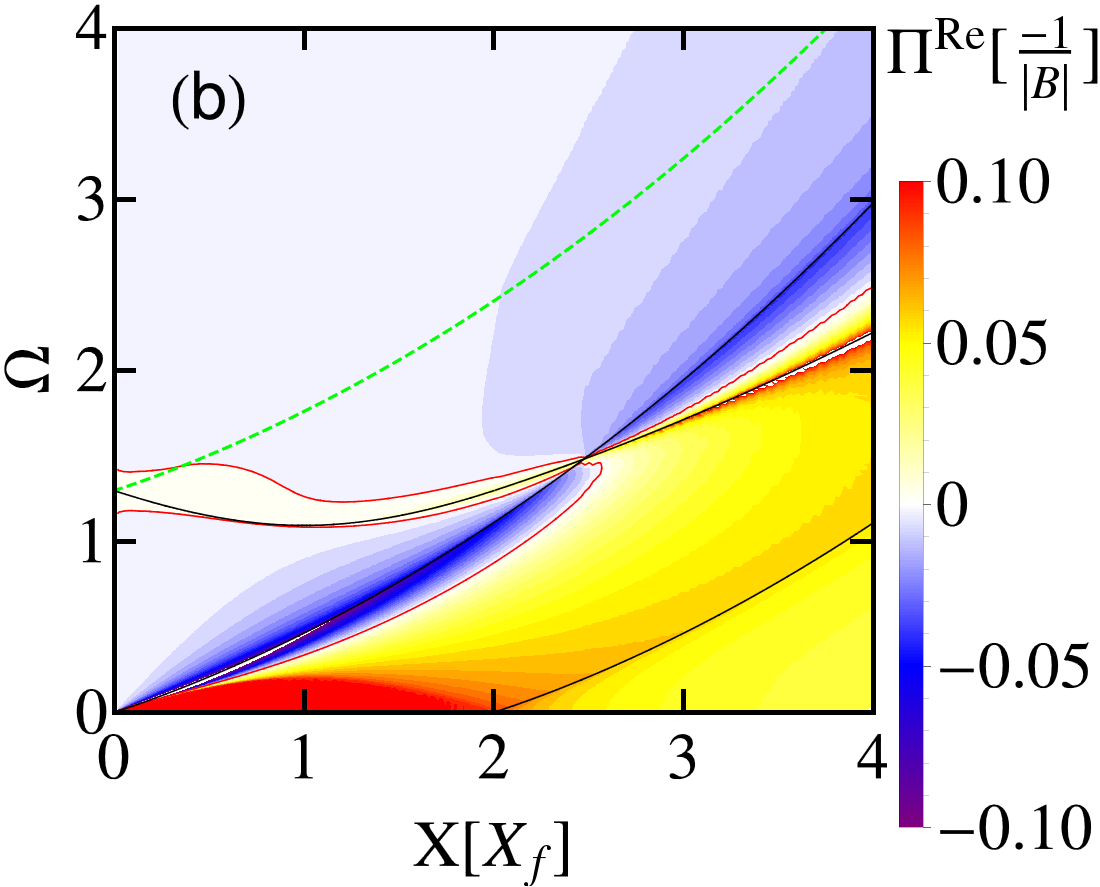}
\includegraphics[width=4.2cm]{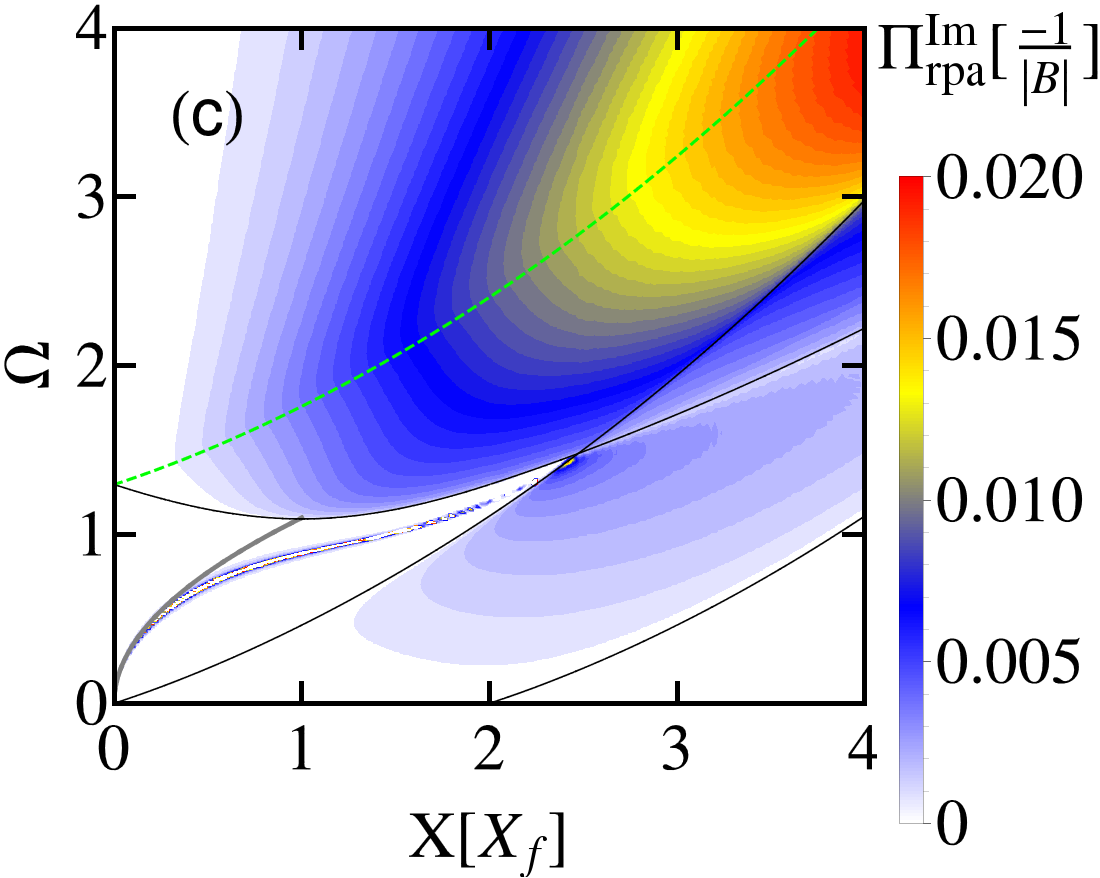}
\includegraphics[width=4.2cm]{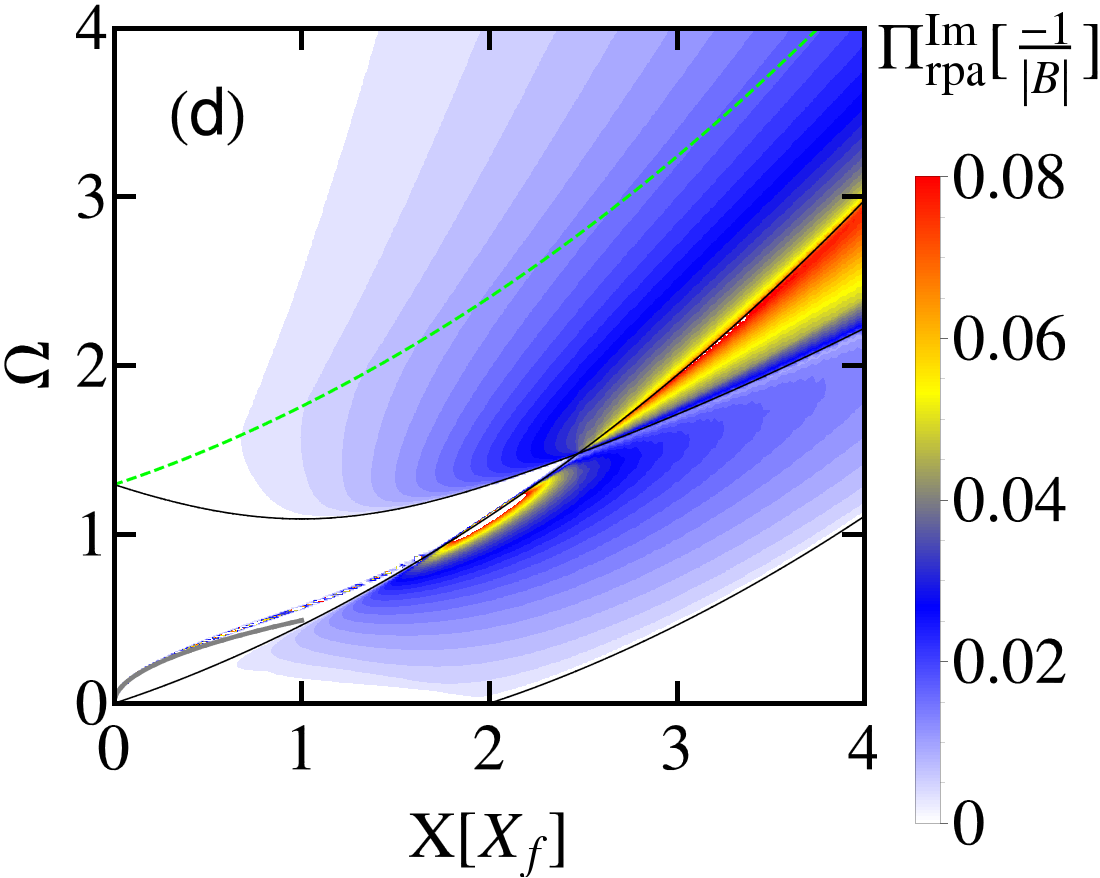}

\caption{(Color online) (a) Imaginary and (b) real part of the polarization $\Pi^{R}$. (c) and (d)
show $\Pi^{Im}_{rpa}$ with $\alpha=10$ and $\alpha=2$, respectively.
$\xi_{M}=\frac{4}{9}$ and $X_{f}=0.33$ in all plots.  \label{fig:Im epsi BHZ MAB Xf033}}

\end{figure}

For the parameters $X_{f}=0.33$ and $\xi_{M}=\frac{4}{9}$ we plot
the polarization in Fig.~\ref{fig:Im epsi BHZ MAB Xf033} (a) and (b).
Compared to the negative mass, the interband spectrum is much weaker. This is a
result of the lower overlap factor and the higher chemical potential (the bandstructure is not as flat as in the TI phase), leading to a stronger
Fermi blockade. For the real part of the polarization, this has the effect that
the two former distinct areas of sign reversal now almost merge. The interband excitations
are so weak that the minimum $-\Pi^{Re}$ always lies closely above the intraband
SPE region - indicating that it is the main source for plasmons.

In panel (c) for $\alpha=10$, one can identify both inter- and intraband plasmon. Interestingly, the polarization
is clearly higher in the pure interband SPE region than in the mixed inter- and intraband SPE spectrum, suggesting
that the latter one serve as an additional damping for the interband plasmon. Going to the smaller interaction strength 
$\alpha=2$ in panel (d), one finds just a single resonance following the upper boundary of the intraband SPE spectrum.
Thus we conclude that the interacting spectrum for moderate interaction strength is governed by just intraband plasmons. 
The interband excitations are too weak to support an additional plasmon but for very high interactions - a consequence 
of the effective decoupling of the bands by the overlap factor. 


  \section{Conclusion} \label{sec_con}

  We have analyzed the dynamical and static polarization properties in random phase approximation of Hg(Cd)Te quantum wells 
  described by the Bernevig-Hughes-Zhang (BHZ) model.
  In the static undoped limit, due to the presence of quadratic terms in the BHZ model and hence to the natural length scale $B/A$, 
  the induced charge density in response to a test charge has a finite spatial extent. 
  This is in contrast to the point-like screening charge obtained with the continuous Dirac model of graphene.  
  In the doped regime, we have observed Friedel oscillations with an intermediate decay behavior between the Dirac ($r^{-3}$) and the 2DEG ($r^{-2}$) cases.   

  The discussion of the full dynamical polarization function has been focused on the appearance of new interband plasmons due to the interplay of Dirac and Schr{\"o}dinger physics. 
  In principle, we expect these plasmons to appear in multiband systems where the imaginary part of the polarization function decays faster with energy than 
  the one in the Dirac case ($\omega^{-1}$), which is the case for the BHZ model (decay as $\omega^{-2}$). 
  These plasmons appear already in the undoped system at experimentally relevant parameters, but it is also possible to observe them in the doped regime, 
  where they coexists with the usual intraband plasmons. 
  This is favored by broken particle hole symmetry in the BHZ model, which allows for the presence of both a Dirac point and an inflection point in the bandstructure. 
  The behavior of these two collective modes is also influenced by the topology of the bandstructure.
  Indeed the two plasmons tend to merge into one another in a gapped trivial insulator, while they remain distinct resonances in the topological insulator phase. 
  We have shown that these new plasmons should appear for momenta and energies on the right order of magnitude for experimental techniques like Raman spectroscopy or
  electron loss spectroscopy on Hg(Cd)Te quantum wells. 
  The wide range of parameters considered in this paper, including the regime of topological trivial and non-trivial insulators, should make our results applicable 
  to all kinds of materials described by phenomenological models interpolating between Dirac and Schr{\"o}dinger fermion physics. 
  
  Throughout this article, we have only discussed bulk excitations of this peculiar two-dimensional system. 
  Hence, we have totally ignored the influence of edge states in the topologically non-trivial regime of the 
  model in the presence of physical boundaries. 
  An extension of our analysis to finite size systems might yield exciting new physics, where we expect an interplay of one-dimensional 
  and two-dimensional collective charge excitations.


  \begin{acknowledgements}

  We acknowledge interesting discussions with E. Hankiewicz, M. Polini, T. Stauber and financial support by the DFG (SPP1666 and the DFG-JST research unit {\it Topotronics}), 
  the Helmholtz Foundation (VITI), and the ENB Graduate School on Topological Insulators.

  \end{acknowledgements}



\end{document}